\documentclass[review,10pt]{elsarticle}

\usepackage{lineno}
\modulolinenumbers[5]

\usepackage{hyperref} 
\hypersetup{ 
  colorlinks   = true, 
  urlcolor     = blue, 
  linkcolor    = blue, 
  citecolor    = blue, 
} 
\usepackage[fleqn]{amsmath}
\usepackage{amsfonts}
\usepackage{amssymb}
\usepackage{amsthm}

\usepackage{pifont}
\usepackage{natbib}
\usepackage{geometry}
\usepackage{graphicx}
\usepackage{txfonts}

\usepackage[utf8]{inputenc}
\usepackage{caption}
\usepackage{subcaption}
\usepackage[usenames, dvipsnames]{color}
\usepackage{color,soul}
\setstcolor{red}

\biboptions{sort&compress}

\begin{document}

\begin{frontmatter}

\title{A Volume-of-Fluid method for variable-density, two-phase flows at supercritical pressure}

\author{Jordi Poblador-Ibanez\fnref{myfootnote1}\corref{mycorrespondingauthor}}
\ead{poblador@uci.edu}
\author{William A. Sirignano\fnref{myfootnote3}}
\address{University of California, Irvine, CA 92697-3975, United States}
\fntext[myfootnote1]{Graduate Student Researcher, Department of Mechanical and Aerospace Engineering.}
\fntext[myfootnote3]{Professor, Department of Mechanical and Aerospace Engineering.}


\cortext[mycorrespondingauthor]{Corresponding author}


\begin{abstract}
A two-phase, low-Mach-number flow solver is created and verified for variable-density liquid and gas with phase change. The interface is sharply captured using a split Volume-of-Fluid method generalized for a non-divergence-free liquid velocity and with mass exchange across the interface. Mass conservation to machine-error precision is achieved in the limit of incompressible liquid. This model is implemented for two-phase mixtures at supercritical pressure but subcritical temperature conditions for the liquid, as it is common in the early times of liquid hydrocarbon injection under real-engine conditions. The dissolution of the gas species into the liquid phase is enhanced, and vaporization or condensation can occur simultaneously at different interface locations. Greater numerical challenges appear compared to incompressible two-phase solvers that are successfully addressed for the first time: (a) local thermodynamic phase equilibrium (LTE) and jump conditions determine the interface solution (e.g., temperature, composition, surface-tension coefficient); (b) a real-fluid thermodynamic model is considered; and (c) phase-wise values for certain variables (e.g., velocity) are obtained via extrapolation techniques. The increased numerical cost is alleviated with a split pressure-gradient technique to solve the pressure Poisson equation (PPE) for the low-Mach-number flow. Thus, a Fast Fourier Transform (FFT) method is implemented, directly solving the continuity constraint without an iterative process. Various verification tests show the accuracy and viability of the current approach. Then, the growth of surface instabilities in a binary system composed of liquid \textit{n}-decane and gaseous oxygen at supercritical pressures for \textit{n}-decane is analyzed. Other features of supercritical liquid injection are also shown.
\end{abstract}

\begin{keyword}
supercritical pressure \sep phase change \sep phase equilibrium \sep atomization \sep volume-of-fluid \sep low-Mach-number compressible flow
\end{keyword}

\end{frontmatter}


\setlength\abovedisplayshortskip{0pt}
\setlength\belowdisplayshortskip{-5pt}
\setlength\abovedisplayskip{-5pt}

\section{Introduction}
\label{sec:intro}

A need to achieve elevated operating pressures exists in engineering applications involving chemical combustion of fuels (e.g., gas turbines, liquid propellant rockets). Optimization of combustion efficiency and energy conversion per unit mass of fuel points to the design of high-pressure combustion chambers. Examples include diesel or gas turbine engines, which operate in the range of 25 to 40 bar, or rocket engines, whose operating pressures range from 70 to 200 bar. Moreover, the mixing process of the fuel with the surrounding oxidizer, as well as its vaporization in the case of liquid fuels, also dictates the overall performance of the combustion process. \par 

Understanding the physics behind the mixing process of a liquid fuel (i.e., atomization and vaporization) is crucial to design the combustion chamber size and the injectors' shape and distribution properly. Many experimental and numerical studies addressing this issue have been performed at subcritical pressures. In this thermodynamic state, liquid and gas are easily identified, and both fluids can be approximated as incompressible fluids (except in transonic or supersonic regimes). However, well-known fuels such as diesel fuel, Jet-A, or RP-1 are based on hydrocarbon mixtures whose critical pressures are in the 20-bar range. Thus, actual operating conditions occur at near-critical or supercritical pressures for the liquid fuel. \par 

Experimental studies performed at these very high pressures show the existence of a thermodynamic transition where the liquid and gas become difficult to identify. Both phases present similar properties near the liquid-gas interface, which is rapidly affected by turbulence while being immersed in a variable-density layer~\cite{mayer1996propellant,h1998atomization,mayer2000injection,oschwald2006injection,chehroudi2012recent,falgout2016gas,crua2017transcritical,desouza2017sub,ayyappan2020study}. Therefore, this behavior has often been described in the past as a very fast transition of the liquid to a gas-like supercritical state, neglecting any role of two-phase interface dynamics~\cite{spalding1959theory,rosner1967liquid}. Nevertheless, evidence of a two-phase behavior at supercritical pressures exists based on requirements that liquid and gas should be in LTE at the interface~\cite{hsieh1991droplet,delplanque1993numerical,yang1994vaporization,delplanque1995transcritical,delplanque1996transcritical,poblador2018transient}. \par 

The observations from experiments at supercritical pressures (i.e., the resemblance with a gas-like turbulent jet) are consistent with fast atomization caused by the extreme environment and the failure of the experimental techniques to capture liquid structures accurately. Because liquid and gas look more alike near the interface, surface-tension forces are reduced, and gas-like liquid viscosities appear~\cite{yang2000modeling,poblador2019axisymmetric,poblador2021liquidjet}. Therefore, the interface may be subject to the rapid growth of small surface instabilities and fast distortion, which can cause an early breakup of very small droplets. Although some progress is being made toward new experimental techniques to capture the two-phase behavior in supercritical pressure environments~\cite{minniti2018ultrashort,minniti2019femtosecond,traxinger2019experimental,klima2020quantification}, experimental methods relying on traditional visualization techniques (e.g., shadowgraphy) might suffer from scattering and refraction issues due to the presence of a cloud of small droplets submerged in a variable-density fluid. Numerical results of incompressible liquid round jets and planar sheets under conditions similar to those found at supercritical pressures also support this reasoning~\cite{jarrahbashi2014vorticity,jarrahbashi2016early,zandian2017planar,zandian2018understanding,zandian2019length}. \par 

At supercritical pressures, diffusion layers with strong variations of fluid properties grow on both sides of the interface. To maintain phase equilibrium at the interface, the lighter gas species dissolution into the liquid is enhanced. Thus, mixture critical properties change near the interface and, in general, the mixture critical pressure is above the chamber pressure~\cite{poblador2018transient}. Delplanque and Sirignano~\cite{delplanque1993numerical,delplanque1995transcritical,delplanque1996transcritical} were among the first to adopt the term ``transcritical" to characterize situations in which two phases coexist because the fuel's pressure is supercritical, but the interface temperature is subcritical. These works looked at droplet vaporization at supercritical pressures and the implications on fuel combustion. Other researchers, such as Yang and Shuen~\cite{yang1994vaporization}, followed suit with similar studies. A two-phase interface can be maintained until sufficient heating increases the droplet temperature, at which point the liquid surface reaches the mixture critical point and transitions to diffusive-controlled phase mixing. Consequently, since critical pressure depends on composition, there can easily exist a fluid region where a subcritical domain is neighbored by a supercritical domain, with near-uniform pressure across both domains. \par

Dahms and Oefelein~\cite{dahms2013transition,dahms2015liquid,dahms2015non} and Dahms~\cite{dahms2016understanding} have extensively discussed and quantified the interface transition from a two-phase behavior to a continuum at supercritical conditions. At supercritical pressures, but subcritical temperatures, the interface phase transition region is only of a few nanometers thickness, which widens as temperature increases toward the mixture critical point. The examples provided do not show interface thicknesses larger than 8 nm for typical hydrocarbon-nitrogen mixtures. At lower interface temperatures, LTE is well established and diffusion layers of the order of micrometers quickly form around the interface~\cite{poblador2018transient,davis2019development,poblador2021selfsimilar}. As the interface temperature nears the mixture critical temperature, the interface enters the continuum domain if the molecular mean free path is substantially smaller than the interface thickness (i.e., Knudsen number below 0.1). Although larger temperatures increase the molecular mean free path, the high-pressure environment may reduce it. Once a continuum regime is achieved, the proper model for the interface is a diffuse region with sharp gradients, similar to the supercritical state, where classical LTE is no longer valid. Nonetheless, an analogy with compressive shocks in supersonic flows suggests that the interface can be considered a discontinuity under LTE or the proper set of boundary conditions with a jump in fluid properties across it. Note that the non-equilibrium layer thickness for a shock is at least an order of magnitude greater than the phase ``non-equilibrium" transition region, and a shock is treated as a discontinuity for practical purposes. The transition layer is described as a phase ``non-equilibrium" domain following the literature even though the thermodynamic equilibrium of the continuous fluid is maintained across the layer. The distinction between the two phases disappears, thereby breaking down classical phase-equilibrium laws. The ``non-equilibrium" terminology arises from the use of non-equilibrium molecular dynamics to model and study the thickening of the phase transition layer. Therefore, the existing semantics in the literature is not optimal. \par

Further evidence of this transcritical behavior and the interface transition to a diffuse layer is seen in Crua et al.~\cite{crua2017transcritical}, where experiments of fuel injection at a wide range of high pressures and temperatures relevant to diesel engines show the existence of droplets at supercritical pressures for the liquid fuel. These droplets are strongly affected by the mixing around them and the reduced surface tension before heating eventually causes a transition to a diffusive mixing. \par

The temperature range where two phases coexist at engine-relevant conditions decreases considerably with pressure, either because the mixture critical temperature drops below typical injection temperatures~\cite{poblador2018transient} or because the dense fluid has a molecular mean free path at least an order of magnitude shorter than the phase transition layer. Therefore, other works such as those from Zhang et al.~\cite{wang2017comprehensive,zhang2018supercritical} and Wang et al.~\cite{wang2019three} analyze the injection of liquid fuels at supercritical pressures and high temperatures without identifying a phase interface. The chamber pressure of 253 bar is well above the critical pressure of any of the analyzed fluids and two phases cannot coexist in the range of observed temperatures (i.e., above 490 K). \par 

Other numerical frameworks have recently been developed to address transcritical and supercritical flows based on a diffuse interface approach without surface-tension force~\cite{ruiz2016numerical,ma2017entropy,ma2019numerical,ma2019large,lagarza2019large,rodriguez2019complex,rodriguez2019simulation,matheis2020large,boyd2021diffuse,boyd2021numerical}. These works do not identify a phase interface under the assumption that the transport between the two phases is driven only by diffusion. Fluid properties vary continuously within a finite region and surface tension is neglected. Instead, these works focus on transcritical issues such as the pressure numerical oscillations generated by the equation of state near the critical point~\cite{boyd2021diffuse}. This diffuse approach could yield inaccurate results since two phases may coexist at transcritical conditions where the pressure is supercritical but not the temperature. Additionally, local variations in composition, temperature and pressure may lead to an unstable mixture and phase separation in a transcritical environment~\cite{jofre2021transcritical}. Therefore, two-phase dynamics (i.e., surface tension and phase change) may be important in certain flow regions. \par 

The intrinsic complexity of supercritical fluids presents a challenge to the scientific community. A modeling and numerical framework are needed to address liquid fuel injection at supercritical pressures. Evidence points to a transcritical two-phase behavior of the atomization process, at least at the liquid-core level and in a wide range of supercritical pressures~\cite{delplanque1995transcritical,delplanque1996transcritical,poblador2018transient,crua2017transcritical,jofre2021transcritical}. It may be possible that small droplets or liquid regions at very high temperatures do experience a transition to a supercritical fluid state. Some of the challenges of a two-phase model include: (a) the non-ideal fluid behavior at supercritical states must be captured via a thermodynamic model~\cite{yang2000modeling,bellan2000supercritical} (e.g., a real-gas equation of state); (b) LTE governs the state of the liquid-gas interface, while both phases exchange heat and mass. Moreover, the interface should be treated as a sharp discontinuity, where fluid properties such as density or viscosity are discontinuous, as well as the heat fluxes and concentration gradients into the interface; (c) the interface location must be appropriately captured; and (d) a computationally-efficient method is desired since the additional requirements to solve supercritical flows are costly. \par

Various interface-capturing methods exist, such as the Level-Set (LS) method by Sussman et al.~\cite{sussman1994level,sussman1998improved} and Osher and Fedkiw~\cite{osher2001level} and the Volume-of-Fluid (VOF) method~\cite{hirt1981volume,scardovelli1999direct}. A more comprehensive review of interface-capturing and interface-tracking methods can be found in Elghobashi~\cite{elghobashi2019direct}. The LS method has been widely used in the literature. It tracks a distance function to the interface and a sharp representation is achieved by combining it with a ghost fluid method (GFM)~\cite{fedkiw1999non,fedkiw1999ghost}. However, the GFM applied to compressible non-ideal fluids at supercritical pressures is not straightforward and the LS method suffers from numerical mass loss. That is, numerical errors in both the advection and reinitialization of the LS artificially displace the interface. This problem worsens in high-curvature regions relative to the mesh size. Therefore, it becomes difficult to address this issue in atomization simulations where a cascade process toward smaller liquid structures occurs. Even if the LS transport equation is discretized with high-order numerical schemes (e.g., WENO) or using the more accurate Gradient Augmented Level-Set method~\cite{nave2010gradient,anumolu2013gradient,trujillo2017distortion,anumolu2018gradient}, some degree of mass loss exists. \par

On the other hand, VOF methods that conserve mass to machine error exist~\cite{baraldi2014mass}. The VOF method tracks the volume occupied by the reference phase (i.e., the liquid) in all computational cells, and it handles vaporization or condensation naturally. The governing equations are solved with a sharp interface approach that only diffuses the interface in a region of the order of \(\mathcal{O}(\Delta x)\) by volume-averaging fluid properties at interface cells and including jump conditions as localized body forces~\cite{dodd2014fast,dodd2021vof}. The VOF is a better option than other diffuse-interface approaches based on the LS method, which impose a numerical interface thickness of \(\mathcal{O}(3\Delta x)\) that can overlap with the actual diffusion layers~\cite{sussman1994level,sussman1998improved,osher2001level}. Therefore, the VOF method is preferred here. \par 

VOF methods such as those developed by Baraldi et al.~\cite{baraldi2014mass}, Dodd and Ferrante~\cite{dodd2014fast} and Dodd et al.~\cite{dodd2021vof} are good starting points to develop numerical tools for transcritical atomization. Even though these works are developed for incompressible liquids with or without phase change, they are computationally efficient and satisfy mass conservation while keeping a sharp interface. Recently, this methodology has also been extended to two-phase flows with phase change where the gas phase is compressible~\cite{dodd2021analysis}. Few works address the extension of VOF methods to compressible liquids~\cite{bo2014volume,duret2018pressure,denner2018pressure}, much less address the non-ideal thermodynamics at high pressures. \par 

This paper introduces a numerical methodology to solve low-Mach-number, compressible two-phase flows at supercritical pressures. A discussion on the applicability and limitations of the proposed model is presented in Section~\ref{sec:descr}. The necessary governing equations are presented in Section~\ref{sec:governing} and a summary of the thermodynamic model used to represent non-ideal fluids is presented in Appendix~\ref{apn:thermo}. In Section~\ref{sec:interface}, we extend the VOF method from Baraldi et al.~\cite{baraldi2014mass} to track compressible liquids with phase change. The main bulk of the numerical approach and algorithm to solve the governing equations is presented in Section~\ref{sec:numerical}. To ease the computational cost, the pressure-correction method by Dodd and Ferrante~\cite{dodd2014fast} for incompressible flows is extended to low-Mach-number flows with phase change. Thus, an FFT method can be used to solve the PPE. Finally, Section~\ref{sec:results} presents some test problems to verify the methodology and determine its viability to simulate liquid injection at supercritical pressures. \par

\section{Model description and physical limitations}
\label{sec:descr}

The theoretical model and the numerical approach described in this paper aim to address the solution of weakly-compressible two-phase flows in a high-pressure thermodynamic regime (i.e., supercritical for the liquid), but still below the mixture critical point. That is, the interface between both fluids is at a sufficiently low temperature relative to the mixture critical temperature. As previously highlighted, two phases can be sustained in this transcritical domain as LTE defines the interface state and the enhanced mixing in the liquid phase modifies the critical properties of the mixture. A comprehensive analysis of the thermodynamic complexity of this injection environment representative of real engines is presented in Jofre and Urzay~\cite{jofre2021transcritical}. As shown in Figure 1 of Jofre and Urzay~\cite{jofre2021transcritical}, a two-phase problem with diminished surface tension and finite energy of vaporization may exist near the injector before the interface temperature reaches the mixture critical point (i.e., diffusional critical point). Close to the mixture critical point, the interface enters a diffuse and continuous transition from the liquid to the gas phase with sharp gradients confined in the nanoscale (i.e., works by Dahms and Oefelein~\cite{dahms2013transition,dahms2015liquid,dahms2015non,dahms2016understanding}). Beyond the mixture critical point, no further distinction between liquid and gas can be made and diffusive mixing between both fluids occurs. \par

The proposed model does not address the interface transition to a supercritical state or the interface behavior near the mixture critical point where classical LTE does not apply. Also, the combustion chemical reaction between fuel and oxidizer is not considered. Instead, it focuses only on the early stages of the fuel injection process, where two phases coexist. Previous works have shown that the interface equilibrium temperature is very close to the liquid bulk temperature~\cite{poblador2018transient,davis2019development,poblador2021selfsimilar}, while typical liquid hydrocarbon fuels are injected at relatively low temperatures. Heavy hydrocarbon mixtures have high critical temperatures (e.g., \(T_c=617.7\) K for \textit{n}-decane); therefore, the interface can stay away from the mixture critical point before substantial mixing and heating occurs, either caused by a sufficiently hot oxidizer stream like in Jofre and Urzay~\cite{jofre2021transcritical} or by downstream combustion. In that case, the fuel-oxidizer mixing can be driven by two-phase atomization under high pressures before the interface transition to a supercritical state occurs. Thus, such a resolved two-phase model is a powerful tool to analyze the physical phenomena driving the liquid fuel early mixing stage at engine-relevant conditions and better understand the physical setup of downstream phenomena. \par

Note that turbulence models are not considered in the governing equations presented in Section~\ref{sec:governing}. Liquid atomization involves a transition from laminar to turbulent flow, but where the early times of the liquid deformation cascade can be modeled following a direct numerical approach. A sufficiently fine mesh may capture the quasi-laminar flow as liquid structures form, and ligaments and droplets break up. \par 

The thermodynamic range of two-phase coexistence is a particular feature of a given fuel-oxidizer mixture and a careful individual analysis to determine whether a two-phase solver can represent reality and under what conditions is needed. Further justification of the proposed approach for the type of mixtures analyzed in Section~\ref{sec:results} is shown in Figure~\ref{fig:diagrams}. Using the thermodynamic model presented in Appendix~\ref{apn:thermo}, phase-equilibrium diagrams are plotted for the binary mixture of \textit{n}-decane/oxygen as a function of interface temperature and pressure (see Figure~\ref{subfig:pheq_diagram}), which show the thermodynamic region of two-phase interface coexistence. The \textit{n}-decane/oxygen mixture has been chosen as representative of hydrocarbon-based liquid fuels used in many engineering applications and other fuel-oxidizer configurations should behave similarly (e.g., diesel-air). The interface temperature range analyzed in this work is bounded by the bulk temperature of each phase without chemical reaction. The examples provided in Section~\ref{sec:results} for the binary mixture of \textit{n}-decane/oxygen have a liquid bulk temperature of 450 K and a gas bulk temperature of 550 K. Therefore, the interface equilibrium temperature is expected to remain close to 450 K and, in cases with strong mixing, the temperature upper bound is limited by the gas bulk temperature. For all analyzed pressures, the interface cannot reach a supercritical state. Even at very high pressures (i.e., 150 bar), the mixture critical temperature is very close to the pure hydrocarbon critical temperature~\cite{poblador2018transient,davis2019development,poblador2021selfsimilar}. \par

\begin{figure}[h!]
\centering
\begin{subfigure}{.5\textwidth}
  \centering
  \includegraphics[width=1.0\linewidth]{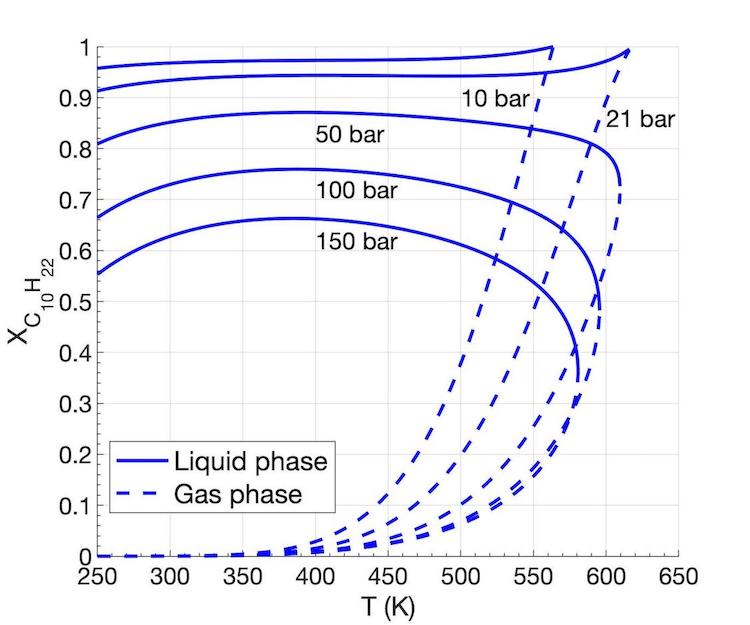}
  \caption{}
  \label{subfig:pheq_diagram}
\end{subfigure}%
\begin{subfigure}{.5\textwidth}
  \centering
  \includegraphics[width=1.0\linewidth]{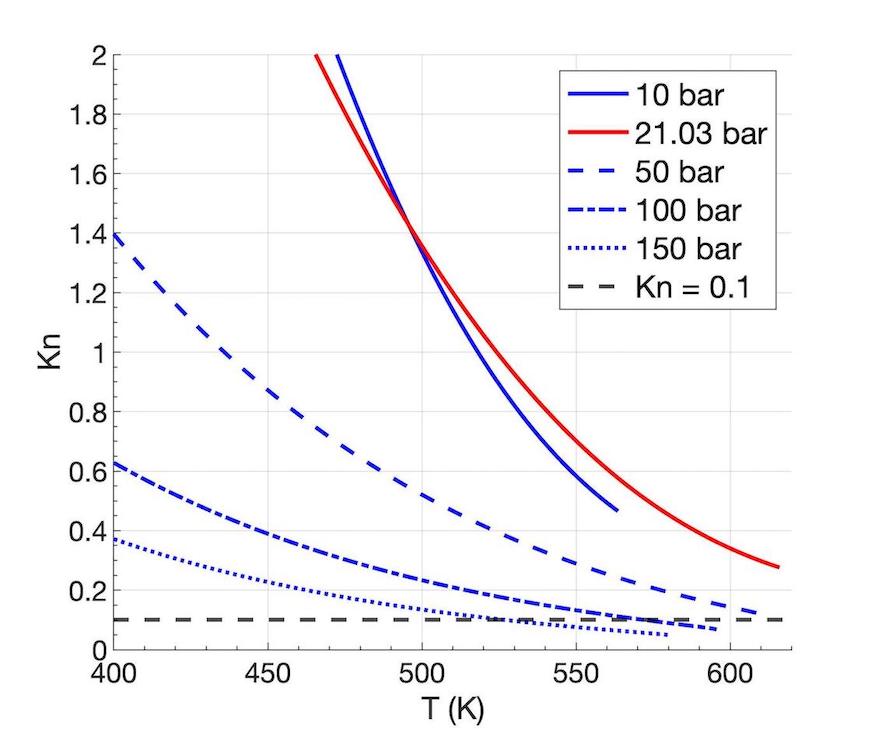}
  \caption{}
    \label{subfig:Kn_diagram}
\end{subfigure}%
\caption{Phase equilibrium solution for the binary mixture of \textit{n}-decane/oxygen as a function of interface temperature and pressure. (a) \textit{n}-decane mole fraction at the interface; and (b) estimated Knudsen number at the interface using the vapor equilibrium composition.}
\label{fig:diagrams}
\end{figure}

Moreover, the Knudsen number criteria detailed in the works by Dahms and Oefelein~\cite{dahms2013transition,dahms2015liquid,dahms2015non} and Dahms~\cite{dahms2016understanding} has to be analyzed to determine whether the interface has entered or not the continuum domain. Here, a rough estimate of the interface thickness, \(l_\Gamma\), growth with temperature is obtained by assuming an exponential growth similar to Figure 6 from Dahms and Oefelein~\cite{dahms2015liquid}. The molecular mean free path of the vapor equilibrium solution is estimated using~\cite{dahms2013transition} \(\Lambda = (k_BT)/(\sqrt{2}\pi p d^2)\) where \(k_B\) is the Boltzmann constant, \(T\) the interface temperature, \(p\) the interface pressure and \(d=\sum_{i=1}^{N=2}X_i d_i\) the average molecule kinetic diameter. The kinetic diameter is chosen as the representative molecule diameter since its definition is directly linked to the molecular mean free path. For oxygen, \(d_{O_2}=0.346\) nm~\cite{ismail2015gas}, and for \textit{n}-decane, \(d_{C_{10}H_{22}}=0.485\) nm~\cite{pazzona2009comparative}. \par 

The Knudsen number, Kn\(=\Lambda/l_\Gamma\), is plotted in Figure~\ref{subfig:Kn_diagram} as a function of the interface equilibrium solution at various temperatures and pressures. The continuum criteria defined in Dahms and Oefelein~\cite{dahms2013transition} of Kn \(<0.1\) is only troublesome for the 150 bar case, where the estimates suggest the interface might enter the continuum domain at temperatures above 525 K. Nevertheless, the interface equilibrium temperature remains well below this threshold in the cases presented in Section~\ref{sec:results}, and the LTE interface model can be justified. As seen in Figure~\ref{fig:150_30A_int_4mus} where some results of a three-dimensional symmetric planar jet at 150 bar are shown, only a few interface locations are close to the estimated theoretical limit where the interface is in phase ``non-equilibrium". \par

Other reported concerns for this type of flow are also considered. Stierle et al.~\cite{stierle2020selection} have shown that the interface thermal resistivity or heat transfer efficiency has to be considered. For a large thermal resistivity, a substantial temperature jump exists across the interface and the phase non-equilibrium transition region must be modeled. Note that this condition does not imply that the transition layer has entered the continuum as in the works by Dahms and Oefelein~\cite{dahms2013transition,dahms2015liquid,dahms2015non} and Dahms~\cite{dahms2016understanding}. In this work, thermal conductivities are small at the interface, but the expected temperature jump across the interface is negligible when compared to the interface equilibrium temperature~\cite{davis2019development}. Moreover, the stability of the mixture (e.g., diffusional stability) for non-ideal mixtures at high pressures is responsible for phase separation and the reappearance of a two-phase interface~\cite{jofre2021transcritical}. No issues have been found during the simulation of various tests, which suggests that the composition obtained from the LTE interface model provides a stable boundary for the mixing occurring in both phases in the low-Mach-number environment for which the thermodynamic pressure remains constant. However, phase separation is possible in more complex scenarios where the local temperature and pressure change sharply and the local mixture composition becomes unstable. \par 

Lastly, various issues can be addressed in future works to improve the model and its performance, as well as widen the thermodynamic domain where it can be applied. For instance, the spurious currents generated around the interface under the VOF framework must be addressed, find means to reduce the added computational cost and handle higher interface temperatures with a diffuse phase transition model near the mixture critical point or a transition to a supercritical interface, similar to how Zhu and Aggarwal~\cite{zhu2000transient} and Aggarwal et al.~\cite{aggarwal2002transcritical} handle the transition from a two-phase interface to supercritical diffuse mixing in transcritical droplet studies. Nonetheless, the paper aims to lay out a clear framework to study two-phase flows at supercritical pressures and the early atomization of liquid fuels at engine-relevant conditions. \par

\section{Governing equations}
\label{sec:governing}

The governing equations of fluid motion for compressible two-phase flows are the continuity equation
\begin{equation}
\label{eqn:cont}
\frac{\partial \rho}{\partial t} + \nabla \cdot (\rho\vec{u})=0
\end{equation}

\noindent
and the momentum equation
\begin{equation}
\label{eqn:mom}
\frac{\partial}{\partial t}(\rho \vec{u})+\nabla \cdot (\rho \vec{u}\vec{u}) = -\nabla p + \nabla \cdot \bar{\bar{\tau}}
\end{equation}


\noindent
where \(\rho\) and \(\vec{u}\) are the fluid density and velocity, respectively, and \(p\) is the pressure. \(\bar{\bar{\tau}}=\mu [\nabla\vec{u}+\nabla\vec{u}^\text{T}-\frac{2}{3}(\nabla\cdot\vec{u})\bar{\bar{I}}]\) is the viscous stress dyad, where \(\mu\) represents the dynamic viscosity of the fluid and \(\bar{\bar{I}}\) represents the identity dyad. For simplicity, a Newtonian fluid under Stokes' hypothesis is assumed. However, fluid behavior is far from ideal at very high pressures, where both liquid and gas are very dense fluids and compressible. Thus, models to estimate the bulk viscosity or second coefficient of viscosity (e.g., Jaeger et al.~\cite{jaeger2018bulkvisc}) might be considered in future works to revise the use of the Stokes' hypothesis. A comparison with the current fluid modeling will shed more light on the issue. \par 

Furthermore, governing equations for the species concentration and for the energy of the fluid are needed. Only binary mixtures are considered in this work for simplicity, but the numerical framework can easily be extended to multi-component mixtures. For a binary mixture, only one species continuity equation is required. As shown in Section~\ref{sec:results}, the focus is on problems where the liquid phase begins as a pure hydrocarbon fuel (i.e., \(Y_2=Y_F=1\)) while the gas phase initially is pure oxygen (i.e., \(Y_1=Y_O=1\)). Choosing the oxidizer species, where \(\sum_{i=1}^{N=2} Y_i=Y_O+Y_F = 1\), the species transport equation is
\begin{equation}
\label{eqn:spcont}
\frac{\partial}{\partial t}(\rho Y_O) + \nabla\cdot(\rho Y_O \vec{u}) = \nabla \cdot (\rho D_m \nabla Y_O)
\end{equation}

Here, a mass-based Fickian diffusion coefficient, \(D_m\), is chosen to model the diffusion flux due to concentration gradients. Thermo-diffusion (i.e., Soret effect) is neglected. Although a high-pressure model is used to estimate this transport coefficient (see Appendix \ref{apn:thermo}), the use of more complex models to evaluate mass diffusion will be investigated in the future (i.e., generalized Maxwell-Stefan formulation for multicomponent mixtures).  \par 

The energy equation is written as an enthalpy transport equation as 
\begin{equation}
\label{eqn:energy}
\frac{\partial}{\partial t}(\rho h) + \nabla\cdot(\rho h \vec{u}) = \nabla \cdot \bigg(\frac{\lambda}{c_p}\nabla h \bigg) + \sum_{i=1}^{N=2} \nabla \cdot \Bigg(\bigg[\rho D_m - \frac{\lambda}{c_p}\bigg]h_i \nabla Y_i\Bigg)
\end{equation}

\noindent
where \(h\) is the mixture specific enthalpy, \(\lambda\) is the thermal conductivity, and \(c_p\) is the specific heat at constant pressure. Pressure terms and viscous dissipation in the energy equation are neglected under the low-Mach-number configuration analyzed in this work. The substitution \(\lambda \nabla T = (\lambda/c_p)\nabla h - \sum_{i=1}^{N=2}(\lambda/c_p)h_i\nabla Y_i\) is made and Fickian diffusion is considered for the energy transport via mass diffusion. Moreover, this term demands the partial derivative of mixture enthalpy with respect to mass fraction, \(h_i\equiv\partial h/\partial Y_i\). Note that this nomenclature must not be confused with the standard definition of partial enthalpy (i.e., species' enthalpy at the same temperature and pressure as the mixture). Only for the ideal case, both approaches would be equivalent, and the mixture enthalpy would be equal to the weighted sum of the individual enthalpies of each species at the same temperature and pressure. For the convection and conduction terms, the proper formulation for mixture enthalpy at high pressures is used. \par

\subsection{Interface matching relations}
\label{subsec:matching}

The solution of the governing equations is valid within each phase. However, a discontinuity in thermodynamic and transport properties exists across the interface. Thus, interface matching relations must be defined and embedded into the solution of each governing equation to connect both phases. This section defines such relations, while Section~\ref{sec:numerical} addresses the integration of these matching relations into the numerical method. \par 

The interface, denoted by \(\Gamma\), separates the liquid and gas domains. The subscripts \(l\) and \(g\) refer to the liquid phase and the gas phase, respectively. The normal and tangential unit vectors at any interface location are represented by \(\vec{n}\) and \(\vec{t}\), respectively, with \(\vec{n}\) defined positive pointing toward the gas phase. The mass flux per unit area across the interface, \(\dot{m}'\), is positive when vaporization occurs and negative when condensation occurs. The vaporization or condensation rate is computed from
\begin{equation}
\label{eqn:massflux}
\dot{m}' = \rho_l (\vec{u}_l-\vec{u}_{\Gamma})\cdot \vec{n}=\rho_g (\vec{u}_g-\vec{u}_{\Gamma})\cdot \vec{n}
\end{equation}

\noindent
where \(\vec{u}_\Gamma\) is the interface velocity, which can vary along the interface. \par 

If \(\dot{m}'\) is non-zero, then the interface moves with respect to the fluid. In this case, the normal component of the velocity field is discontinuous across the interface while the tangential component is continuous. These conditions are given by
\begin{equation}
\label{eqn:veljump1}
(\vec{u}_g-\vec{u}_l) \cdot \vec{n} = \bigg(\frac{1}{\rho_g}-\frac{1}{\rho_l}\bigg)\dot{m}' \quad ; \quad \vec{u}_g \cdot \vec{t} = \vec{u}_l \cdot \vec{t}
\end{equation}

A pressure jump across the interface is caused by a combination of surface-tension force, mass exchange across the interface and a mismatch in the normal viscous stresses, as seen in Eq.~(\ref{eqn:momjump1}). \(\sigma\) represents the surface-tension coefficient and \(\kappa\) is the interface curvature, defined positive with a convex liquid shape (i.e., \(\kappa=\nabla\cdot\vec{n}\)).
\begin{equation}
\label{eqn:momjump1}
p_l - p_g = \sigma \kappa + (\bar{\bar{\tau}}_l \cdot \vec{n}) \cdot \vec{n} - (\bar{\bar{\tau}}_g \cdot \vec{n} ) \cdot \vec{n}+\bigg(\frac{1}{\rho_g}-\frac{1}{\rho_l}\bigg)(\dot{m}')^2
\end{equation}

Because the interface properties may vary along the interface, the shear stress across the interface will not be continuous in the presence of a surface-tension coefficient gradient. The tangential stress balance is given by
\begin{equation}
(\bar{\bar{\tau}}_g \cdot \vec{n})\cdot \vec{t}-(\bar{\bar{\tau}}_l \cdot \vec{n}) \cdot \vec{t} = \nabla_s \sigma \cdot \vec{t}
\label{eqn:momjump2}
\end{equation}

\noindent 
where \(\nabla_s=\nabla-\vec{n}(\vec{n}\cdot\nabla)\) is the surface gradient. While the normal force \(\sigma \kappa\) in Eq.~(\ref{eqn:momjump1}) tends to minimize surface area per unit volume and smooth the liquid surface in two-dimensional structures, the surface-tension coefficient gradient along the interface drives the flow toward regions of higher surface-tension coefficient. Smoothing can also occur in three dimensions, but surface tension is also responsible for ligament thinning and neck formation leading to liquid breakup. \par 

Matching conditions for the species continuity equation and the energy equation become, respectively,
\begin{equation}
\label{eqn:spcontmatch}
\dot{m}'(Y_{O,g}-Y_{O,l}) = (\rho D_m \nabla Y_O)_g \cdot \vec{n} - (\rho D_m \nabla Y_O)_l \cdot \vec{n}
\end{equation}
\noindent
and
\begin{equation}
\label{eqn:energymatch}
\begin{split}
\dot{m}'(h_g-h_l) = \bigg(\frac{\lambda}{c_p}\nabla h\bigg)_g \cdot \vec{n} - \bigg(\frac{\lambda}{c_p}\nabla h\bigg)_l \cdot \vec{n} &+ \Bigg[\bigg(\rho D_m - \frac{\lambda}{c_p}\bigg)(h_O-h_F) \nabla Y_O\Bigg]_g \cdot \vec{n} \\
&- \Bigg[\bigg(\rho D_m - \frac{\lambda}{c_p}\bigg)(h_O-h_F) \nabla Y_O\Bigg]_l \cdot \vec{n}
\end{split}
\end{equation}

\noindent
where the energy matching equation has been simplified for the binary-mixture configuration. \par 

Phase-equilibrium relations provide a necessary thermodynamic closure for the interface matching. LTE is imposed through an equality in chemical potential for each species \(i\) on both sides of the interface. This condition can be expressed in terms of an equality in fugacity~\cite{soave1972equilibrium,poling2001properties}, \(f_i\), as
\begin{equation}
\label{eqn:pheq}
f_{li}(T_l,p_l,X_{li}) = f_{gi}(T_g,p_g,X_{gi})
\end{equation}

\noindent
where fugacity is a function of temperature, pressure and mixture composition. From a thermodynamic point of view, the pressure jump across the interface due to surface tension is negligible and pressure is treated constant and equal to the chamber value for phase-equilibrium purposes (i.e., \(p_l \approx p_g \approx p_{\text{ch}}\)). As explained later in Section~\ref{sec:numerical}, the thermodynamic pressure is assumed to be constant for low-Mach-number compressible flows and dynamic pressure variations are related to fluid motion but have little effect on fluid properties. Under this assumption, phase equilibrium can be expressed using the fugacity coefficient, \(\Phi_i \equiv f_i/pX_i\), as \(X_{li}\Phi_{li}=X_{gi}\Phi_{gi}\) where \(X_i\) represents the mole fraction of species \(i\). \par 

Furthermore, the interface presents a negligible thickness of the order of nanometers~\cite{dahms2013transition,dahms2015liquid,dahms2016understanding} while mass, momentum and energy quickly diffuse across regions of the order of micrometers around the interface~\cite{poblador2018transient,davis2019development,poblador2021selfsimilar}. Thus, the interface thickness is neglected in the present work and temperature is assumed continuous (i.e., \(T_g=T_l=T_\Gamma\)). This assumption simplifies the LTE solution and a mixture composition can readily be obtained on each side of the interface. The validity and limitations of this interface model for the mixtures considered in this paper have been discussed in Section~\ref{sec:descr}. \par

\subsection{Thermodynamic model}
\label{subsec:thermo}

The previous set of governing equations are coupled to a thermodynamic model based on a volume-corrected Soave-Redlich-Kwong (SRK) cubic equation of state~\cite{lin2006volumetric} and various models and correlations to estimate fluid and transport properties for the non-ideal fluid~\cite{poling2001properties,chung1988generalized,leahy2007unified}. For the low-Mach-number flows analyzed in this work, the thermodynamic pressure is assumed uniform. A summary of the thermodynamic model is presented in Appendix~\ref{apn:thermo} (i.e., details on the SRK equation of state, models to evaluate transport properties and the surface-tension coefficient) and extensive details are available in Davis et al.~\cite{davis2019development}. \par

\section{Interface model}
\label{sec:interface}

The accurate solution of the location and geometrical properties of the interface separating two immiscible fluids is of utmost importance in a two-phase fluid solver. In this work, a compressible extension of the VOF method is used to advect and capture the interface (Subsection~\ref{subsec:VoF}) and evaluate its geometrical properties (Subsection~\ref{subsec:VoFgeom}). The method applies to cases where density on both sides of the interface is variable without regard to whether the variations are dependent on pressure, composition or temperature. At high pressures, the dissolution of lighter gas species into the liquid phase is enhanced, thus causing the liquid volume to expand near the interface. That is, in addition to thermal expansion. Moreover, phase change is an essential feature of high-pressure environments, where vaporization or condensation can occur simultaneously at different locations along the interface~\cite{poblador2019axisymmetric}. This behavior depends on the LTE and the balancing of the mass, momentum and energy fluxes at the liquid-gas interface, as described in Subsection~\ref{subsec:matching}. \par 

\subsection{The Volume-of-Fluid method for compressible liquids}
\label{subsec:VoF}

The VOF method~\cite{hirt1981volume,scardovelli1999direct} advects a characteristic function, \(\chi(\vec{x},t)\), with the fluid velocity, \(\vec{u}\), following Eq.~(\ref{eqn:characVoF}). \(\chi=1\) in the reference phase (i.e., liquid phase in the present work) and \(\chi=0\) in the other phase (i.e., gas phase). The volume fraction, \(C = \frac{1}{V_0}\iiint_{V_0} \chi dV\), represents the volume occupied by the reference fluid in a cell with respect to the total cell volume, \(V_0\).
\begin{equation}
\label{eqn:characVoF}
\frac{D\chi}{Dt}=\frac{\partial \chi}{\partial t} + \vec{u} \cdot \nabla\chi=0
\end{equation}


The advection of Eq.~(\ref{eqn:characVoF}) is performed by extending the algorithm and VOF tools proposed in Baraldi et al.~\cite{baraldi2014mass} to compressible liquids with phase change. A three-step split advection algorithm is implemented, consisting of an Eulerian Implicit, an Eulerian Algebraic and a Lagrangian Explicit steps (EI-EA-LE algorithm). Details about this algorithm and its extension to compressible liquids are explained in the following paragraphs and shown in Eqs.~(\ref{eqn:EILEcomp}),~(\ref{eqn:EILEcorrection}) and~(\ref{eqn:EIEALEsteps}). Compared to the original EI-EA-LE method proposed by Scardovelli et al.~\cite{scardovelli2002marker}, the algorithm from Baraldi et al.~\cite{baraldi2014mass} is wisp-free and mass-conserving to machine-error precision for incompressible flows. Yet, numerical errors exist and are bounded by the accuracy to which \(\nabla\cdot\vec{u}=0\) is satisfied and other errors introduced by the geometrical operations of the VOF method, which can be expected to increase when the liquid structure is under-resolved. \par

During the split advection of Eq.~(\ref{eqn:characVoF}), the interface is geometrically reconstructed between steps using the Piecewise Linear Interface Construction (PLIC) method by Youngs~\cite{youngs1982time}. Although the reconstruction process is computationally expensive, the method presented in Baraldi et al.~\cite{baraldi2014mass} is computationally more efficient than other higher-order VOF methods (e.g., 3D-ELVIRA~\cite{miller2002conservative}) and ensures that the volume fraction obtained by solving Eq.~(\ref{eqn:characVoF}) is conserved. Therefore, mass is conserved to machine-error precision when the fluid density is constant. As reported in Haghshenas et al.~\cite{haghshenas2017algebraic}, this is only achieved by low-order convective schemes as the one used in Baraldi et al.~\cite{baraldi2014mass}. This low-order advection scheme causes smearing of the solution around the interface, introducing geometrical errors as the interface is advected. Nevertheless, volume-conservation properties are favored over higher-order advection schemes and a sufficiently low CFL (i.e., Courant-Friedrichs-Lewy) condition is used to limit the magnitude of such errors. \par 

Eq.~(\ref{eqn:characVoF}) is rewritten in conservative form accounting for mass exchange and fluid compressibility as
\begin{equation}
\label{eqn:characVoF2}
\frac{\partial \chi}{\partial t}  + \nabla \cdot (\chi \vec{u}_l) = \chi \nabla \cdot \vec{u}_l - \frac{\dot{m}}{\rho_l}
\end{equation}

\noindent
where \(\dot{m}\) is the mass flux per unit volume added to (condensation with \(\dot{m}<0\)) or subtracted from (vaporization with \(\dot{m}>0\)) the liquid phase. \(\dot{m}\) is evaluated as \(\dot{m} = \dot{m}'\delta_\Gamma\), where \(\dot{m}'\) is the mass flux per unit area across the interface and \(\delta_\Gamma\) activates the phase-change term only at the interface cells. The value of \(\dot{m}'\) is a result of the solution of the system of interface matching conditions discussed in Subsection~\ref{subsec:matching} and Subsection~\ref{subsec:phaseequilibrium}. Here, \(\delta_\Gamma\) is obtained from the concept of interfacial surface area density as given in Palmore and Desjardins~\cite{palmore2019volume}, whereby in a given region \(\Omega\) of the domain, \(\delta_\Gamma = 
(\int_{\Gamma\cap\Omega}dS)/(\int_{\Omega}dV)\). This term is non-zero only at interface cells, where \(\delta_\Gamma = A_\Gamma/V_0\), with \(V_0\) being the cell volume and \(A_\Gamma\) the area of the interface plane crossing the cell (i.e., obtained from the PLIC). \par 

Notice the use in Eq.~(\ref{eqn:characVoF2}) of the reference phase density, \(\rho_l\), and a liquid phase velocity, \(\vec{u}_l\). Because of the different fluid compressibilities and the velocity jump across the interface in the presence of phase change (see Subsection~\ref{subsec:matching}), the liquid phase has to be advected using a velocity field only representative of the liquid. Extrapolation techniques dealing with this issue are explained in Subsection~\ref{subsec:extrapolation}. For the compressible liquid, \(\rho_l\) is the interface liquid density. \par

Integrating Eq.~(\ref{eqn:characVoF2}) over the volume of the cell and in time with a first-order forward Euler scheme, an equation to update the volume fraction of a given cell is obtained as
\begin{equation}
\label{eqn:volumeVoF}
C^{n+1} = C^n - \sum_{i=1}^{N_{\text{faces}}} F_i + \tilde{C}(\nabla\cdot\vec{u}_l )\Delta t- \frac{\dot{m}}{\rho_l} \Delta t
\end{equation}

The term \(\sum_{i=1}^{N_{\text{faces}}} F_i \) represents the sum of the signed fluxes of the reference phase in and out of the cell, which are evaluated geometrically within the EI-EA-LE split advection algorithm coupled with PLIC~\cite{baraldi2014mass}. \(\tilde{C}\) is the volume fraction of the cell, but where the choice of implicit (\(\tilde{C}=C^{n+1}\)) or explicit (\(\tilde{C}=C^n\)) evaluation could be made. However, only the implicit formulation, \(\tilde{C}=C^{n+1}\) has been found to provide consistent results with the split advection method used here. In summary, Eq.~(\ref{eqn:volumeVoF}) accounts for the variation of the volume fraction at a given cell caused not only by convective fluxes in and out of the cell, but also by the local volume expansion of the reference phase and the volume of the reference phase added/subtracted due to phase change. In these ways, it differs from prior approaches that treated incompressible liquids with or without phase change. \par 

The three-step EI-EA-LE split advection is constructed such that the terms \(C^{n+1} = C^n - \sum_{i=1}^{N_{\text{faces}}} F_i \) from Eq.~(\ref{eqn:volumeVoF}) are recovered for an incompressible fluid without phase change. The EI and LE steps consider the local non-zero divergence in the advection direction while the EA step is designed and only used to satisfy the incompressible three-dimensional version of Eq.~(\ref{eqn:volumeVoF}). In a two-dimensional code, only an EI-LE split advection algorithm is needed and it already satisfies \(C^{n+1} = C^n - \sum_{i=1}^{N_{\text{faces}}} F_i \) when \(\nabla \cdot \vec{u}_l=0\) and \(\dot{m}=0\). \par

The following lines illustrate an example of the split advection consecutive steps. The nomenclature follows that \(u\), \(v\) and \(w\) represent the liquid velocity components in \(x\)-, \(y\)- and \(z\)-directions and \(E\), \(W\), \(N\), \(S\), \(T\) and \(B\) define the East-West (\(x\)-direction), North-South (\(y\)-direction) and Top-Bottom (\(z\)-direction) cell faces, respectively. \par 

For a two-dimensional compressible liquid without phase change (\(\dot{m}=0\)), the EI-LE steps yield, with the EI step in the \(x\)-direction and the LE step in the \(y\)-direction,
\begin{subequations}
\label{eqn:EILEcomp}
\begin{equation}
\label{subeqn:EIcomp}
C^{EI}=\frac{C^{n} + F_W^u - F_E^u}{1-\frac{u_E-u_W}{\Delta x}\Delta t}
\end{equation}
\begin{equation}
\label{subeqn:LEcomp}
C^{LE} = C^{EI}\bigg(1+\frac{v_N-v_S}{\Delta y}\Delta t\bigg) + F_S^v - F_N^v = C^n - \sum_{i=1}^{N_{\text{faces}}} F_i + C^{EI}(\nabla\cdot\vec{u}_l )\Delta t \neq C^{n+1}
\end{equation}
\end{subequations}

\noindent
which does not immediately satisfy the form of Eq.~(\ref{eqn:volumeVoF}). Thus, a corrective step is needed after the LE step, as defined in Eq.~(\ref{eqn:EILEcorrection}). In a three-dimensional compressible flow, the EA step is designed such that Eq.~(\ref{eqn:EILEcomp}) and the correction shown in Eq.~(\ref{eqn:EILEcorrection}) are still valid.
\begin{equation}
\label{eqn:EILEcorrection}
C^{n+1} = C^{LE} + (\tilde{C}-C^{EI})(\nabla\cdot\vec{u}_l)\Delta t
\end{equation}

The present algorithm implements the volume addition or subtraction caused by mass exchange before advecting the interface. On a uniform mesh and with the EI step in the \(x\)-direction, the EA step in the \(y\)-direction and the LE step in the \(z\)-direction, the advection steps shown in Baraldi et al.~\cite{baraldi2014mass} now follow Eq.~(\ref{eqn:EIEALEsteps}), including the preliminary step to address phase change and the final correction step to match the form of Eq.~(\ref{eqn:volumeVoF}). In the code, the algorithm alternates the direction of the EI-EA-LE steps to minimize directional bias.
\begin{subequations}
\label{eqn:EIEALEsteps}
\begin{equation}
\label{subeqn:PCstep}
C^{PC}=C^n - \frac{\dot{m}}{\rho_l} \Delta t
\end{equation}
\begin{equation}
\label{subeqn:EIstep}
C^{EI}=\frac{C^{PC} + F_W^u - F_E^u}{1-\frac{u_E-u_W}{\Delta x}\Delta t}
\end{equation}
\begin{equation}
\label{subeqn:EAstep}
C^{EA} = \frac{C^{EI}\big[1-\frac{u_E-u_W}{\Delta x}\Delta t + (\nabla\cdot\vec{u}_l)\Delta t\big] + F_S^v - F_N^v }{1+\frac{w_T-w_B}{\Delta z}\Delta t}
\end{equation}
\begin{equation}
\label{subeqn:LEstep}
C^{LE} = C^{EA}\bigg(1+\frac{w_T-w_B}{\Delta z}\Delta t\bigg) + F_B^w - F_T^w
\end{equation}
\begin{equation}
\label{eqn:correctionstep}
C^{n+1} = C^{LE} + (\tilde{C}-C^{EI})(\nabla\cdot\vec{u}_l)\Delta t
\end{equation}
\end{subequations}

The definition of the EI, EA and LE steps shown in Eq.~(\ref{eqn:EIEALEsteps}) ensures that the solution of the volume fraction, \(C\), stays bounded (i.e., \(0\leq C \leq 1\)) and \(C=1\) within the reference phase. However, as mentioned earlier, small errors may exist due to inaccuracies in the evaluation of geometrical fluxes and how well the velocity field satisfies \(\nabla\cdot\vec{u}_l\). As a result of the finite precision when evaluating the geometrical fluxes, wisps or residual values of \(C\) are left in the domain. The wisp-suppression algorithm from Baraldi et al.~\cite{baraldi2014mass} is used to limit and control the number of wisps. \par

Moreover, the EA step, Eq.~(\ref{subeqn:EAstep}), might introduce small undershoots (i.e., \(C<0\)) and overshoots (i.e., \(C>1\)) in its incompressible form~\cite{baraldi2014mass}. For compressible flow, the same problem exists. Additionally, phase change and volume dilation in a compressible framework may also cause undershoots and overshoots of \(C\) in interface cells where almost no liquid is present or where the liquid occupies almost the entire cell volume. To eliminate these issues, a redistribution algorithm following that of Baraldi et al.~\cite{baraldi2014mass} and Harvie and Fletcher~\cite{harvie2000new} is used. However, the present work adds directionality following the interface normal unit vector, \(\vec{n}\), to the redistribution algorithm whenever possible. Since most of the undershoots and overshoots in \(C\) will be caused by phase change and volume expansion in the direction perpendicular to the interface, this approach becomes more consistent and preserves the interface shape better. \par 

The authors acknowledge that this VOF method for compressible liquids is not mass-conserving to machine error. Using a volume-preserving algorithm does not ensure mass conservation when density is not constant. Thus, mass conservation will improve as the mesh is refined and a lower time step is used, better capturing the density field. This is no different than other available VOF methods for compressible flows~\cite{bo2014volume,duret2018pressure,denner2018pressure}. Nevertheless, two main reasons motivate the use of this approach: (a) to maintain a sharp interface; and (b) the method simplifies to the mass-conserving VOF method from Baraldi et al.~\cite{baraldi2014mass} when \(\nabla\cdot\vec{u}=0\) and \(\dot{m}=0\). \par

\subsection{Evaluation of interface geometry}
\label{subsec:VoFgeom}

The normal unit vector, \(\vec{n}\), is evaluated using the Mixed-Youngs-Centered (MYC) method~\cite{aulisa2007interface} and curvature is computed using an improved Height Function (HF) method~\cite{lopez2009improved}. The HF method is second-order accurate but presents considerable curvature errors whenever the normal unit vector of the interface is not aligned with the coordinate axes~\cite{baraldi2014mass}. This issue contributes, among other factors, to the generation of spurious velocity currents around the interface due to a lack of an exact interfacial pressure balance. This issue is a reason for caution in liquid injection problems where the growth of instabilities along the liquid-gas interface must be only related to physical phenomena. This problem is more important at supercritical pressures where the liquid and gas phases look more alike near the interface and the surface-tension force that would stabilize these numerical instabilities caused by spurious currents is reduced. \par

Efforts have been made to develop more accurate methods to evaluate the interface geometry under the VOF framework. For instance, Popinet~\cite{popinet2009accurate} presents an adaptative scheme to enhance the accuracy of curvature computations for under-resolved interfaces using the HF method. This modification to the HF method is not implemented in the present work, but may be considered in the future. Other works combine the VOF and LS methods to use the smoother LS distribution to obtain a better estimate of the interface geometry~\cite{sussman2000coupled,sussman2003second}. However, a key step whereby the LS function is re-distanced with respect to the PLIC interface reconstruction does not ensure a mesh-converging curvature. Together with the intrinsic complexity of combining the VOF and LS methods, the HF method is preferred. \par

\section{Numerical solution of the governing equations}
\label{sec:numerical}

\begin{figure}[h!]
\centering
\includegraphics[width=0.8\linewidth]{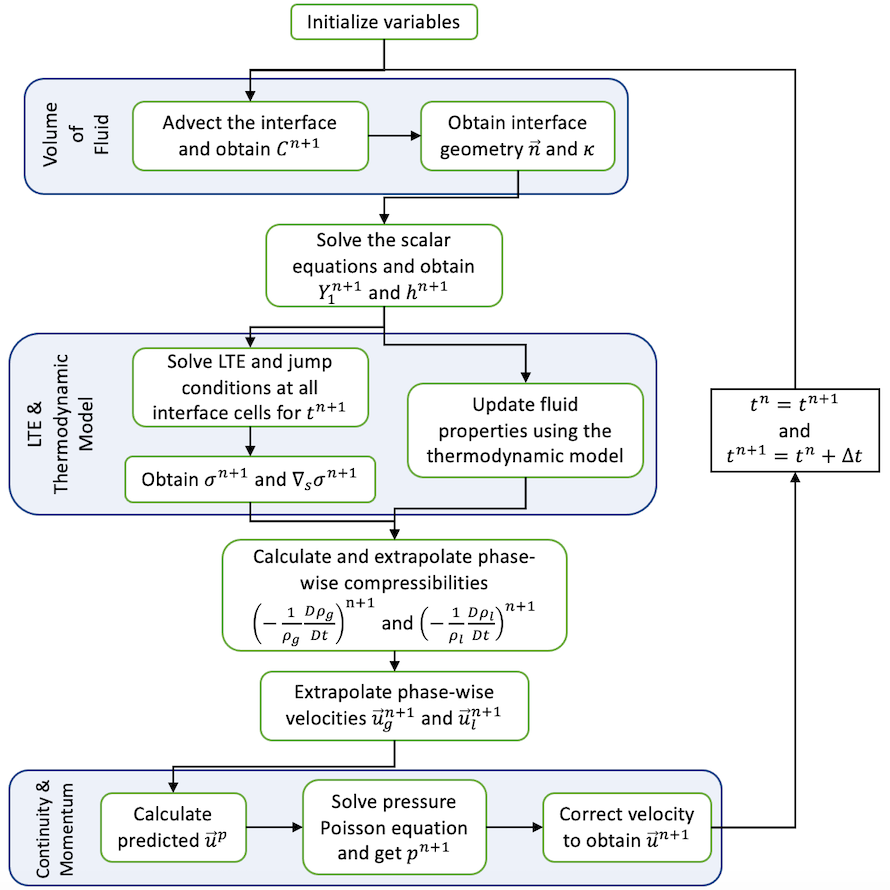}
\caption{Algorithm flowchart at every time step to solve low-Mach-number, two-phase flows at supercritical pressures.}
\label{fig:algorithm}
\end{figure}

The main algorithm steps at every time step are shown as a flowchart in Figure~\ref{fig:algorithm}. The goal here is to provide some context on the necessary steps to solve the governing equations before proceeding with the individual details. \par 

The simulation is initialized by assigning initial conditions to all variables involved in the solution process. First, the bulk of the VOF method is used, which includes the interface advection, the evaluation of the interface normal unit vector using the MYC method, the PLIC interface reconstruction and the HF method to evaluate the interface curvature. Once the interface has been updated at \(t^{n+1}\), the governing equations for species continuity and energy are solved. \par 

When the entire domain has an updated solution for the interface location and its geometry, as well as the mixture composition and enthalpy in both phases, the LTE and jump conditions are solved at each interface cell. At the same time, since pressure is fixed in the thermodynamic model for the low-Mach-number configuration, the fluid properties are updated everywhere (e.g., \(\rho^{n+1}\)). \par  

Before solving the continuity and momentum equations, the new fluid compressibilities are calculated and extrapolated. After that, the phase-wise velocities are obtained from the extrapolated fluid compressibilities and used at the next time step. Then, a predictor-projection method is used to solve the momentum equation, which splits the pressure gradient into an implicit constant-coefficient term and an explicit variable-coefficient term to solve a low-Mach-number PPE using an FFT methodology. \par 

The VOF methodology has been presented in Section~\ref{sec:interface}. The following sections address the rest of the main blocks of the solution algorithm in order. Subsection~\ref{subsec:scalareqs} discusses how the governing equations for the scalar variables (i.e., species continuity and energy) are solved. Then, the methodology to obtain the interface properties is presented in Subsection~\ref{subsec:phaseequilibrium} and the evaluation of fluid compressibilities, as well as the extrapolation techniques used to obtain phase-wise compressibilities and velocities, are discussed in Subsection~\ref{subsec:extrapolation}. The solution method of the continuity and momentum equations is presented in Subsection~\ref{subsec:contmom}, where a low-Mach-number Poisson equation for the pressure field is developed. Finally, Subsection~\ref{subsec:timestep} provides information on the evaluation of the time step, \(\Delta t\), and some final remarks about the algorithm. \par

The proposed methodology is presented for a computational domain discretized with a Cartesian uniform staggered mesh. Control volumes or cells are defined, where velocity components are located at the center of the faces of the control volume and the rest of the variables (e.g., pressure, mass fraction, fluid properties, volume fraction occupied by the liquid phase) are defined at the center of the cell. Despite this simplified mesh configuration, the proposed method could be extended to non-uniform meshes or orthogonal meshes~\cite{aithal2020fast}. We focus on the modeling and numerical difficulties of high-pressure transcritical flows rather than particular details associated with specific and more complex mesh configurations. \par

\subsection{Discretization of the species continuity and energy equations}
\label{subsec:scalareqs}

The governing equations for species mass fraction, Eq.~(\ref{eqn:spcont}), and energy, Eq.~(\ref{eqn:energy}), are solved differently than the continuity and momentum equations presented in Subsection~\ref{subsec:contmom}. Here, the non-conservative forms of both equations are discretized using finite differences. The reasons for this discretization choice are to obtain a better control on numerical stability and to directly include the interface solution in the discretization. Thus, these equations are solved in each phase independently using phase-wise variables. For low-Mach-number flows, the solution of LTE at the interface is directly coupled to the energy and species mass balances across the interface. Once the interface solution is known, it is imposed as a phase boundary condition.  \par 

Other researchers solve the conservative form of the energy equation in terms of the fluid temperature using finite-volume techniques~\cite{juric1998computations,tryggvason2015direct}. In that case, fluid properties are volume-averaged at interface cells and a source term is included in the energy equation to describe the energy jump across the interface (i.e., Eq.~(\ref{eqn:energymatch})). This one-fluid approach is suitable to solve for the temperature, since it is continuous across the interface. However, one would still have to extrapolate the temperature field on both sides of the interface or use one-sided (phase-wise) stencils to obtain the correct temperature gradients numerically for each phase. As seen in Subsection~\ref{subsec:contmom}, a similar approach is used to address the solution of the continuity and momentum equations. However, the mixture composition and enthalpy present sharp discontinuities across the interface. Therefore, a phase-wise approach whereby the interface solution is embedded in the discretization is preferred to avoid further costly extrapolations. The finite-difference method applied to the species and energy transport equations is an adequate choice, as used in other two-phase works~\cite{anumolu2018gradient,dodd2021analysis,dodd2021vof}. \par 

The non-conservative forms of Eqs.~(\ref{eqn:spcont}) and~(\ref{eqn:energy}) are integrated in time using an explicit first-order step as 
\begin{equation}
\label{eqn:spcont_disc}
\frac{DY_O}{Dt}^{n+1}=\frac{Y_{O}^{n+1}-Y_{O}^{n}}{\Delta t} + (\vec{u}_{f}\cdot \nabla Y_O)^n = \frac{1}{\rho^n} \bigg[ \nabla \cdot (\rho D_m \nabla Y_O)^n \bigg]
\end{equation}
\begin{equation}
\label{eqn:energy_disc}
\frac{Dh}{Dt}^{n+1}=\frac{h^{n+1}-h^n}{\Delta t} + (\vec{u}_{f}\cdot \nabla h)^n = \frac{1}{\rho^n} \Bigg[ \nabla \cdot \bigg(\frac{\lambda}{c_p}\nabla h \bigg)^n + \sum_{i=1}^{N=2} \nabla \cdot \Bigg(\bigg[\rho D_m - \frac{\lambda}{c_p}\bigg]h_i \nabla Y_i\Bigg)^n \Bigg]
\end{equation}

\noindent
where the convective and diffusive terms are evaluated explicitly at time \(n\) and the phase-wise velocity \(\vec{u}_f\) is used. Here, \(f\) can refer to the gas phase (i.e., \(f=g\)) or to the liquid phase (i.e., \(f=l\)). The term \(\vec{u}_{f}\) is evaluated at the cell center using a linear average from cell face values. The choice of a first-order integration in time is made in line with the VOF split-advection algorithm. The time step value is already restricted to ensure numerical stability and minimize the geometrical errors introduced when advecting the volume-fraction field. Any influence of the low-order temporal scheme in the solution of Eqs.~(\ref{eqn:spcont_disc}) and~(\ref{eqn:energy_disc}) is therefore limited. A higher-order temporal integration could be considered in future works. \par 

\begin{figure}[h!]
\centering
\includegraphics[width=0.5\linewidth]{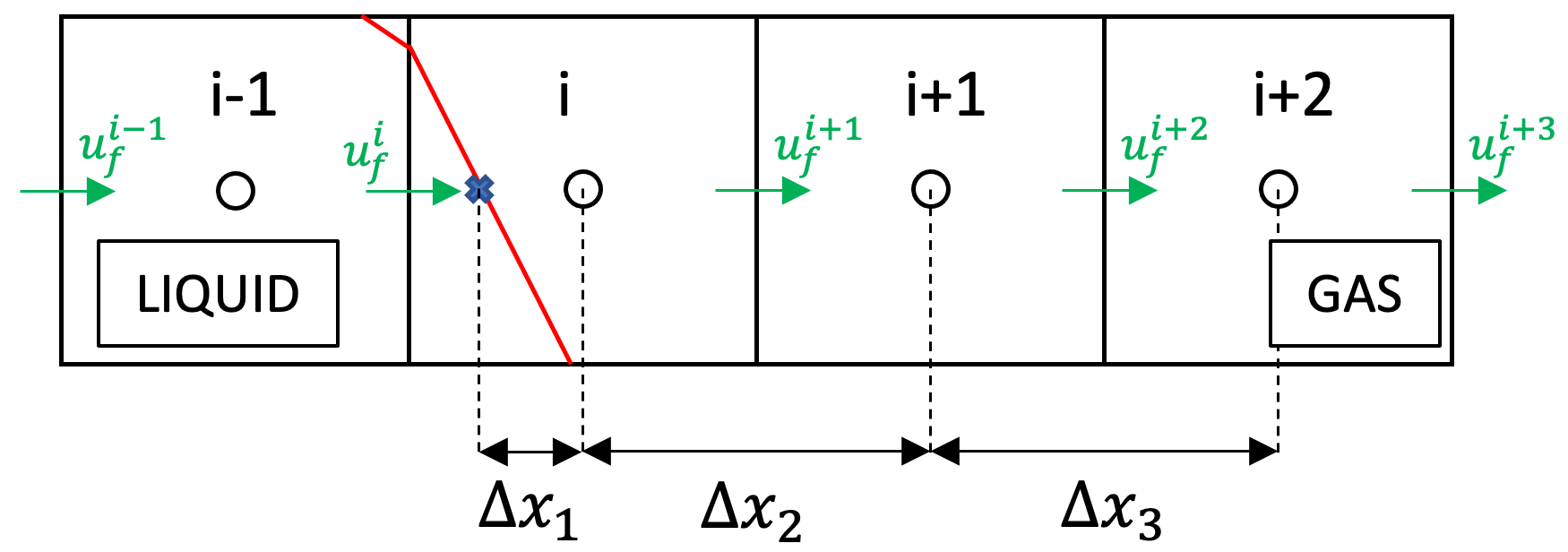}
\caption{Sketch showing the interface intersecting the numerical stencil in the \(x\)-direction and how its location is included in the numerical discretization of the scalar equations.}
\label{fig:scalar_disc}
\end{figure}

Figure~\ref{fig:scalar_disc} shows a sketch of a two-dimensional mesh where the interface is identified (i.e., in cells \(i\) and \(i\)-1). The interface can intersect the numerical stencil used to evaluate \(\partial Y_O/\partial x\) depending on the cell and discretization order. Thus, \(\Delta x_1 < \Delta x\) must be determined. Notice here \(\Delta x_2 = \Delta x_3 = \Delta x\), being \(\Delta x\) the mesh size. The PLIC interface reconstruction at cell \(i\) could be used to obtain \(\Delta x_1\). However, the approach used in Dodd et al.~\cite{dodd2021vof} to estimate \(\Delta x_1\) is faster and more stable. The volume of liquid occupying the space between node \(i\) and \(i\)-1 is used to estimate \(\Delta x_1\). That is, a staggered value of the volume fraction, \(C^{i-1/2}\), is obtained from \(C^{i-1}\) and \(C^i\), as well as from the respective PLIC interface reconstructions. Depending on the interface configuration, this approach becomes exact (i.e., equivalent to using the location obtained with the PLIC interface). Even when \(C^{i-1/2}>0\), \(\Delta x_1\neq\Delta x\) is only evaluated if nodes \(i\) and \(i\)-1 belong to different phases. For instance, following Figure~\ref{fig:scalar_disc}, the staggered volume fraction is used to estimate \(\Delta x_1 \approx (1-C^{i-1/2})\Delta x\). \par 

The convective terms are discretized using an adaptive first-/second-order upwinding scheme to maintain numerical stability and boundedness (i.e., \(Y_O \leq 1\)). Within the limits of the CFL conditions, only a first-order upwind discretization of the convective term is unconditionally bounded~\cite{herrmann2006flux}. On the other hand, diffusive terms are discretized using second-order central differences. Some examples and specific details regarding the discretization of convective and diffusive terms and the inclusion of the interface in the numerical stencils are provided in Appendix~\ref{apn:discr}. \par 

The discretization proposed here is at most second-order accurate in space and may decrease to first order near the interface or when boundedness problems arise. Overall, the convective term will be discretized with a second-order scheme. The boundedness condition becomes important only during the early times if a sharp initial condition has been imposed in each phase. Once the mesh captures the mixing regions well, the second-order upwinding scheme should be bounded. Then, the interface proximity to the grid nodes only occurs for certain cells at each time step. Similarly, the diffusive term are second order except when the interface is too close to a grid node. Nevertheless, stability concerns prompt the usage of this low-order schemes. Moreover, the mesh is very fine to capture the interface properly and to obtain a smooth and converged solution of the extrapolations discussed in Subsection~\ref{subsec:extrapolation}. Thus, a low-order spatial accuracy in the discretization of the scalar equations in some regions is not concerning. \par

\subsection{Interface local phase equilibrium and jump conditions}
\label{subsec:phaseequilibrium}

\begin{figure}[h!]
\centering
\begin{subfigure}{.5\textwidth}
  \centering
  \includegraphics[width=0.7\linewidth]{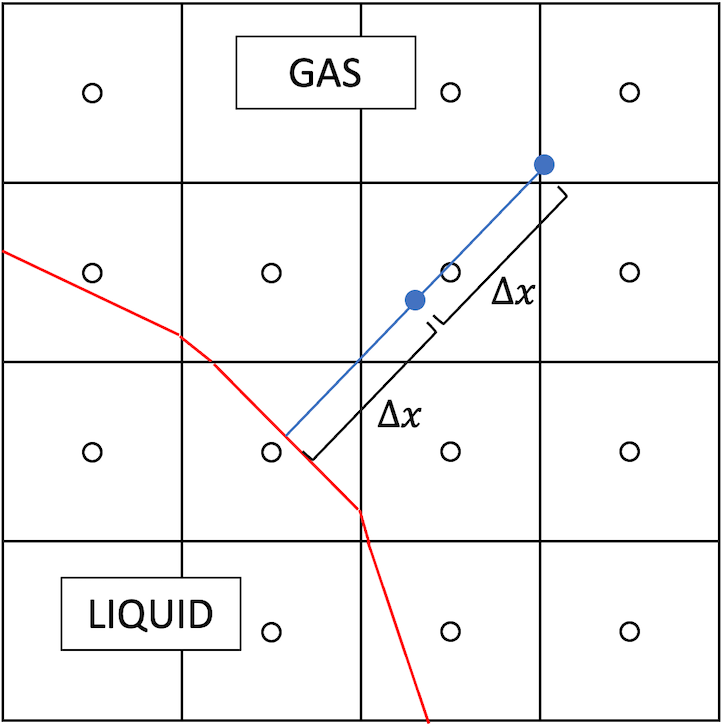}
  \caption{}
  \label{subfig:probe_out}
\end{subfigure}%
\begin{subfigure}{.5\textwidth}
  \centering
  \includegraphics[width=0.7\linewidth]{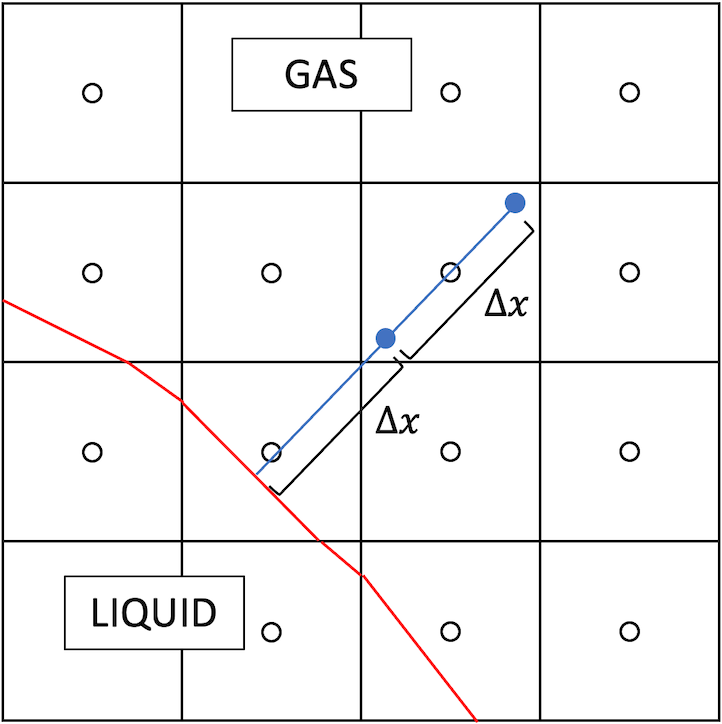}
  \caption{}
  \label{subfig:probe_in}
\end{subfigure}%
\caption{Construction of the normal probe used to evaluate jump conditions and LTE at the interface. For simplicity, only the probe extending into the gas phase is represented in a two-dimensional configuration. The liquid-phase normal probe is constructed in a similar manner and the extension to three dimensions is straightforward. (a) Case where a constant spacing of \(\Delta x\) results in the probe node closest to the interface being defined by grid nodes belonging to different phases; and (b) Case where a constant spacing of \(\Delta x\) results in a well-defined probe.}
\label{fig:normalprobe}
\end{figure}

As discussed in Subsection~\ref{subsec:matching}, the interface is assumed to be in LTE. To obtain the interface equilibrium state at interface cells, the normal-probe technique is used~\cite{tryggvason2015direct,poblador2019axisymmetric,dodd2021vof}. A line perpendicular to the interface plane is drawn extending into both the liquid and the gas phases. The centroid of the interface plane at a given cell is chosen as the starting point of the probe. On this line, two nodes are created in each phase where the mass fraction and enthalpy values are linearly interpolated (i.e., bilinear interpolation in two dimensions and trilinear in three dimensions). Thus, the normal gradients to the interface needed in Eqs.~(\ref{eqn:spcontmatch}) and (\ref{eqn:energymatch}) can be evaluated. Ideally, the nodes on the probe are equally spaced with \(\Delta x\). However, some situations require a larger spacing to avoid using grid nodes in opposite phases when interpolating the mass fraction or the enthalpy values (see Figure~\ref{fig:normalprobe}). This situation must be avoided since there is a sharp jump of these variables across the interface. \par 

In general, a second-order, one-sided, finite-difference method can be used to evaluate the perpendicular gradient at each side of the interface. To do so, the values of each variable in the two nodes on the probe and the interface value are used. Notice it is assumed that the entire interface plane has the same equilibrium solution. As the interface deforms and thin ligaments form, the normal gradients may be calculated using a first-order, one-sided finite difference method if the normal probe crosses the interface again. However, at this point the mesh is under-resolving the interface and its solution might already be poorly defined. \par 

The interface solution is unknown and an iterative process is needed to solve the system of equations formed by the jump conditions and LTE. With the simplifications introduced in this work (e.g., low-Mach-number flows, binary mixture), the interface matching relations for species continuity and energy, Eqs.~(\ref{eqn:spcontmatch}) and (\ref{eqn:energymatch}), together with phase equilibrium, Eq.~(\ref{eqn:pheq}), are decoupled from the momentum matching relations, Eqs.~(\ref{eqn:momjump1}) and (\ref{eqn:momjump2}). Therefore, the same iterative solver discussed in Poblador-Ibanez and Sirignano~\cite{poblador2018transient} is used to obtain the interface solution at each interface cell. The solution of this system of equations defines the properties of the local interface plane: mass flux and heat flux across the interface, temperature, surface-tension coefficient, composition and fluid properties on each side of the interface and, in turn, the pressure jump to be imposed in the momentum equation. \par

\subsection{Evaluation of fluid compressibilities and phase-wise velocities}
\label{subsec:extrapolation}

Each phase's compressibility has to be determined in order to evaluate phase-wise velocities and solve the momentum equation as presented in Subsection~\ref{subsec:contmom}. Under the low-Mach-number constraint, it is sufficient to know the density variations caused by temperature and concentration changes as the thermodynamic pressure is assumed constant in open-boundary problems. In this work, the material derivative of density is related to the material derivatives of mixture enthalpy and mass fraction of each species as
\begin{equation}
\label{eqn:rhochange1}
\frac{D\rho}{Dt} = \frac{\frac{\partial \rho}{\partial T}\bigg|_{Y_i}}{\frac{\partial h}{\partial T}\bigg|_{Y_i}}\frac{Dh}{Dt} + \sum_{i=1}^{N}\Bigg( \frac{\partial \rho}{\partial Y_i}\bigg|_{T,Y_{j\neq i}}-\frac{\frac{\partial \rho}{\partial T}\bigg|_{Y_i}}{\frac{\partial h}{\partial T}\bigg|_{Y_i}}\frac{\partial h}{\partial Y_i}\bigg|_{T,Y_{j\neq i}}\Bigg)\frac{DY_i}{Dt}
\end{equation} 

For a binary mixture, Eq.~(\ref{eqn:rhochange1}) is simplified to
\begin{equation}
\label{eqn:rhochange2}
-\frac{1}{\rho}\frac{D\rho}{Dt} = \frac{1}{c_p\bar{v}}\frac{\partial \bar{v}}{\partial T}\bigg|_{Y_i}\frac{Dh}{Dt} + \Bigg(\frac{\rho}{W_O}\frac{\partial \bar{v}}{\partial X_O}\bigg|_{T,X_{j\neq i}}-\frac{\rho}{W_F}\frac{\partial \bar{v}}{\partial X_F}\bigg|_{T,X_{j\neq i}} - \frac{h_O-h_F}{c_p\bar{v}}\frac{\partial \bar{v}}{\partial T}\bigg|_{Y_i}\Bigg)\frac{DY_O}{Dt}
\end{equation}

\noindent
where \(\bar{v}\) is the mixture molar volume and \(W_O\) and \(W_F\) are the molecular weights of the oxidizer species and the fuel species, respectively. All coefficients are evaluated at constant pressure and at time \(n\)+1, although it is not shown for a simpler notation. The thermodynamic partial derivatives that appear in Eq.~(\ref{eqn:rhochange2}) are obtained using the thermodynamic model described in Appendix~\ref{apn:thermo}. Detailed expressions to evaluate these thermodynamic terms are available in Davis et al.~\cite{davis2019development}. Eq.~(\ref{eqn:rhochange2}) is only evaluated at single-phase cells once the interface location and the scalar fields have been updated in time. The material derivatives \(Dh/Dt\) and \(DY_O/Dt\) are obtained from the solution of the respective non-conservative governing equations, Eqs.~(\ref{eqn:spcont_disc}) and~(\ref{eqn:energy_disc}). At interface cells, a similar evaluation of \(-\frac{1}{\rho}\frac{D\rho}{Dt} \) is not straightforward, especially for the phase occupying less volume. \par 

Note that each fluid compressibility can be associated with the divergence of phase-wise velocities (i.e., \(\nabla\cdot\vec{u}_g=-\frac{1}{\rho_g}\frac{D\rho_g}{Dt}\) and \(\nabla\cdot\vec{u}_l=-\frac{1}{\rho_l}\frac{D\rho_l}{Dt}\)). Therefore, knowing the divergence of each phase-wise velocity in a narrow band of cells around the interface, including the interface cells, is necessary to determine the phase-wise velocities used in the VOF advection algorithm and in the governing equations, as well as the one-fluid velocity divergence from Eq.~(\ref{eqn:divnewtime}) used to solve the pressure-velocity coupling. To do so, the phase-wise velocity divergences are extrapolated from the real phase into a thin ghost region across the interface. \par 

\begin{figure}[h!]
\centering
\includegraphics[width=0.5\linewidth]{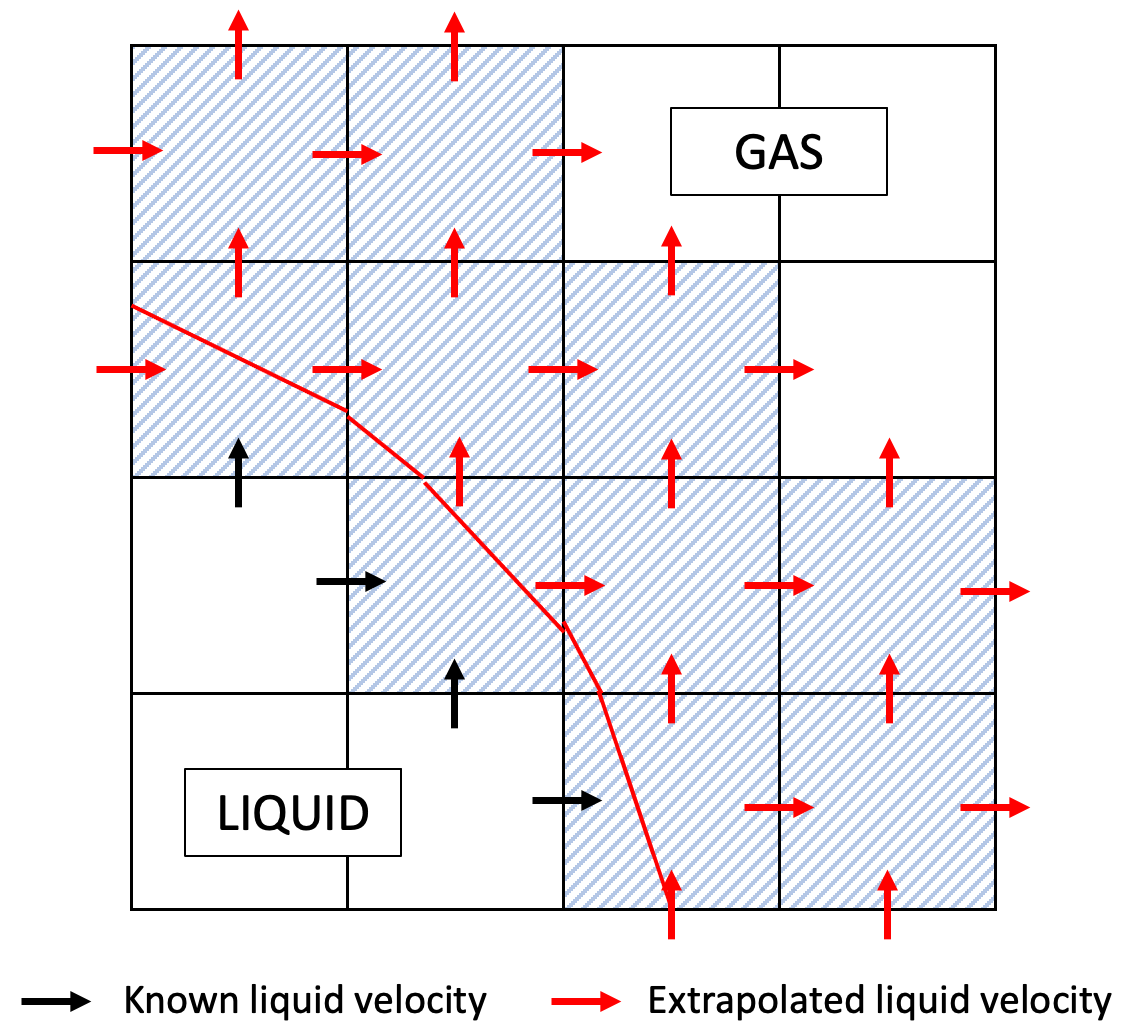}
\caption{Definition of the extrapolation region (dashed cells) for liquid-phase values of fluid compressibility and phase-wise velocity in a two-dimensional mesh.}
\label{fig:extrap}
\end{figure}

The multidimensional extrapolation techniques presented by Aslam~\cite{aslam2004partial} are used to populate this narrow band of cells with characteristic values of the compressibility of each fluid. For instance, Figure~\ref{fig:extrap} shows the two-dimensional extrapolation region for liquid-based values. The extension to a three-dimensional configuration is straightforward and a similar definition is done to define the extrapolation region for gas-based values. Even though the details shown in Aslam~\cite{aslam2004partial} are based on an implementation of the extrapolation across regions defined by a LS function, the same methodology can be adapted to a VOF framework. Then, phase-wise velocities are obtained extending the extrapolation method discussed in Dodd et al.~\cite{dodd2021vof} to compressible flows. Appendix~\ref{apn:extrap} provides more details on the extrapolation equations and how we implement them in a VOF framework. \par  

\subsection{Discretization of the momentum equation and predictor-projection method}
\label{subsec:contmom}

A one-fluid approach is used to solve the two-phase continuity and momentum equations in conservative form (i.e., Eqs.~(\ref{eqn:cont}) and~(\ref{eqn:mom})). This approach, as well as the proposed method to solve the species and energy transport equations, have a discontinuity in the velocity field perpendicular to the interface in the presence of phase change. However, this discontinuity is mild compared to the velocity magnitude around the interface. Still, phase-wise values for the velocity field need to be used in certain terms of the momentum equation as shown in the following lines. This methodology is a standard approach used in the literature~\cite{juric1998computations,dodd2014fast,dodd2021analysis} and we favor a conservative method for global mass and momentum. \par 

Following the work by Dodd and Ferrante~\cite{dodd2014fast}, fluid properties are volume-averaged at each cell using the volume fraction occupied by each phase as \(\phi=\phi_g+(\phi_l-\phi_g)C\), where \(\phi\) is any fluid property such as density or viscosity. The one-fluid property diffuses the sharpness of the interface within a region of \(\mathcal{O}(\Delta x)\). To satisfy the normal and tangential momentum jumps across the interface (i.e., Eqs.~(\ref{eqn:momjump1}) and~(\ref{eqn:momjump2})), the surface-tension force is added by means of a body force active only at the interface, \(\vec{F}_\sigma=\vec{f}_\sigma\delta_\sigma(\vec{x}-\vec{x}_\Gamma)\). \par 

The Continuum Surface Force (CSF) approach from Brackbill et al.~\cite{brackbill1992continuum} extended to flows with variable surface tension~\cite{kothe1996volume,seric2018direct} is used to replace \(\vec{f}_\sigma=-\sigma\kappa\vec{n} + \nabla_s\sigma\) and the Dirac \(\delta\)-function as \(\delta_\sigma(\vec{x}-\vec{x}_\Gamma)=|\nabla C|\). The gradient of the surface-tension coefficient tangent to the interface is evaluated using the method described in Seric et al.~\cite{seric2018direct}, which takes advantage of the HF technique to evaluate \(\nabla_s\sigma = \frac{\partial \sigma}{\partial s_1}\vec{t}_1+\frac{\partial \sigma}{\partial s_2}\vec{t}_2\) in a three-dimensional configuration. That is, the gradient at a given interface cell is directly evaluated along two orthogonal tangential directions, \(s_1\) and \(s_2\). The reduction to a two-dimensional configuration is readily available. Similar to the evaluation of \(\kappa\), a minimum resolution of the interface is needed to obtain accurate results~\cite{seric2018direct}. Even though the MYC method is used to evaluate the interface normal unit vector, the approximation \(\vec{n}=-\nabla C/|\nabla C|\) is taken in the modeling of the surface-tension force in the momentum equation. Therefore, the gradient of the volume fraction provides directionality and locality to the surface-tension force. Finally, a density scaling is used to obtain a body force per unit volume which is independent of the fluid density~\cite{brackbill1992continuum,kothe1996volume}. This modification generates a uniformly-distributed surface-tension force, which improves the performance of the CSF approach and reduces the magnitude of spurious currents at high density ratios. \par 

Under all these considerations, the momentum equation is rewritten as
\begin{equation}
\label{eqn:momwithsurf}
\frac{\partial}{\partial t}(\rho \vec{u})+\nabla \cdot (\rho \vec{u}\vec{u}) = -\nabla p + \nabla \cdot \bar{\bar{\tau}} + \frac{\rho}{\langle\rho\rangle}\bigg(\sigma\kappa\nabla C + \nabla_s\sigma|\nabla C|\bigg)
\end{equation}

\noindent
with \(\langle\rho\rangle=\frac{1}{2}(\rho_G+\rho_L)\), where \(\rho_G\) and \(\rho_L\) are the freestream gas and liquid densities, respectively. Similar to the normal force term \(\sigma\kappa\nabla C\), the tangential force term \(\nabla_s\sigma|\nabla C|\) is further simplified once the tangential unit vectors, \(\vec{t}_1\) and \(\vec{t}_2\), are evaluated from the normal unit vector, \(\vec{n}\). \par

The continuity-momentum coupling is addressed by using the predictor-projection method by Chorin~\cite{chorin1967numerical}. The predictor step consists of a first-order, semi-explicit time integration of Eq.~(\ref{eqn:momwithsurf}) without the pressure gradient, given by
\begin{equation}
\label{eqn:mom_pred}
\vec{u}^p = \frac{\rho^n\vec{u}^n}{\rho^{n+1}} + \frac{\Delta t}{\rho^{n+1}}\bigg[ -\nabla\cdot \big(\rho\vec{u}\vec{u}\big)^n + \nabla\cdot\bar{\bar{\tau}}^n + \frac{\rho^{n+1}}{\langle\rho\rangle}\bigg(\sigma^{n+1}\kappa^{n+1}\nabla C^{n+1} + \nabla_s\sigma^{n+1}|\nabla C^{n+1}|\bigg)\bigg]
\end{equation}

\noindent
where the surface-tension force term is evaluated implicitly. As shown in Figure~\ref{fig:algorithm}, the interface location, the scalar fields and the interface equilibrium solution are updated before solving the Navier-Stokes equations. This way, the density at the new time, \(\rho^{n+1}\), can be evaluated, as well as the interface curvature and surface-tension coefficient, \(\kappa^{n+1}\) and \(\sigma^{n+1}\). Since the advection of the interface is performed with first-order temporal accuracy, the global temporal accuracy of \(\vec{u}\) and \(p\) is limited to first order as well except in the limit where the CFL number tends to zero~\cite{baraldi2014mass,dodd2014fast}. Thus, higher-order temporal integrations in Eq.~(\ref{eqn:mom_pred}) (e.g., Adams-Bashforth scheme) might not add any major improvement to the flow solver global performance, but may be considered in the future. This issue is also discussed in Subsection~\ref{subsec:scalareqs} for the solution of the species and enthalpy transport equations. \par 

After the predictor step, the projection step includes the pressure gradient term to correct \(\vec{u}^p\) in order to satisfy the continuity equation and provide \(\vec{u}^{n+1}\) as shown in Eq.~(\ref{eqn:mom_proj1}).
\begin{equation}
\label{eqn:mom_proj1}
\vec{u}^{n+1} = \vec{u}^p - \Delta t \frac{\nabla p^{n+1}}{\rho^{n+1}}
\end{equation}

An equation for the pressure field is constructed by taking the divergence of Eq.~(\ref{eqn:mom_proj1}) as 
\begin{equation}
\label{eqn:ppe1}
\nabla\cdot\bigg(\frac{\nabla p^{n+1}}{\rho^{n+1}}\bigg)=\frac{1}{\Delta t}\bigg(\nabla\cdot\vec{u}^p-\nabla\cdot\vec{u}^{n+1}\bigg)
\end{equation}

\noindent
where the resulting pressure field satisfies the continuity constraint embedded in the term \(\nabla\cdot\vec{u}^{n+1}\). \par 

Following Duret et al.~\cite{duret2018pressure}, \(\nabla\cdot\vec{u}^{n+1}\) is evaluated by constructing a mass conservation equation for each phase. Substituting \(\rho=\rho_g+(\rho_l-\rho_g)C=\rho_g(1-C)+\rho_lC\) into Eq.~(\ref{eqn:cont}) and including phase change, the following relation is obtained
\begin{equation}
\label{eqn:divnewtime}
\nabla\cdot\vec{u}^{n+1}=-(1-C)\frac{1}{\rho_g}\frac{D\rho_g}{Dt}-C\frac{1}{\rho_l}\frac{D\rho_l}{Dt} + \dot{m}\bigg(\frac{1}{\rho_g}-\frac{1}{\rho_l}\bigg)
\end{equation}

\noindent
where the implicit notation has been dropped for simplicity. Eq.~(\ref{eqn:divnewtime}) reduces to \(\nabla\cdot\vec{u}^{n+1}=-\frac{1}{\rho}\frac{D\rho}{Dt}\) or Eq.~(\ref{eqn:cont}) away from the interface. At interface cells, the divergence of the one-fluid velocity field becomes a volume-averaged fluid compressibility plus the volume expansion (or compression) caused by the change of phase. For low-Mach-number flows, the terms \(\frac{1}{\rho_g}\frac{D\rho_g}{Dt}\) and \(\frac{1}{\rho_l}\frac{D\rho_l}{Dt}\) are assumed to be independent of pressure. The evaluation of the fluid compressibilities has been discussed in Subsection~\ref{subsec:extrapolation}. \par 

The split pressure-gradient method for two-phase flows proposed by Dodd and Ferrante~\cite{dodd2014fast} is used, where the pressure gradient is split into a constant-coefficient implicit term and a variable-coefficient explicit term as 
\begin{equation}
\label{eqn:psubs}
\frac{1}{\rho^{n+1}}\nabla p^{n+1} \rightarrow \frac{1}{\rho_0}\nabla p^{n+1} + \bigg(\frac{1}{\rho^{n+1}}-\frac{1}{\rho_0}\bigg)\nabla \hat{p}
\end{equation}

\noindent
with \(\hat{p}=2p^n-p^{n-1}\) being an explicit linear extrapolation in time of the pressure field and \(\rho_0 = \text{min}(\rho) \equiv \rho_G \). Notice that for the type of problems analyzed in this work, the lowest density in the domain will always be the freestream gas density, \(\rho_G\). Eqs.~(\ref{eqn:mom_proj1}) and~(\ref{eqn:ppe1}) can be rewritten as
\begin{equation}
\label{eqn:mom_proj2}
\vec{u}^{n+1} = \vec{u}^p - \Delta t \bigg[ \frac{1}{\rho_0}\nabla p^{n+1} + \bigg(\frac{1}{\rho^{n+1}}-\frac{1}{\rho_0}\bigg)\nabla \hat{p} \bigg]
\end{equation}

\noindent
and
\begin{equation}
\label{eqn:ppe2}
\nabla^2 p^{n+1} = \nabla \cdot \bigg[ \bigg( 1-\frac{\rho_0}{\rho^{n+1}}\bigg)\nabla \hat{p} \bigg] + \frac{\rho_0}{\Delta t}\bigg(\nabla\cdot\vec{u}^p-\nabla\cdot\vec{u}^{n+1}\bigg)
\end{equation}

Dodd and Ferrante~\cite{dodd2014fast} and Dodd et al.~\cite{dodd2021vof} validate the substitution from Eq.~(\ref{eqn:psubs}) with various benchmark tests. The substitution is exact when \(\nabla \hat{p} \equiv \nabla p^{n+1}\) and approximate when \(\nabla \hat{p} \approx \nabla p^{n+1}\). The accuracy of this method in predicting the pressure field is very good as long as the pressure is smooth in time (i.e., incompressible or low-Mach-number compressible flows). In two-phase flows, the pressure jump across the interface might become problematic in situations of combined high surface tension, curvature and density ratio (i.e., \(\rho_l/\rho_g\)), in which case the time step needs to be reduced to ensure sufficient temporal smoothness in \(\hat{p}\) and obtain a stable solution~\cite{dodd2014fast}. This issue is not expected to have a significant impact on the type of flows that this model aims to analyze, however it may deteriorate the computational efficiency of the proposed method. Cifani~\cite{cifani2019analysis} and Turnquist and Owkes~\cite{turnquist2021fast} address this issue and improve the performance of the split pressure-gradient method at high density ratios. These works may be considered in the future to adapt the methodology for low-pressure configurations where mixing and phase change are relevant. \par

The main advantage of the split pressure-gradient method is that Eq.~(\ref{eqn:ppe2}) becomes a constant-coefficient PPE under the low-Mach-number assumption (i.e., decoupled density and pressure). Combined with a uniform mesh, this equation can be solved using a fast Poisson solver based on performing a series of Discrete Fourier Transforms or FFT~\cite{dodd2014fast,costa2018fft}. This pressure solver can be adapted to various sets of boundary conditions~\cite{costa2018fft} (e.g., periodic or homogeneous Neumann boundary conditions) and provides a direct solution of the pressure field without an iterative process, achieving computational speed-ups orders of magnitude larger than iterative solvers based on Gauss elimination (i.e., \(\mathcal{O}(10^2)\)) or multigrid solvers (i.e., \(\mathcal{O}(10)\)). \par

This sharp, one-fluid method is affected by the presence of spurious currents around the interface. These oscillations are numerical and are caused by various factors that induce a lack of an exact interfacial pressure balance: (a) the lack of a smooth curvature distribution obtained with the HF method; (b) the sharp volume-averaging used to estimate fluid properties at interface cells; and (c) the lack of a smooth distribution of localized interfacial source terms related to mass exchange and fluid compressibilities. How these issues impact other parts of the computational model needs to be investigated (e.g., solution of the energy and species transport equation or the extrapolation of phase-wise velocities). Some insights are provided in Section~\ref{sec:results} regarding the mesh convergence of the solution and how it is affected by this strong coupling. \par 

Eqs.~(\ref{eqn:mom_pred}) and (\ref{eqn:ppe2}) are discretized using standard finite-volume techniques. The viscous term, \(\nabla \cdot \bar{\bar{\tau}}\), is discretized with a second-order central-difference method using phase-wise velocities. If the one-fluid velocity were used in this term, an artificial pressure spike would exist across the interface due to the velocity jump in the presence of phase change, as discussed in Dodd et al.~\cite{dodd2021vof}. To maintain numerical stability, accuracy and boundedness, the convective term, \(\nabla\cdot \big(\rho\vec{u}\vec{u}\big)^n\), is discretized using the SMART algorithm by Gaskell and Lau~\cite{gaskell1988curvature}. The SMART algorithm is up to third-order accurate in space. At interface cells, however, a hybrid method is used which alternates between the second-order central differences and first-order upwind schemes depending on the cell Peclet number (i.e., \(Pe<2\) to use central differences). For the convective term, the one-fluid velocity must be used to capture the momentum jump caused by vaporization or condensation as seen in Eq.~(\ref{eqn:momjump1}). \par

Density and viscosity are volume-averaged only at interface cells where \(0<C<1\). In the compressible framework, the gas and liquid interface properties are chosen as representative values for the averaging. Similarly, \(\rho_g\) and \(\rho_l\) appearing in the fluid expansion (or compression) term due to phase change in Eq.~(\ref{eqn:divnewtime}) are also obtained from the local interface solution, as well as the mass flux, \(\dot{m}'\), used to evaluate \(\dot{m}\). Any interface property, \(\phi\), (e.g., curvature) is obtained in the staggered velocity cell from the following average~\cite{dodd2014fast}
\begin{equation}
\label{eqn:faceavg}
\phi_{i+1/2,j,k}=
\begin{cases}
\phi_{i+1,j,k} & \text{if $\phi_{i,j,k}=0$} \\
\phi_{i,j,k} & \text{if $\phi_{i+1,j,k}=0$} \\
\frac{1}{2}(\phi_{i+1,j,k}+\phi_{i,j,k}) & \text{otherwise}
\end{cases}
\end{equation}
\noindent
which considers the fact that two adjacent interface cells may not exist to evaluate the average property. Here, the location of a \(u\)-node is addressed and the same method can be used in all the other velocity nodes. Eq.~(\ref{eqn:faceavg}) is used to evaluate \(\sigma\), \(\kappa\), \(\frac{\partial \sigma}{\partial s_1}\) and \(\frac{\partial \sigma}{\partial s_2}\), which only have a non-zero value at interface cells. \par

\subsection{Time step criteria and final notes on the algorithm}
\label{subsec:timestep}

The time step must satisfy the CFL condition for numerical stability in an explicit solver. A CFL condition similar to Kang et al.~\cite{kang2000boundary} is used here, which has been applied successfully in other works~\cite{duret2018pressure,anumolu2018gradient}. The following conditions are defined to determine the time step magnitude
\begin{equation}
\begin{cases}
\tau_{\vec{u}} = \frac{|u|_{\text{max}}}{\Delta x} + \frac{|v|_{\text{max}}}{\Delta y} + \frac{|w|_{\text{max}}}{\Delta z} \\
\tau_\mu =\bigg( \frac{2}{\Delta x^2} + \frac{2}{\Delta y^2} + \frac{2}{\Delta z^2} \bigg)\bigg(\frac{\mu}{\rho}\bigg)_{\text{max}} \\
\tau_\sigma = \sqrt{\frac{\sigma_{\text{max}} \kappa_{\text{max}}}{\rho_{\text{min}} \text{min}(\Delta x^2,\Delta y^2,\Delta z^2)}}
\end{cases}
\quad \text{and} \quad
\begin{cases}
\Delta t_{\vec{u}} = \frac{2}{\tau_{\vec{u}} + \tau_\mu + \sqrt{(\tau_{\vec{u}} + \tau_\mu)^2 + 4\tau_{\sigma}^{2}}} \\
\Delta t_h = \frac{\text{min}(\Delta x^2,\Delta y^2,\Delta z^2)}{2\alpha_{\text{max}}} \\
\Delta t_Y = \frac{\text{min}(\Delta x^2,\Delta y^2,\Delta z^2)}{2(D_m)_{\text{max}}}
\end{cases}
\end{equation}

\noindent
where \(\alpha = \lambda/(\rho c_p)\). The time step is evaluated as \(\Delta t = C_{\text{CFL}} \text{min}(\Delta t_{\vec{u}},\Delta t_h,\Delta t_Y)\) where \(C_{\text{CFL}}=0.1-0.2\) is chosen conservatively low. The choice of \(C_{\text{CFL}}\) is a numerical compromise between numerical stability, accuracy and computational cost. \par 

The computational cost of this algorithm is larger than similar algorithms for incompressible flows without phase change. Usually, the main cost of any fluid dynamics simulation is linked to the pressure solver. However, the split pressure-gradient method is a very efficient tool to solve incompressible and low-Mach-number configurations. Three major necessary steps are responsible for at least 40\% of the computational cost per time step: the solution of the local interface state, the update of fluid properties using the thermodynamic model and the extrapolation of phase-wise fluid compressibilities and velocities. \par 

Moreover, scalability is a concern in configurations where the interface deforms considerably, such as those aiming to study liquid jet injection. As the interface surface area grows, more interface nodes are added and the thermodynamic and topology complexity of the interface increases. Moreover, the convergence rate of phase-wise extrapolations might be reduced. Thus, the computational cost per time step may increase considerably over time. Computational implementation details are briefly discussed in Appendix~\ref{apn:implementation} (e.g., parallel code implementation). \par

\section{Results and verification}
\label{sec:results}

The results presented in this work cover the relevant characteristics seen in liquid injection environments at supercritical pressure where two phases still coexist. Simple analytical solutions including all the physics are not available. Moreover, experiments at these conditions are sparse and they either focus on the full-scale injection problem or the evaporation of isolated droplets. Due to the lack of detailed experimental data (i.e., showing surface topology, instability growth rates or detailed mixing), we focus on verifying and assessing the numerical consistency of the proposed model (e.g., grid convergence), and address the impact of known issues such as the effect of spurious currents around the interface or mass conservation. \par 

Previous codes following a similar methodology have been tested and validated in simpler scenarios (e.g., incompressible flow with or without phase change~\cite{baraldi2014mass,dodd2014fast,palmore2019volume,dodd2021vof}). Thus, the validation of the numerical method in these cases is not shown in this section. Two validation tests in the incompressible limit are presented in Appendix~\ref{apn:nume_two}. Each part of the code (e.g., thermodynamic model, VOF advection algorithm) has been validated independently. \par

Subsection~\ref{subsec:1Dres} verifies the code against a previous one-dimensional study. Appendix~\ref{apn:extrap} discusses that the extrapolation of the fluid compressibilities may be done in a constant fashion or in a linear fashion. The results presented in Subsection~\ref{subsec:1Dres} are obtained with a linear extrapolation, whereas the results from Subsections~\ref{subsec:2Dres} to~\ref{subsec:jet_res3D} follow a constant extrapolation to ensure a stable solution. \par

\subsection{One-dimensional transient flow near the liquid-gas interface}
\label{subsec:1Dres}

The transient behavior around a liquid-gas interface with zero curvature has been analyzed at various pressures using a two-dimensional configuration with an initially straight interface and no shear flow. A thorough analysis of this problem is presented in Poblador-Ibanez and Sirignano~\cite{poblador2018transient}, where a simpler one-dimensional code is used. The main differences between both approaches are the following. Here, the interface is allowed to move as mass exchange and volume expansion occur while the full set of governing equations are solved. On the other hand, Poblador-Ibanez and Sirignano~\cite{poblador2018transient} solve the diffusion-driven problem relative to the interface by fixing its location and assuming pressure to be constant throughout the entire domain. Thus, the momentum equation is not solved and the velocity field is directly obtained from the continuity equation. \par 

The problem configuration consists of a liquid \textit{n}-decane at \(T_L=450\) K sitting on a wall surrounded by a hotter gas (i.e., pure oxygen) at \(T_G=550\) K. Without an energy source, the highest temperature in the domain is bounded by \(T_G\), which is below the critical temperature of \textit{n}-decane (i.e., approximately 617.7 K). With both fluids initially at rest, the liquid-gas interface will reach a state of thermodynamic equilibrium as oxygen dissolves into the liquid and \textit{n}-decane vaporizes. The initial interface location is 50 \(\mu\)m away from the wall. Volume expansion or compression due to the mixing process and phase change generates a velocity field perpendicular to the interface. Periodic boundary conditions are imposed in the \(y\)-direction, while no-slip wall boundary conditions are imposed at \(x=0\) and an open-boundary is imposed sufficiently far away from the interface in the gas phase. With no interface perturbation or shear flow, the two-dimensional code must predict a one-dimensional solution. Four different pressures are analyzed: one subcritical case at 10 bar and three supercritical cases with 50, 100 and 150 bar~\cite{poblador2018transient}. Note that the critical pressure for \textit{n}-decane is approximately 21.03 bar. A mesh size of \(\Delta x = 200\) nm and time step of \(\Delta t = 2\) ns are used for all pressures. \par

A direct comparison between the results from the present work and the results shown in Poblador-Ibanez and Sirignano~\cite{poblador2018transient} is not possible because the thermodynamic model is slightly different. Here, a volume correction is implemented to enhance the accuracy of the SRK equation of state, whereas this correction is not added in Poblador-Ibanez and Sirignano~\cite{poblador2018transient}. Thus, different fluid properties are predicted, especially in the liquid phase. Nevertheless, a qualitative comparison is possible with reasonable agreement. \par 

\begin{figure}[h!]
\centering
\begin{subfigure}{.5\textwidth}
  \centering
  \includegraphics[width=1.0\linewidth]{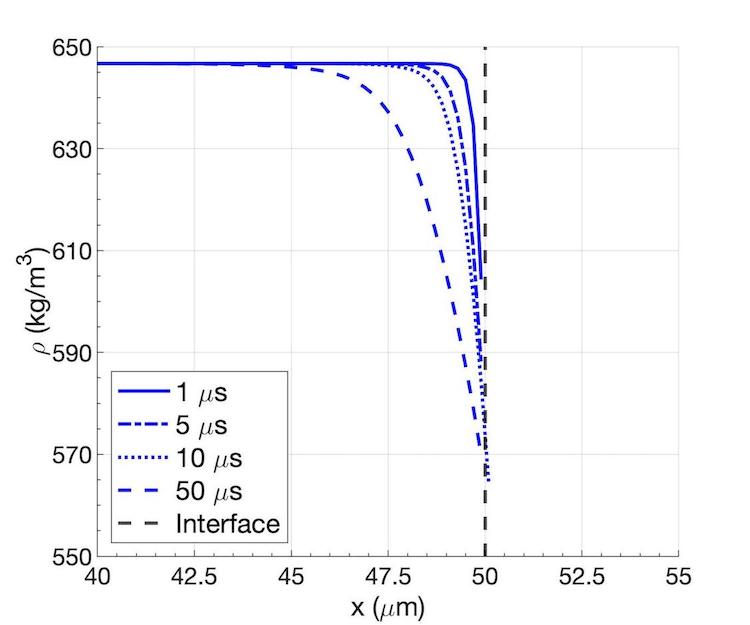}
  \caption{}
  \label{subfig:1D_denliq}
\end{subfigure}%
\begin{subfigure}{.5\textwidth}
  \centering
  \includegraphics[width=1.0\linewidth]{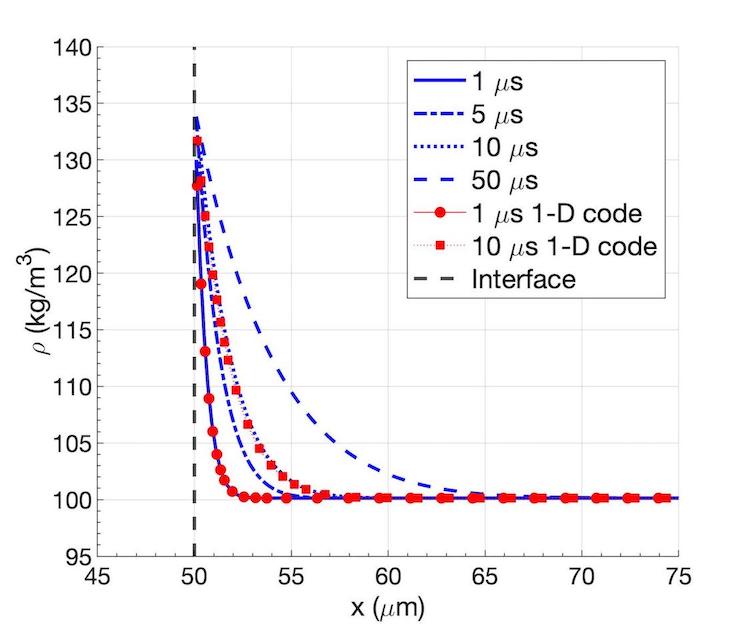}
  \caption{}
    \label{subfig:1D_dengas}
\end{subfigure}%
\caption{Temporal evolution of the density profiles for the one-dimensional transient flow near a liquid-gas interface for an oxygen/\textit{n}-decane binary mixture at \(p=150\) bar. Verification against the one-dimensional solution from Poblador-Ibanez and Sirignano~\cite{poblador2018transient} is shown. (a) liquid density; and (b) gas density.}
\label{fig:1D_den}
\end{figure}

\begin{figure}[h!]
\centering
\begin{subfigure}{.5\textwidth}
  \centering
  \includegraphics[width=1.0\linewidth]{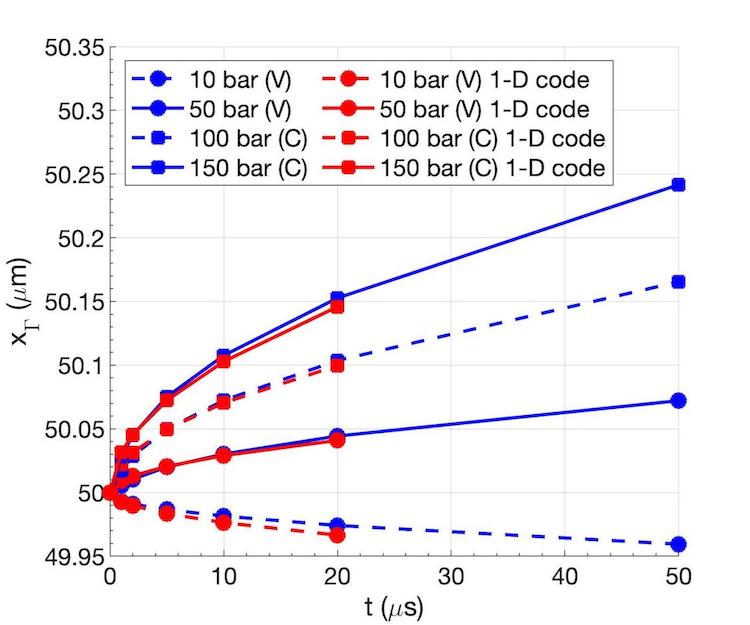}
  \caption{}
  \label{subfig:1D_Xint}
\end{subfigure}%
\begin{subfigure}{.5\textwidth}
  \centering
  \includegraphics[width=1.0\linewidth]{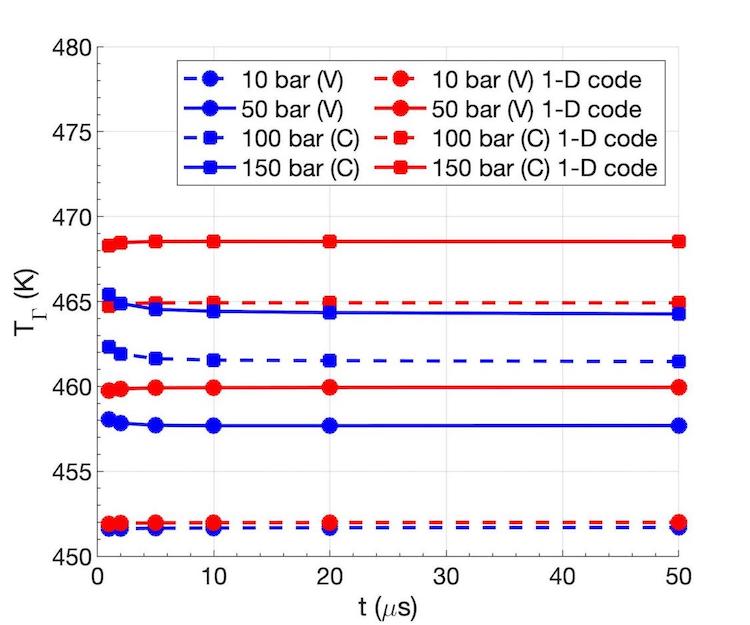}
  \caption{}
    \label{subfig:1D_Tint}
\end{subfigure}%
\caption{Temporal evolution of the interface location and temperature at different pressures for the one-dimensional transient flow near a liquid-gas interface for an oxygen/\textit{n}-decane binary mixture. Depending on the ambient pressure, net vaporization (V) or net condensation (C) occurs across the interface. Verification against the one-dimensional solution from Poblador-Ibanez and Sirignano~\cite{poblador2018transient} is shown. (a) interface location; and (b) interface temperature.}
\label{fig:1D_int}
\end{figure}

The results are one-dimensional. No indication of a deviation or instability is found. Figure~\ref{fig:1D_den} shows the temporal evolution of density profiles at 150 bar. Overall, the results look very similar to those shown in Poblador-Ibanez and Sirignano~\cite{poblador2018transient} except for minor differences caused by the improvement in the thermodynamic model. Mixing in the gas phase agrees in both works, where the density profile (see Figure~\ref{subfig:1D_dengas}) extends about 7-9 \(\mu\)m into the gas phase at \(t=10\) \(\mu\)s. As the mixing layers grow, the interface tends to a steady-state solution as reported in Poblador-Ibanez and Sirignano~\cite{poblador2018transient} (see Figure \ref{subfig:1D_Tint}). Similar works dealing with a two-dimensional laminar mixing layer show the same trend~\cite{davis2019development,poblador2021selfsimilar}. Nevertheless, this behavior may not be true in more complex flows where the interface deforms. \par 

The volume correction in the SRK equation of state is negligible in the gas phase, whereas a stronger impact is seen in the liquid phase. Moreover, this improved thermodynamic model results in slightly different interface solutions compared to Poblador-Ibanez and Sirignano~\cite{poblador2018transient}. For instance, the predicted liquid density (Figure~\ref{subfig:1D_denliq}) is larger than the one shown in Poblador-Ibanez and Sirignano~\cite{poblador2018transient}, but it is also more accurate when compared to reference data from NIST. Without the volume correction in the SRK equation of state, the pure liquid density at 150 bar is closer to 545 kg/m\(^3\). Additionally, lower interface temperatures are predicted once liquid fluid properties are evaluated more accurately. This change in the interface solution results in a slightly different evolution of the energy and species mixing. \par

Lastly, the two main assumptions considered in Poblador-Ibanez and Sirignano~\cite{poblador2018transient} are verified. Throughout the simulation, pressure remains nearly constant and the velocity field is mainly driven by density changes caused by mixing. Moreover, Figure \ref{subfig:1D_Xint} shows the interface location as time marches. Because under this problem configuration mass exchange weakens as mixing occurs, the interface displacement is actually negligible. The maximum interface displacements after 50 \(\mu\)s are of the order of 100 nm, which are similar to the grid spacing used in this problem and negligible compared to the thickness of the diffusion layers. In Figure~\ref{subfig:1D_Xint}, the interface location according to the results from Poblador-Ibanez and Sirignano~\cite{poblador2018transient} has been estimated by integrating the interface velocity, \(u_\Gamma\), over time. This velocity is evaluated by shifting the velocity field to satisfy \(u(x=0)=0\). Then, \(u_\Gamma\) is obtained from the mass balance across the interface (i.e., Eq.~(\ref{eqn:massflux})). On the other hand, the present work uses information of the volume fraction distribution to obtain the interface location at any given time. Because of the changes in the interface solution, the two approaches show slightly different results. \par 

The direction of the interface displacement discussed in Poblador-Ibanez and Sirignano~\cite{poblador2018transient} is also confirmed. At 10 bar, net vaporization is strong with very little dissolution of oxygen into the liquid phase. Thus, the interface recedes and the overall liquid volume decreases. However, as pressure increases, the dissolution of oxygen into \textit{n}-decane is enhanced, the interface temperature is higher and liquid volume expansion occurs near the interface. At 50 bar, even though the interface presents net vaporization, it is not strong enough to compensate for the liquid volume expansion. At 100 bar and 150 bar, both local volume expansion and net condensation contribute to the overall increase in liquid volume. This feature of high-pressure, two-phase flows may cause the interface to present net condensation and net vaporization simultaneously at different locations depending on its deformation and the heat flux into the interface~\cite{poblador2019axisymmetric,poblador2021liquidjet}. \par

\subsection{Two-dimensional capillary wave}
\label{subsec:2Dres}

The capillary wave problem is an appropriate test to validate the relaxation time of a perturbed two-phase interface driven by capillary forces. Gravity is neglected here, although the analytical solution of this problem for incompressible flows with infinite depth proposed by Prosperetti~\cite{prosperetti1981motion} may include it. Dodd and Ferrante~\cite{dodd2014fast} validate their two-phase code for incompressible flows without phase change and analyze the capillary wave problem with evaporation to verify the spatial convergence of their two-phase code~\cite{dodd2021vof}. \par 

A similar analysis is performed for the same binary mixture of oxygen and \textit{n}-decane seen in Subsection \ref{subsec:1Dres}. Each phase has the same initial temperature and composition, but the analyzed thermodynamic pressures vary between 10, 50, 100 and 150 bar. The two-dimensional domain is a rectangular box of 30 \(\mu\)m wide with a liquid layer of 20 \(\mu\)m depth. The liquid-gas interface is spatially perturbed with a sinusoidal wave of 1 \(\mu\)m amplitude and 30 \(\mu\)m wavelength, while both fluids are initially at rest. The amplitude-to-wavelength ratio of 1/30 can be considered a small interface perturbation. The height of the domain must contain the mixing region in both phases during the analyzed times. For 100 and 150 bar, a height of 60 \(\mu\)m is enough. For 50 bar, it is increased to 70 \(\mu\)m and, for 10 bar, to 100 \(\mu\)m. Periodic boundary conditions are imposed in the \(x\)-direction (i.e., tangential to the interface), an open-boundary (i.e., outflow conditions) is imposed in the \(y\)-direction at the end of the gas-phase domain and no-slip wall boundary conditions are imposed at the bottom of the liquid layer. \par 

The mixing in both phases is similar to the one-dimensional case since the interface perturbation is small and there is no shear flow. The relaxation of the interface deformation caused by capillary forces drives the overall picture further toward a one-dimensional solution. Nevertheless, this transient process allows us to study the accuracy and resolution of the numerical model. \par

\subsubsection{Spatial convergence}
\label{subsubsec:2Dspatial}

\begin{figure}[h!]
\centering
\begin{subfigure}{.33\textwidth}
  \centering
  \includegraphics[width=1.0\linewidth]{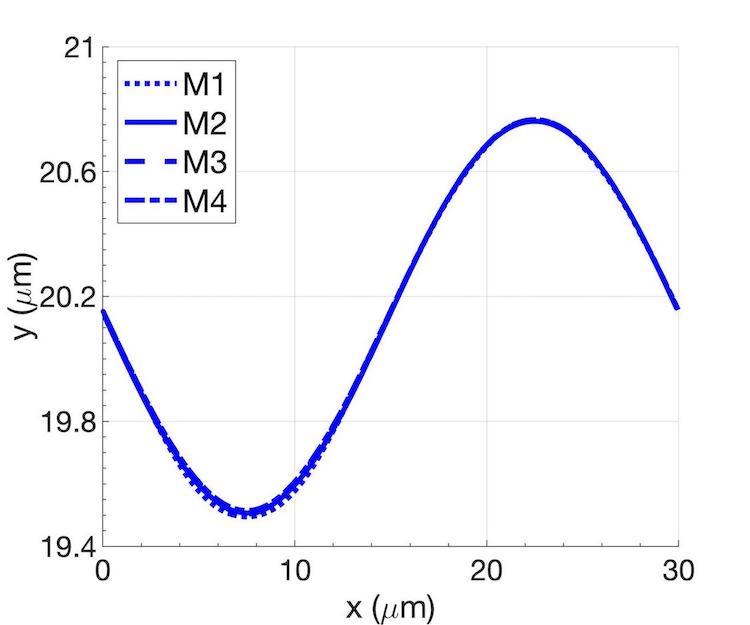}
  \caption{}
  \label{subfig:150mesh_amplitude}
\end{subfigure}%
\begin{subfigure}{.33\textwidth}
  \centering
  \includegraphics[width=1.0\linewidth]{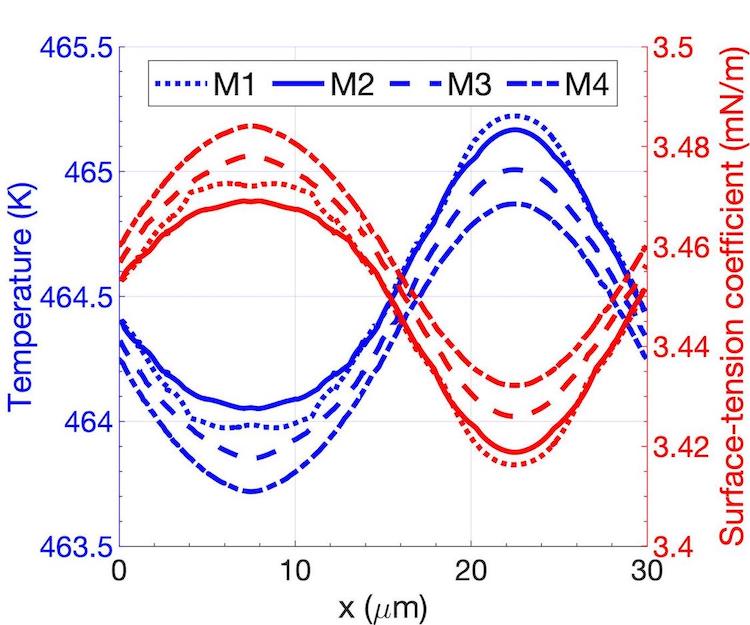}
  \caption{}
    \label{subfig:150mesh_temp_surf}
\end{subfigure}%
\begin{subfigure}{.33\textwidth}
  \centering
  \includegraphics[width=1.0\linewidth]{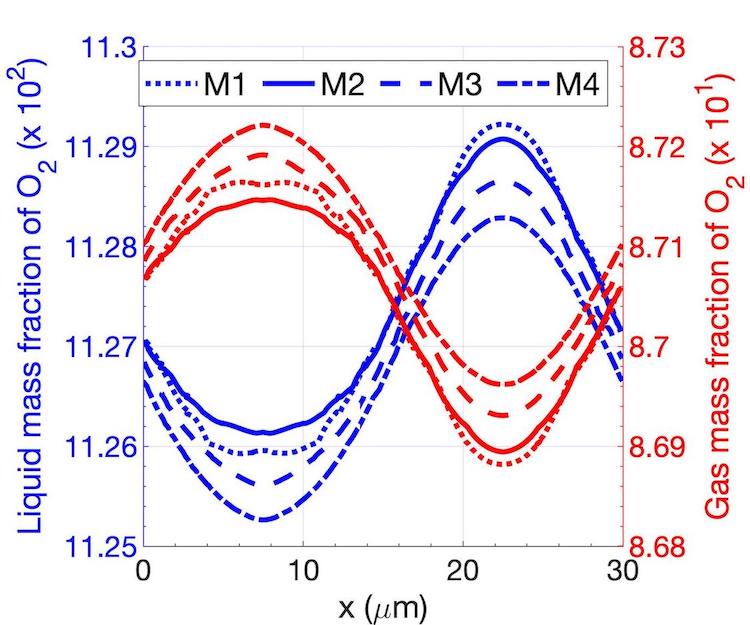}
  \caption{}
    \label{subfig:150mesh_Y1gas_Y1liq}
\end{subfigure}%
\\
\begin{subfigure}{.33\textwidth}
  \centering
  \includegraphics[width=1.0\linewidth]{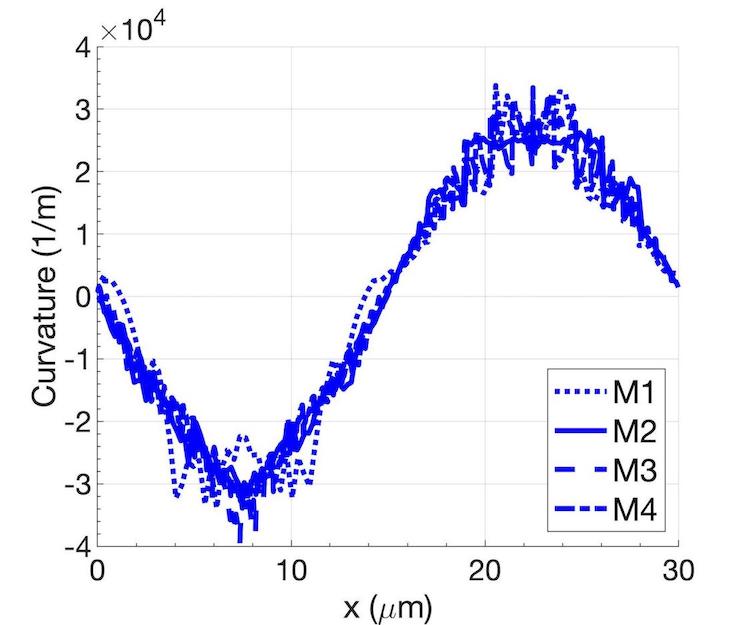}
  \caption{}
  \label{subfig:150mesh_curv}
\end{subfigure}%
\begin{subfigure}{.33\textwidth}
  \centering
  \includegraphics[width=1.0\linewidth]{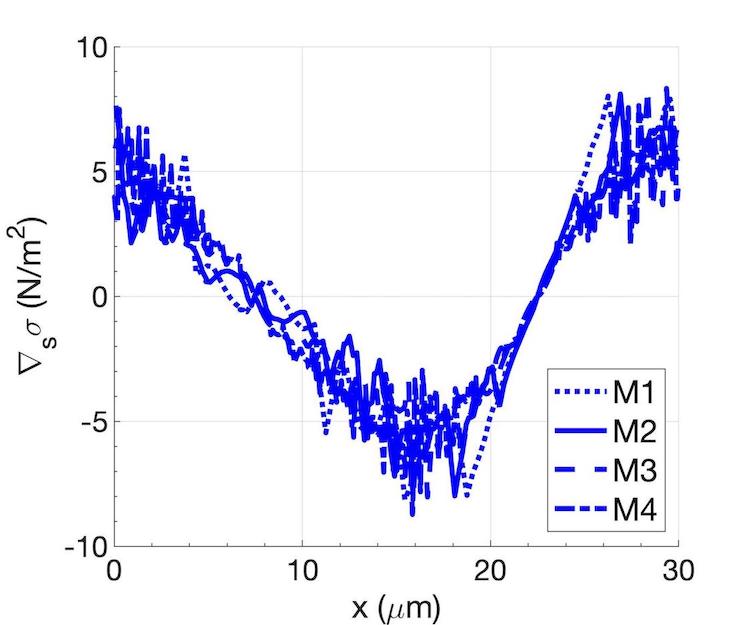}
  \caption{}
    \label{subfig:150mesh_surfgrad}
\end{subfigure}%
\begin{subfigure}{.33\textwidth}
  \centering
  \includegraphics[width=1.0\linewidth]{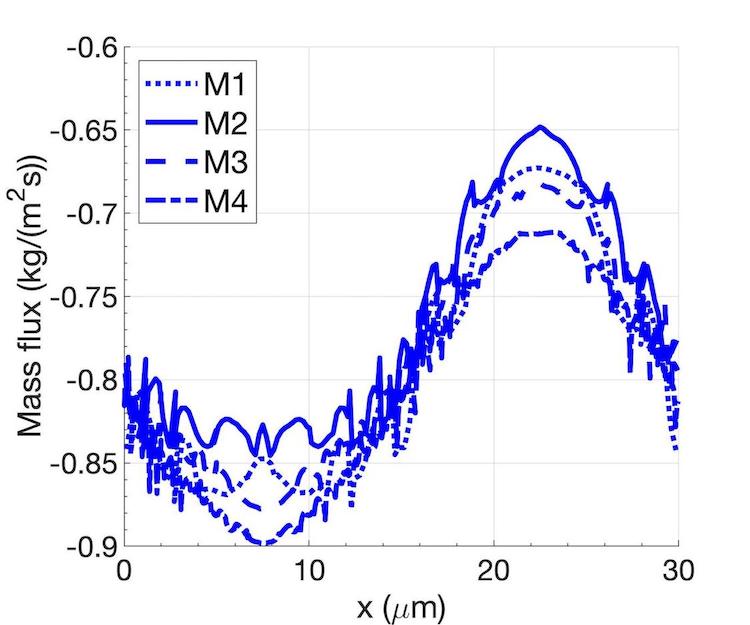}
  \caption{}
    \label{subfig:150mesh_mass}
\end{subfigure}%
\caption{Interface geometry and matching solution of the two-dimensional capillary wave at \(t=19\) \(\mu\)s for the 150 bar case. (a) amplitude; (b) temperature and surface-tension coefficient; (c) oxygen mass fraction in the liquid and gas phases; (d) curvature; (e) gradient of the surface-tension coefficient; and (f) net mass flux.}
\label{fig:150mesh}
\end{figure}

The 150-bar case is chosen to analyze the spatial convergence of the numerical model under interface deformation. The temporal convergence is expected to be first order because of how the equations are discretized. Thus, it is not analyzed here and we rely on a sufficiently small CFL condition to minimize temporal errors. Table~\ref{tab:2Dwavemesh} presents the four different uniform meshes that are studied and Figure~\ref{fig:150mesh} shows results of the interface geometry and equilibrium solution at 19 \(\mu\)s after sufficient relaxation of the diffusion layers. \par 

\begin{table}[!h]
\centering
\begin{tabular}{ c c c c c }
\hline
Mesh & M1 & M2 & M3 & M4 \\
 \hline
\(\Delta x\) (\(\mu\)m) & 3/10 & 2/10 & 1/10 & 1/20 \\
Cells/Wavelength & 100 & 150 & 300 & 600 \\
Cells/Amplitude & 3.33 & 5 & 10 & 20 \\
\hline
\end{tabular}
\caption{Mesh properties used in the analysis of a two-dimensional capillary wave at supercritical pressures. The number of cells per wavelength or amplitude refer to the initial configuration of the liquid-gas interface.}
\label{tab:2Dwavemesh}
\end{table}

The solution of LTE and jump conditions along the interface is sensitive to the mesh resolution. Although other interface properties may be used, the profiles of temperature, net mass flux per unit area and oxygen mass fraction on both sides of the interface are shown. As the mesh is refined, the mixing around the interface is captured better and the enthalpy and concentration gradients obtained using the normal probe technique explained in Subsection~\ref{subsec:phaseequilibrium} become more accurate. Moreover, the PLIC method tends to a more continuous interface reconstruction. These improvements translate into a smoother distribution of fluid properties along the interface as seen in Figures~\ref{subfig:150mesh_temp_surf},~\ref{subfig:150mesh_Y1gas_Y1liq} and~\ref{subfig:150mesh_mass}. \par

The apparent deviation between two consecutive meshes is only enlarged by the scale of the figures. Although mesh M1 (i.e., \(\Delta x=\) 0.3 \(\mu\)m) might look like an outlier due to a poorer resolution, especially of the liquid phase mixing region, the normalized errors or differences in the solution between two consecutive meshes are of the order of 1\% or less. A thorough analysis shows the normalized errors converge with a first-order rate or lower, which is expected given the complexity of the numerical approach. For example, the normalized errors in the interface temperature at \(x=7.5\) \(\mu\)m from M2 to M3 and from M3 to M4 are \(4.526\times 10^{-4}\) and \(2.942\times 10^{-4}\), respectively, which correspond to a convergence rate of 0.62. On the other hand, the convergence rate of the interface temperature at \(x=22.5\) \(\mu\)m is even lower (i.e., 0.22). Other works with simpler models to determine the interface properties also report similar issues~\cite{dodd2021vof,palmore2019volume}. Even though one-dimensional interfaces show good grid independence properties, limited spatial convergence appears once two-dimensional or three-dimensional interfaces are analyzed. \par 

Although the matching solution along the interface looks smoother as the mesh is refined, the net mass flux across the interface, \(\dot{m}'\), is not (see Figure~\ref{subfig:150mesh_mass}). As explained in Poblador-Ibanez et al.~\cite{poblador2021selfsimilar}, \(\dot{m}'\) is very sensitive to small changes in equilibrium composition or temperature. Thus, imperceptible perturbations are captured in \(\dot{m}'\). Moreover, the normal probe used to determine the LTE and jump conditions is built on the PLIC interface. It is not guaranteed that the PLIC interface will be continuous across cells and PLIC loses any information on local interface curvature (i.e., the interface is locally represented by a planar surface). Therefore, the construction of the normal probes is in line with the observed oscillations. \par 

Similar oscillations appear when evaluating the curvature and the gradient of the surface-tension coefficient (see Figures~\ref{subfig:150mesh_curv} and~\ref{subfig:150mesh_surfgrad}). Even though the interface shape and the distribution of the surface-tension coefficient look smooth, as seen in Figures~\ref{subfig:150mesh_amplitude} and~\ref{subfig:150mesh_temp_surf}, \(\kappa\) and \(\nabla_s\sigma\) are evaluated using the HF method. This method is known to generate a non-smooth distribution of \(\kappa\) albeit converging with mesh refinement, as discussed in Subsection~\ref{subsec:VoFgeom}. Moreover, VOF methods using a one-fluid approach to solve the momentum equation present some degree of oscillations in the velocity and pressure fields near the interface due to the sharp treatment of fluid properties and localized body force terms. \par 

Overall, the oscillations of certain interface parameters contribute to the generation of spurious currents around the interface. The effect of these numerical perturbations on the actual development of the liquid surface must be assessed. At this point, we accept this problem as long as it does not cause unstable numerical oscillations. The cases analyzed in this work present an interface evolution that follows a physical explanation. Moreover, in problems involving the injection of liquid jets, the magnitudes of these spurious currents are negligible compared to the jet velocity. \par

\begin{figure}[h!]
\centering
\includegraphics[width=0.5\linewidth]{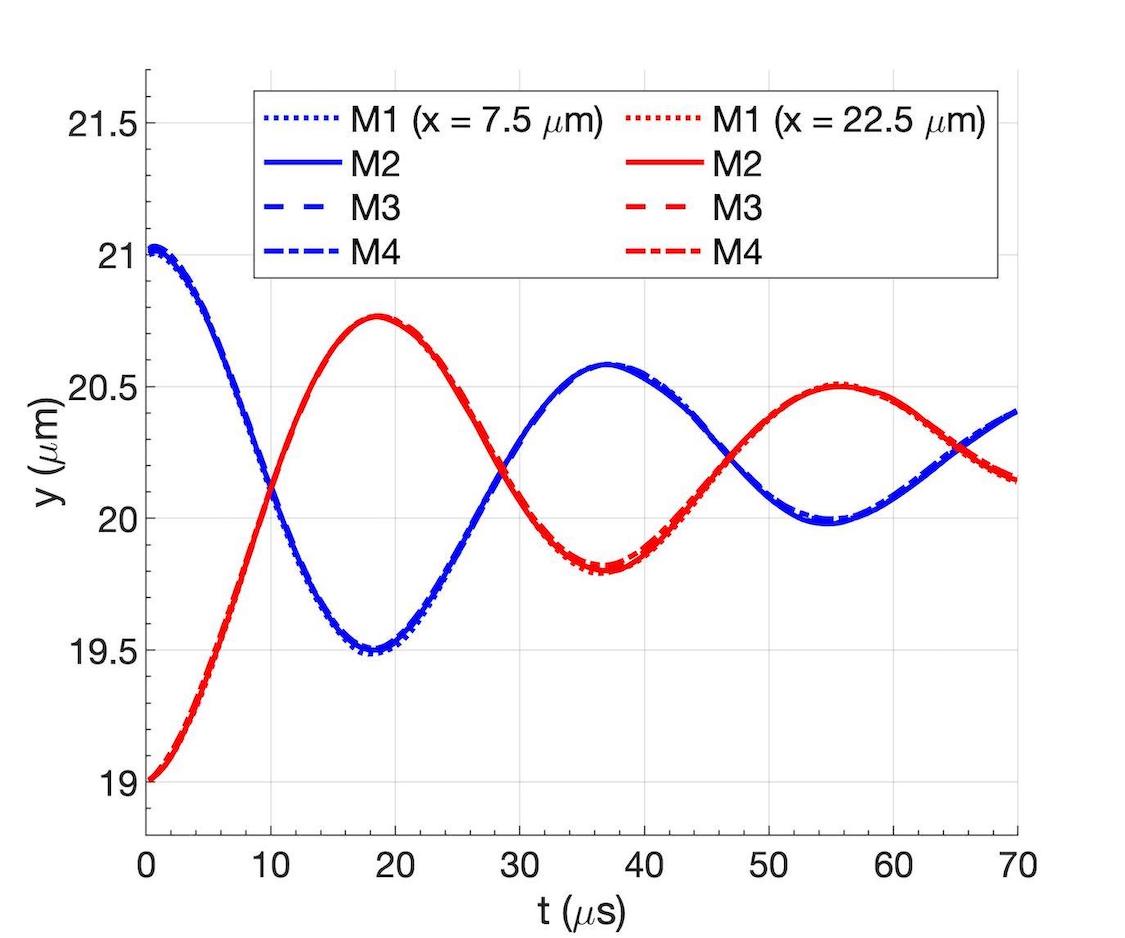}
\caption{Interface amplitude relaxation of the two-dimensional capillary wave at 150 bar. The evolution of the wave's amplitude at \(x=7.5\) \(\mu\)m (initial wave crest location) and at \(x=22.5\) \(\mu\)m (initial wave trough location) is shown.}
\label{fig:150W1W2_amplitude}
\end{figure}

As seen in Figures~\ref{subfig:150mesh_amplitude} and~\ref{fig:150W1W2_amplitude}, the slow spatial convergence of the interface matching solution and the small oscillations in some of the parameters have little effect on the relaxation of the interface amplitude. Figure~\ref{fig:150W1W2_amplitude} shows the temporal evolution of the vertical position of the interface at \(x=7.5\) \(\mu\)m, which corresponds to the initial location of the wave crest, and at \(x=22.5\) \(\mu\)m, which is the location of the initial wave trough. Figures~\ref{subfig:150mesh_amplitude} and~\ref{fig:150W1W2_amplitude} verify the grid convergence of the numerical modeling with regards to the interface displacement as the curves representing the surface's shape and position overlap each other from mesh M1 to M4. \par

\begin{figure}[h!]
\centering
\begin{subfigure}{.5\textwidth}
  \centering
  \includegraphics[width=1.0\linewidth]{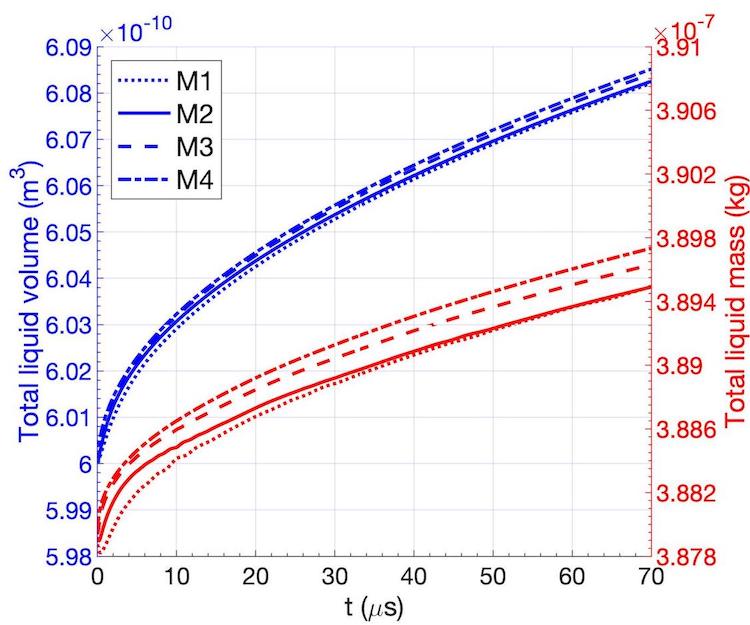}
  \caption{}
  \label{subfig:150mesh_totalvolume_totalmass}
\end{subfigure}%
\begin{subfigure}{.5\textwidth}
  \centering
  \includegraphics[width=1.0\linewidth]{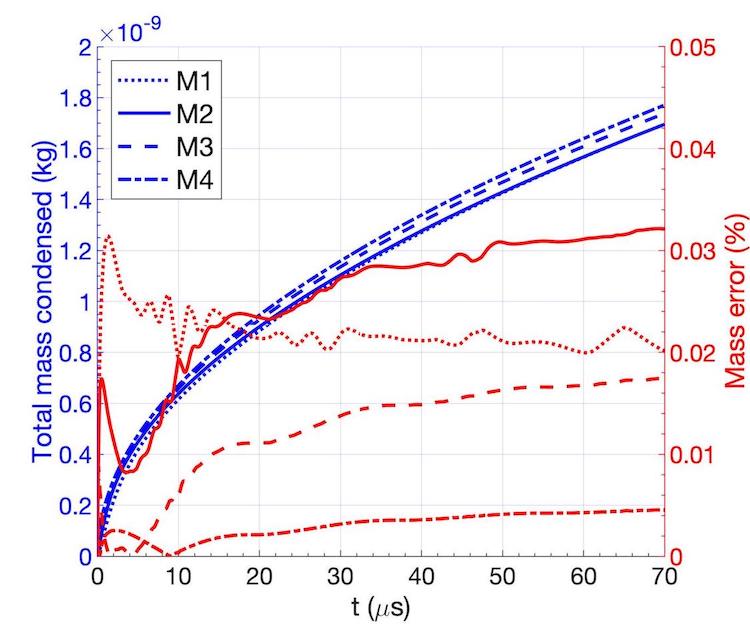}
  \caption{}
    \label{subfig:150mesh_totalmassexchange_masserror}
\end{subfigure}%
\caption{Temporal evolution of the total liquid volume, the total liquid mass, the total net mass exchanged across the interface and the estimated mass error for the two-dimensional capillary wave at 150 bar. (a) total liquid volume and total liquid mass; and (b) total mass condensed and mass error.}
\label{fig:150mesh_totalvolume_mass}
\end{figure}

The total change in liquid volume and mass is shown in Figure~\ref{subfig:150mesh_totalvolume_totalmass}. Note that in the two-dimensional problems analyzed in this paper, a third dimension is considered for convenience with a depth of 1 m. Then, surface areas or liquid volumes are discussed. Convergence with mesh refinement is observed, especially in the total liquid volume. The strong condensation during the relaxation of the sharp initial conditions is well captured with meshes M3 and M4. However, differences in the mass flux and the accuracy to which density variations are captured cause the curves to deviate over time. Grid convergence is observed, but with a first-order rate or lower due to the influence of the interface solution in the mass exchange rates. \par 

Looking at mass errors in the liquid phase, the total amount of mass that condenses at the interface (see Figure~\ref{subfig:150mesh_totalmassexchange_masserror}) also converges with mesh refinement and should be equal to the difference between the total liquid mass at a given time and the initial liquid mass. Here, the total liquid mass is estimated at every time step as \(m_L = \sum\nolimits_{i,j,k} \rho_{i,j,k} C_{i,j,k} \Delta x\Delta y \Delta z\), with \(\rho_{i,j,k}\) being the interface liquid density, \(\rho_l\), at interface cells. This simplification at interface cells introduces small errors when evaluating mass conservation. On the other hand, the total condensed mass is obtained as \(m_\text{cond}=-\sum\nolimits_{t_\text{initial}}^{t_{\text{final}}}\sum\nolimits_{i,j,k}\dot{m}'A_\Gamma\Delta t\). Recall \(\dot{m}'\) is only non-zero at interface cells and \(A_\Gamma\) represents the area of the local interface plane at a given cell. \par 

Even though both approaches should be equivalent, they manifest the issues with mesh resolution when capturing the density field and solving the jump conditions and LTE at the interface. Figure~\ref{subfig:150mesh_totalmassexchange_masserror} shows that the mass error evaluated as \(E_m(\%)=100(m_L-m_\text{initial}-m_\text{cond})/m_\text{initial}\) is reduced with mesh refinement. Although mass errors exist, the overall thermodynamics of the interface can be captured with reasonable accuracy (e.g., identify regions of high or low mass exchange rate or whether condensation or vaporization occurs). These errors are shown to be larger in liquid injection problems as continuous interface deformation generates smaller liquid structures (see Subsection~\ref{subsubsec:2djet_mesh}). Thus, a fixed mesh resolution may lose accuracy in capturing density variations and a smooth interface solution. Note that the total mass exchanged across the interface only represents about 0.46\% of the total liquid mass in the amount of time analyzed in this problem. \par 

The results presented in this section highlight the complexity of the numerical modeling for two-phase flows at supercritical pressures, but at the same time verify the consistency of the methodology implemented in this work. Searching for efficient and reliable methods translates into a limited grid convergence rate for many interface parameters. Therefore, the mesh used in atomization simulations must be a compromise between computational cost and interface accuracy (i.e., geometry and solution of LTE and jump conditions). \par

\subsubsection{Pressure effects}
\label{subsubsec:2Dpressue}

The two-dimensional capillary wave has been analyzed for the same four pressures used in Subsection~\ref{subsec:1Dres} (i.e., 10, 50, 100 and 150 bar) with the uniform mesh M3 (i.e., \(\Delta x = 0.1\) \(\mu\)m). The initial conditions used for all pressures are identical to those previously discussed. The goal here is to present the main differences in the interface thermodynamics as pressure transitions from subcritical to supercritical values. \par 

\begin{figure}[h!]
\centering
\includegraphics[width=0.5\textwidth]{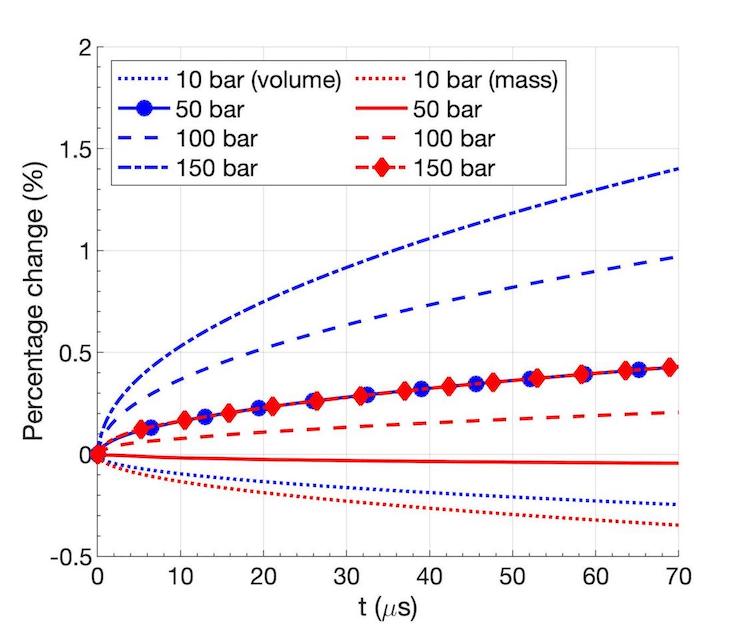}
\caption{Temporal evolution of the change in liquid volume and liquid mass for the two-dimensional capillary wave at 10, 50, 100 and 150 bar using mesh M3.}
\label{fig:pressures_M3_totalvolumeandmass}
\end{figure}

The total volume and mass change of the liquid phase with time is very similar to the one-dimensional problem described in Subsection~\ref{subsec:1Dres} (see Figure~\ref{fig:pressures_M3_totalvolumeandmass}). Note that the curves representing the volume percentage change at 50 bar and the mass percentage change at 150 bar coincidentally overlap. At 50, 100, and 150 bar, the liquid phase expands near the interface and, together with condensation in the 100 and 150 bar cases, the total liquid volume increases with time. On the other hand, vaporization drives the reduction of the liquid volume at 10 bar. Regarding the total liquid mass change over time, the thermodynamic transition whereby net vaporization turns into net condensation as pressure increases is observed. \par 

\begin{figure}[h!]
\centering
\begin{subfigure}{.5\textwidth}
  \centering
  \includegraphics[width=1.0\linewidth]{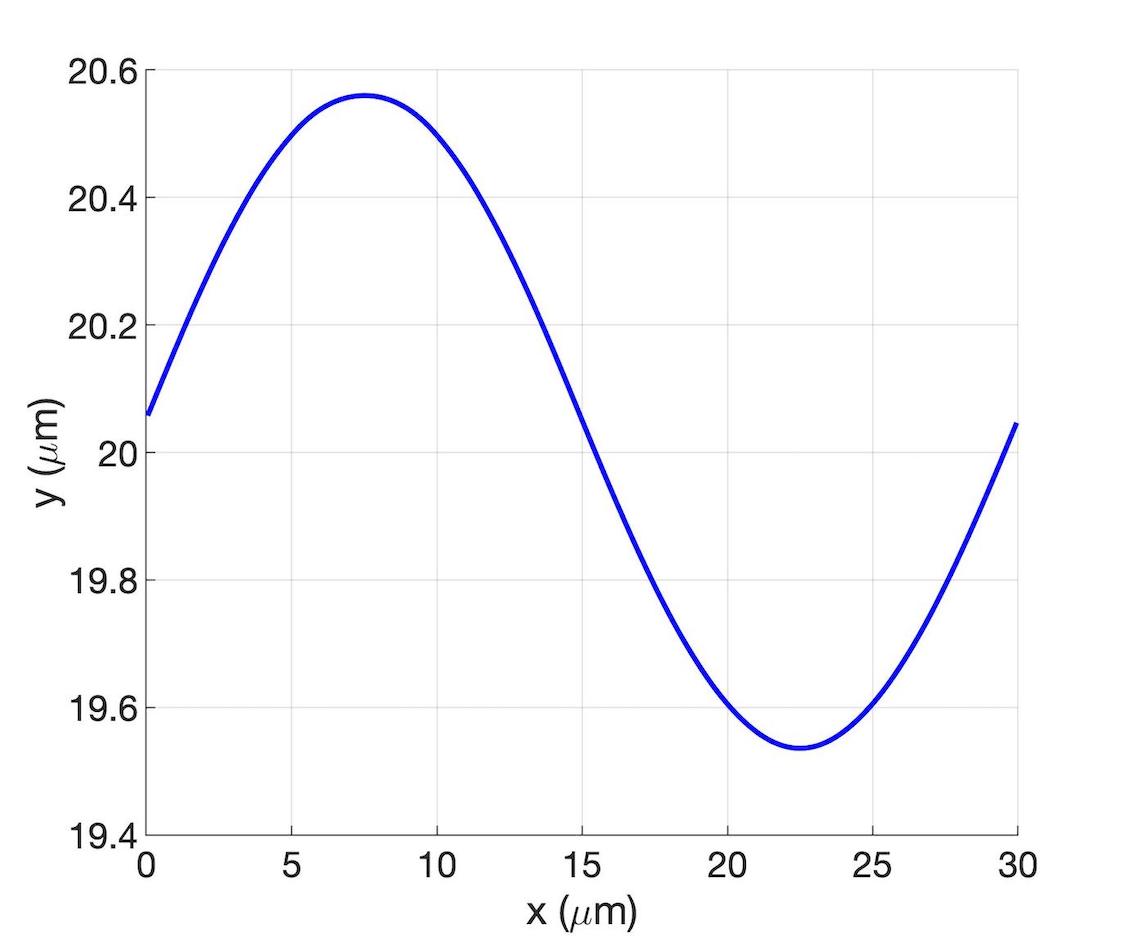}
  \caption{}
  \label{subfig:50_M3_amplitude}
\end{subfigure}%
\begin{subfigure}{.5\textwidth}
  \centering
  \includegraphics[width=1.0\linewidth]{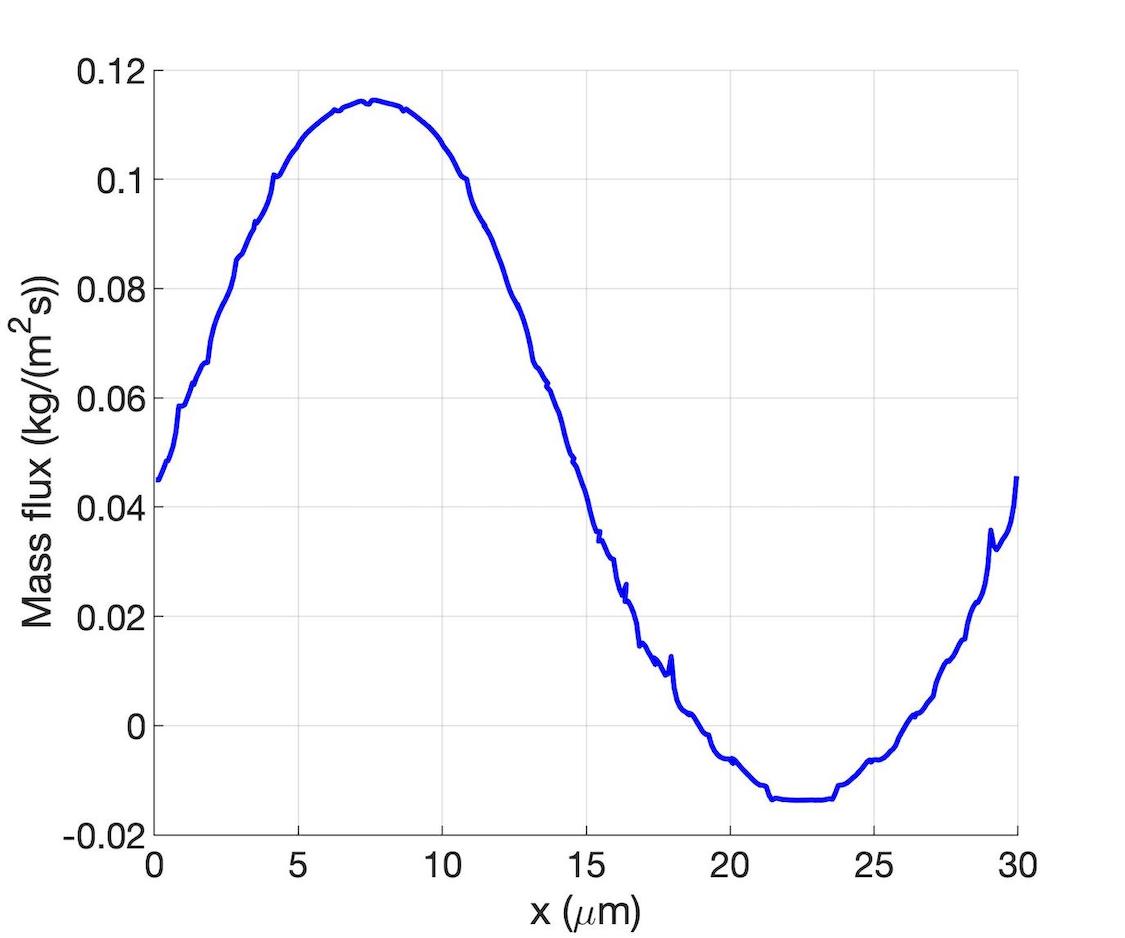}
  \caption{}
    \label{subfig:50_M3_mass}
\end{subfigure}%
\caption{Interface geometry and matching solution of the two-dimensional capillary wave at \(t=24.94\) \(\mu\)s for the 50 bar case using mesh M3. (a) interface amplitude; and (b) net mass flux.}
\label{fig:50_M3_amplitudemass}
\end{figure}

The deformation of the interface in the capillary wave test shows some of the features of the interface varying thermodynamic behavior at high pressures. As seen in Figure~\ref{fig:50_M3_amplitudemass}, the effects of compressed mixing layers can cause the interface to vaporize and condense at different locations simultaneously. As noted in Poblador-Ibanez and Sirignano~\cite{poblador2019axisymmetric}, regions of compressed (heated) gas may present stronger vaporization or weaker condensation, whereas regions of compressed liquid show weaker vaporization or stronger condensation, depending on the ambient pressure. In configurations where mass exchange across the interface is already weak, such as at 50 bar, small interface deformations may trigger this phase-change reversal. \par 

\begin{figure}[h!]
\centering
\begin{subfigure}{.5\textwidth}
  \centering
  \includegraphics[width=1.0\linewidth]{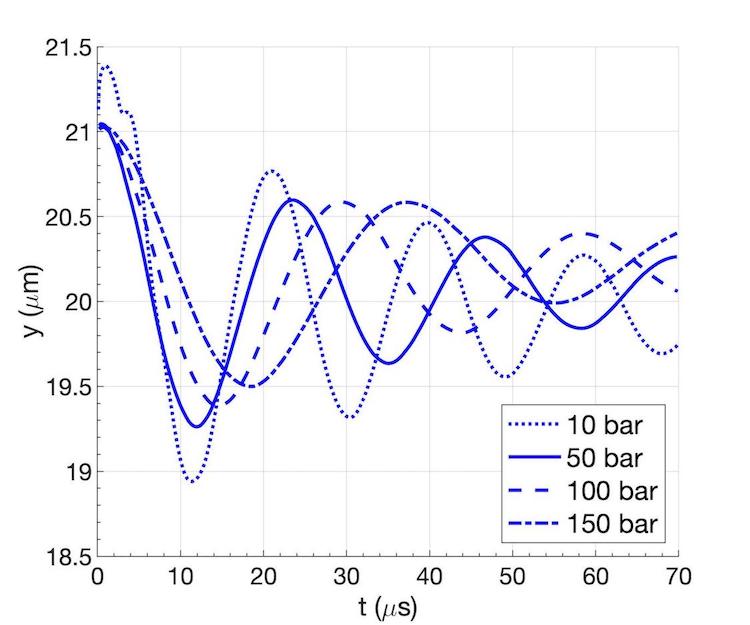}
  \caption{}
  \label{subfig:pressures_M3_W1_amplitude}
\end{subfigure}%
\begin{subfigure}{.5\textwidth}
  \centering
  \includegraphics[width=1.0\linewidth]{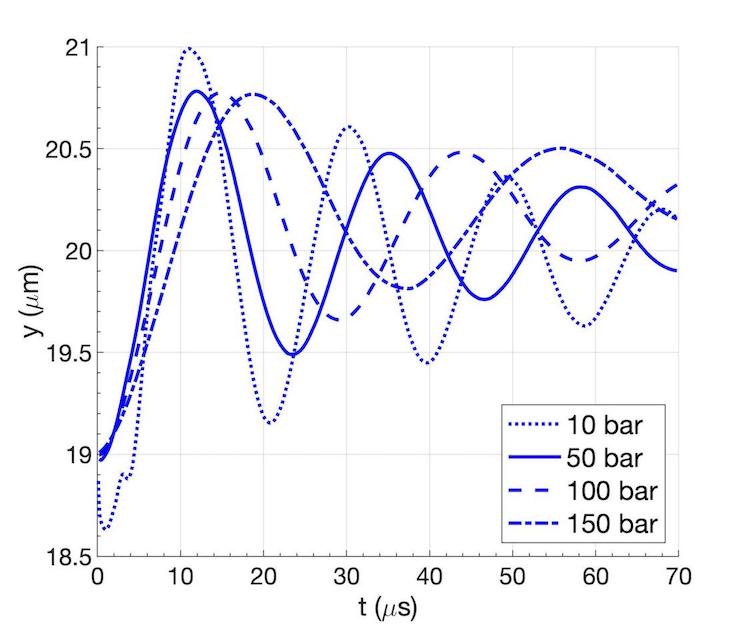}
  \caption{}
    \label{subfig:pressures_M3_W2_amplitude}
\end{subfigure}%
\caption{Interface amplitude relaxation of the two-dimensional capillary wave at 10, 50, 100 and 150 bar using mesh M3. (a) at \(x=7.5\) \(\mu\)m (initial wave crest location); and (b) at \(x=22.5\) \(\mu\)m (initial wave trough location).}
\label{fig:pressures_M3_W1W2_amplitude}
\end{figure}

Another important feature of high-pressure, two-phase flows is the reduction of surface-tension forces with a subsequent increase in the time for dynamic relaxation to a smaller surface area per unit volume. Figure~\ref{fig:pressures_M3_W1W2_amplitude} shows the relaxation of the wave amplitude over time at different pressures. The vertical position of the interface at the initial wave crest location of \(x=7.5\) \(\mu\)m is shown in Figure~\ref{subfig:pressures_M3_W1_amplitude}, and the interface vertical location at the initial wave trough location of \(x=22.5\) \(\mu\)m is shown in Figure~\ref{subfig:pressures_M3_W2_amplitude}. Notice the stronger reaction to the sharp initial conditions at 10 bar, where the interface displacement deviates considerably from the behavior at higher pressures. This pattern may be caused by a combination of strong initial vaporization at 10 bar and the numerical method sensitivity to high density ratios, \(\rho_l/\rho_g\). Nevertheless, clear conclusions can be extracted. \par 

The effect of volume expansion is noticed as the wave amplitude presents a drift to higher \(y\) values as pressure increases. Moreover, higher surface-tension forces at lower pressures cause higher-frequency interface oscillations. The liquid viscosity near the interface drops as pressure increases because of the enhanced dissolution of the lighter gas species. Despite this reduction in viscosity, the coupled dynamic system produces similar damping of the wave amplitude for all analyzed pressures. Reference values of the surface-tension coefficient and the interface liquid and gas densities and viscosities are available in Davis et al.~\cite{davis2019development}. In that work, the same thermodynamic model used here is implemented, but no interface deformation is considered. Still, the small deformations in the capillary wave test do not change the interface solution significantly. \par 

\subsection{Two-dimensional droplet under forced convection}
\label{subsec:drop_res2D}

The temporal evolution of a two-dimensional cylindrical droplet under forced convection at high pressures is considered to verify the dynamical behavior of the two-phase solver and the interface thermodynamics. The same problem is analyzed in Deng et al.~\cite{deng1992two}, but where the liquid droplet is incompressible and the thermodynamic model is built on a Redlich-Kwong (RK) equation of state~\cite{redlich1949thermodynamics}. \par

A pure \textit{n}-heptane droplet with a diameter of 0.2 mm and an initial temperature of 300 K is immersed in a hotter nitrogen gaseous stream at 500 K, moving with a freestream velocity of 4 m/s. The droplet is initially at rest. The ambient pressure is 40 bar, which is supercritical for the fuel species. The critical temperature and pressure of \textit{n}-heptane are, respectively, 540 K and 27.4 bar. Deng et al.~\cite{deng1992two} initialize the gaseous phase everywhere with a uniform velocity equal to the freestream velocity. However, this initialization is physically questionable, especially at the droplet's surface, and, in fact, our pressure solver immediately corrects the velocity field by instantaneously accelerating the droplet. To avoid the issue, we initialize the velocity field in the gas phase with the potential flow solution around a stationary circle and obtain good agreement with the droplet displacement shown in the reference. \par 

The computational domain has a length of 5 mm and a height of 2 mm. Uniform inflow boundary conditions are considered at the right boundary and outflow boundary conditions are assumed across the other three boundaries. The droplet is initially centered at a distance of 1 mm from the inlet and at the midpoint between the top and bottom boundaries. Three different uniform meshes are considered, summarized in Table~\ref{tab:2Ddropmesh}. \par 

\begin{table}[h!]
\centering
\begin{tabular}{ c c c c }
\hline
Mesh & D1 & D2 & D3 \\
 \hline
\(\Delta x\) (\(\mu\)m) & 4.167 & 2.778 & 1.852 \\
Cells/Diameter & 48 & 72 & 108 \\
\hline
\end{tabular}
\caption{Mesh properties used in the analysis of a two-dimensional cylindrical droplet at supercritical pressures. The number of cells per diameter refer to the initial configuration of the liquid-gas interface.}
\label{tab:2Ddropmesh}
\end{table}

\begin{figure}[h!]
\centering
\begin{subfigure}{.33\textwidth}
  \centering
  \includegraphics[width=1.0\linewidth]{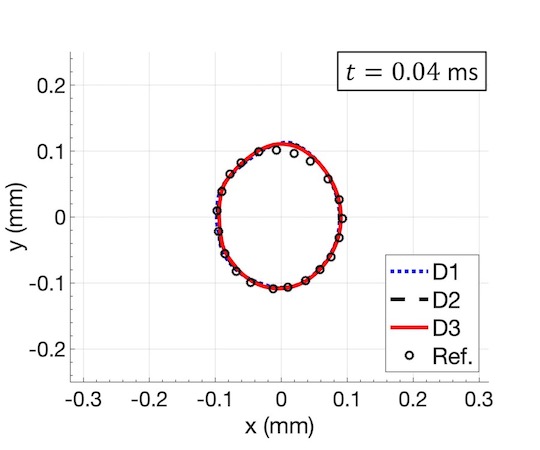}
  \label{subfig:time_004_tag}
\end{subfigure}%
\begin{subfigure}{.33\textwidth}
  \centering
  \includegraphics[width=1.0\linewidth]{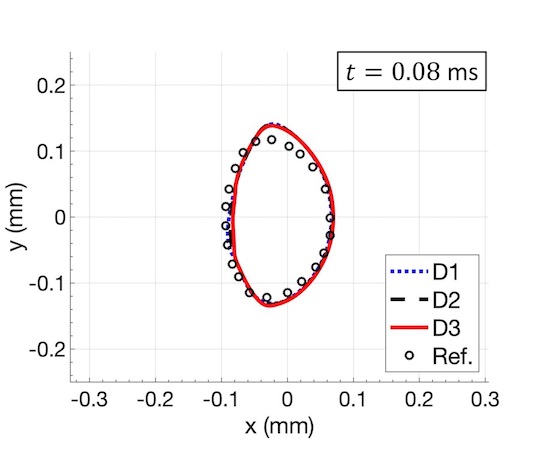}
    \label{subfig:time_008_tag}
\end{subfigure}%
\begin{subfigure}{.33\textwidth}
  \centering
  \includegraphics[width=1.0\linewidth]{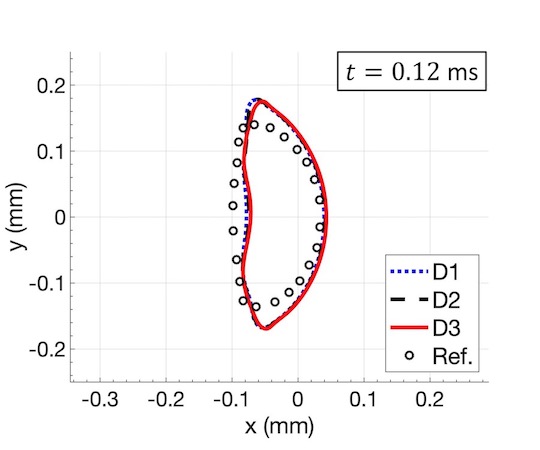}
    \label{subfig:time_012_tag}
\end{subfigure}%
\\[-3ex]
\begin{subfigure}{.33\textwidth}
  \centering
  \includegraphics[width=1.0\linewidth]{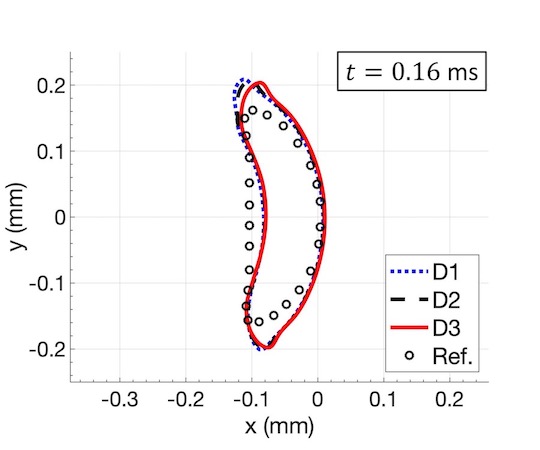}
  \label{subfig:time_016_tag}
\end{subfigure}%
\begin{subfigure}{.33\textwidth}
  \centering
  \includegraphics[width=1.0\linewidth]{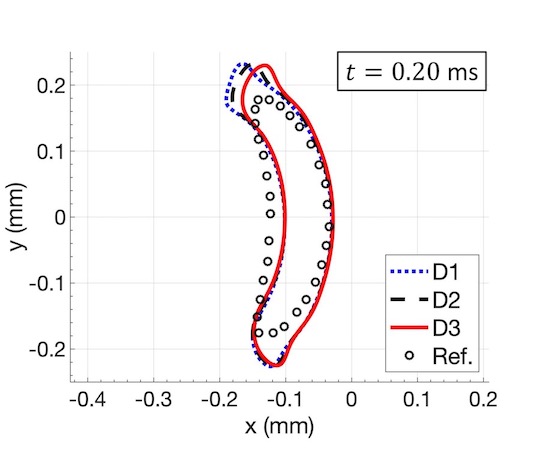}
    \label{subfig:time_020_tag}
\end{subfigure}%
\begin{subfigure}{.33\textwidth}
  \centering
  \includegraphics[width=1.0\linewidth]{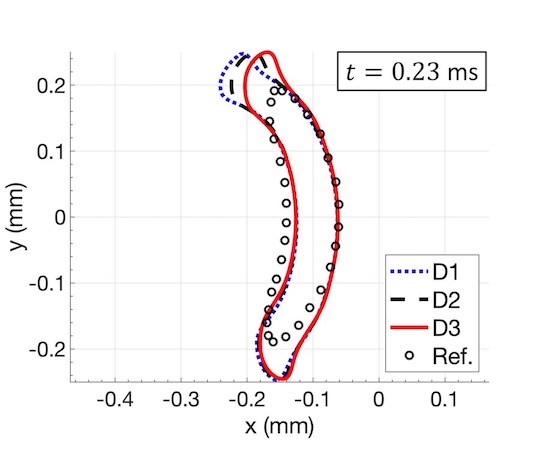}
    \label{subfig:time_023_tag}
\end{subfigure}%
\caption{Surface deformation of the two-dimensional cylindrical droplet with three mesh sizes D1, D2 and D3 compared against the reference solution from Deng et al.~\cite{deng1992two}. Plotted times correspond to 0.04 ms, 0.08 ms, 0.12 ms, 0.16 ms, 0.20 ms and 0.23 ms. The interface shape is represented by the centroid location of each interface plane obtained from PLIC.}
\label{fig:2d_drop_location}
\end{figure}

Figure~\ref{fig:2d_drop_location} shows the displacement of the droplet and its deformation over time compared to the results presented in Deng et al.~\cite{deng1992two}. The incoming gas phase flows from right to left. Laboratory coordinates are used where the droplet is initially centered at \(x=y=0\). Good agreement is found in the displacement of the upstream side of the droplet. The early snapshots show the reference solution moving slightly faster. Such difference may be explained by the different pure liquid density predictions between the equations of state, about 580 kg/m\(^3\) with the RK equation and about 693 kg/m\(^3\) with the volume-corrected SRK equation. For both models, the pure nitrogen gas phase density is about 26 kg/m\(^3\). \par 

\begin{figure}[h!]
\centering
\begin{subfigure}{.5\textwidth}
  \centering
  \includegraphics[width=0.9\linewidth]{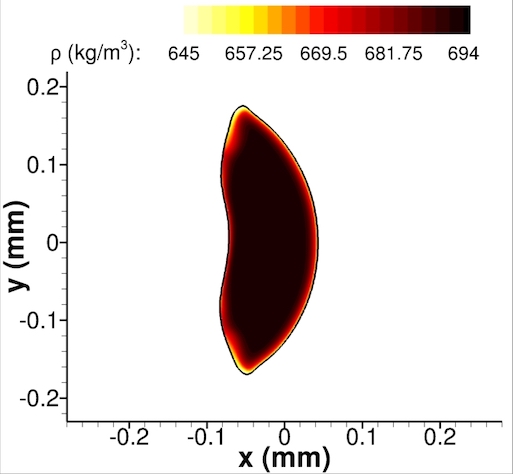}
  \caption{}
  \label{subfig:den_t012_108}
\end{subfigure}%
\begin{subfigure}{.5\textwidth}
  \centering
  \includegraphics[width=0.9\linewidth]{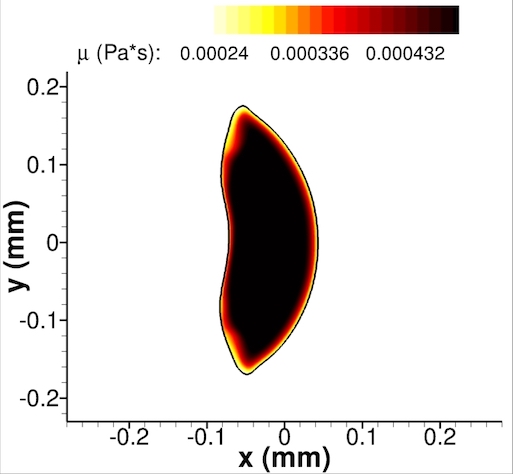}
  \caption{}
    \label{subfig:vis_t012_108}
\end{subfigure}%
\caption{Liquid density and viscosity of the two-dimensional cylindrical droplet at \(t=0.12\) ms. The results obtained with mesh D3 are shown. (a) liquid density; and (b) liquid viscosity.}
\label{fig:2d_drop_vis_and_den}
\end{figure}

The deformation of the droplet is different. This deviation is not unexpected since our model considers varying fluid properties in the liquid phase, where density and viscosity drop considerably as nitrogen dissolves and the liquid heats. Such a scenario is also discussed in Subsection~\ref{subsubsec:2djet_pressure} for the two-dimensional planar jet test. As observed in Figure~\ref{fig:2d_drop_vis_and_den}, the liquid density and viscosity drop across the mixing layer inside the droplet, especially near the top and bottom edges, where higher heat transfer increases the surface temperature and enhances the dissolution of nitrogen. Note that the liquid viscosity may drop by a factor of two. The pure gas viscosity is \(2.605\times10^{-5}\) Pa\(\cdot\)s and the pure liquid viscosity is \(5\times10^{-4}\) Pa\(\cdot\)s. Such variation in fluid properties results in an increased transverse stretching of the droplet compared to the incompressible reference solution. \par

Additionally, some directional bias in the numerical solution affects the evolution of the droplet's top edge. The code alternates sweeping directions to minimize such bias. As the mesh is refined, a symmetric behavior is recovered. Nonetheless, there has not been a similar problem in the two- and three-dimensional planar jet solutions presented in Subsections~\ref{subsec:jet_res2D} and~\ref{subsec:jet_res3D}. \par

The apparent difference in droplet volume (i.e., area in two dimensions) between both solutions is explainable. During the early times, the volume contained by both solutions is approximately the same. Later, greater differences exist. Regardless of the mass-conservation properties of the interface tracking/capturing approach, there exist differences of physical nature. The compressible solution predicts volume expansion up to \(t\approx0.5\) ms despite the droplet's vaporization, caused by the heating of the liquid phase and the dissolution of nitrogen. Moreover, the vaporization rates predicted by our model are substantially smaller. \par

\begin{figure}[h!]
\centering
\includegraphics[width=0.5\textwidth]{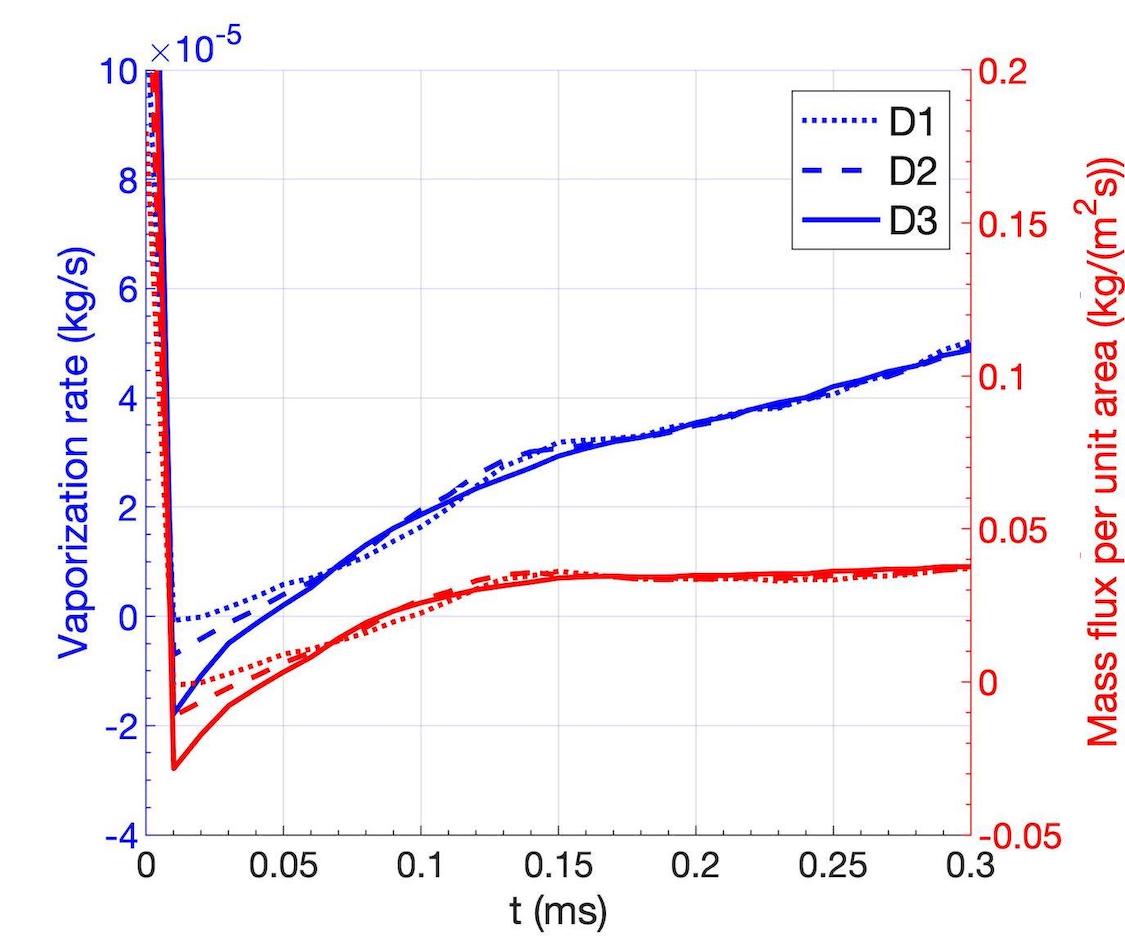}
\caption{Temporal evolution of the vaporization rate and average mass flux per unit area for the two-dimensional cylindrical droplet.}
\label{fig:mesh_massflux}
\end{figure}

Figure~\ref{fig:mesh_massflux} presents the evolution over time of the vaporization rate and an average mass flux per unit area. Substantial vaporization occurs during the adjustment of the initial conditions, but vaporization rates rapidly decrease and even show some transient net condensation due to the high-pressure environment. After some time, the mass flux per unit area stabilizes to a constant value, as reported in Deng et al.~\cite{deng1992two}, and the vaporization rates increase with the growth of the surface area. Nonetheless, the solution of the interface matching relations and LTE, coupled with the volume-corrected SRK equation of state, predicts vaporization rates at least an order of magnitude lower than those reported in Deng et al.~\cite{deng1992two}. \par

This result does not come as a surprise. The RK equation of state predicts vapor pressures of substances with high acentric factors poorly~\cite{soave1972equilibrium} (i.e., like \textit{n}-heptane), which in turn affects the predicted equilibrium solutions for mixtures including such components. Zhu and Aggarwal~\cite{zhu2000transient} detail this problem in binary mixtures of \textit{n}-heptane and nitrogen and highlight the superior performance of the SRK equation of state. The deviations introduced by the RK model overestimate the liquid species' vaporization and the gaseous species' dissolution at high pressures. At the same time, the heat of vaporization necessary for the change of phase is underestimated by a varying factor between 2 and 10 times less than the SRK model~\cite{zhu2000transient}. \par

Further, it is worth highlighting the discrepancies in the reported mixture critical temperature for the considered binary mixture at 40 bar. Deng et al.~\cite{deng1992two} report a critical temperature of 470 K, while the RK model implemented in Zhu and Aggarwal~\cite{zhu2000transient} suggests a critical temperature between 515 K and 540 K for a pressure of 40 bar. Such difference could aggravate the excess vaporization. On the other hand, the critical temperature predicted by our SRK model is about 538 K, in line with previous estimates~\cite{zhu2000transient}.

\subsection{Three-dimensional droplet evaporation}
\label{subsec:drop_3D}

The vaporization of a three-dimensional \textit{n}-heptane droplet immersed in a quiescent environment of hotter gaseous nitrogen is analyzed. Experimental results for the evaporation of this fuel droplet subject to a wide range of ambient conditions are presented in Nomura et al.~\cite{nomura1996experimental}. The analyzed pressures vary from 1 bar to 50 bar, while the nitrogen gas temperature varies from 400 K to 800 K. That is, a wide range of sub- and supercritical conditions is investigated. These experimental results have been used to validate droplet models under transcritical conditions, such as those from Zhu and Aggarwal~\cite{zhu2000transient} and Zhang~\cite{zhang2003evaporation}. \par

Here, we focus on the evaporation of the \textit{n}-heptane droplet at transcritical conditions. The initial droplet diameter and temperature are, respectively, \(D=0.5\) mm and 300 K. The ambient nitrogen is at 50 bar and 493 K. The computational domain is a cubic box of size 6.4\(D\). The computational cost constrains the domain size since long physical times are required to achieve significant volume reduction. Furthermore, the simulations are performed by using the symmetry planes. In other words, only 1/8th of the spherical droplet is considered. Outflow boundary conditions are imposed along the non-symmetric boundaries. Details on the uniform mesh resolution are presented in Table~\ref{tab:3Ddropmesh}, where the resolution recommendation from Baraldi et al.~\cite{baraldi2014mass} has been followed to limit the impact of spurious currents and other geometrical errors on the droplet's surface. Nonetheless, as the liquid vaporizes, such a level of resolution will be lost. \par

\begin{table}[h!]
\centering
\begin{tabular}{ c c c c }
\hline
Mesh & S1 & S2 & S3 \\
 \hline
\(\Delta x\) (\(\mu\)m) & 20.83 & 15.63 & 12.5 \\
Cells/Diameter & 24 & 32 & 40 \\
\hline
\end{tabular}
\caption{Mesh properties used in the analysis of a three-dimensional spherical droplet evaporating at supercritical pressures. The number of cells per diameter refer to the initial configuration of the liquid-gas interface.}
\label{tab:3Ddropmesh}
\end{table}

\begin{figure}[h!]
\centering
\begin{subfigure}{.5\textwidth}
  \centering
  \includegraphics[width=1.0\linewidth]{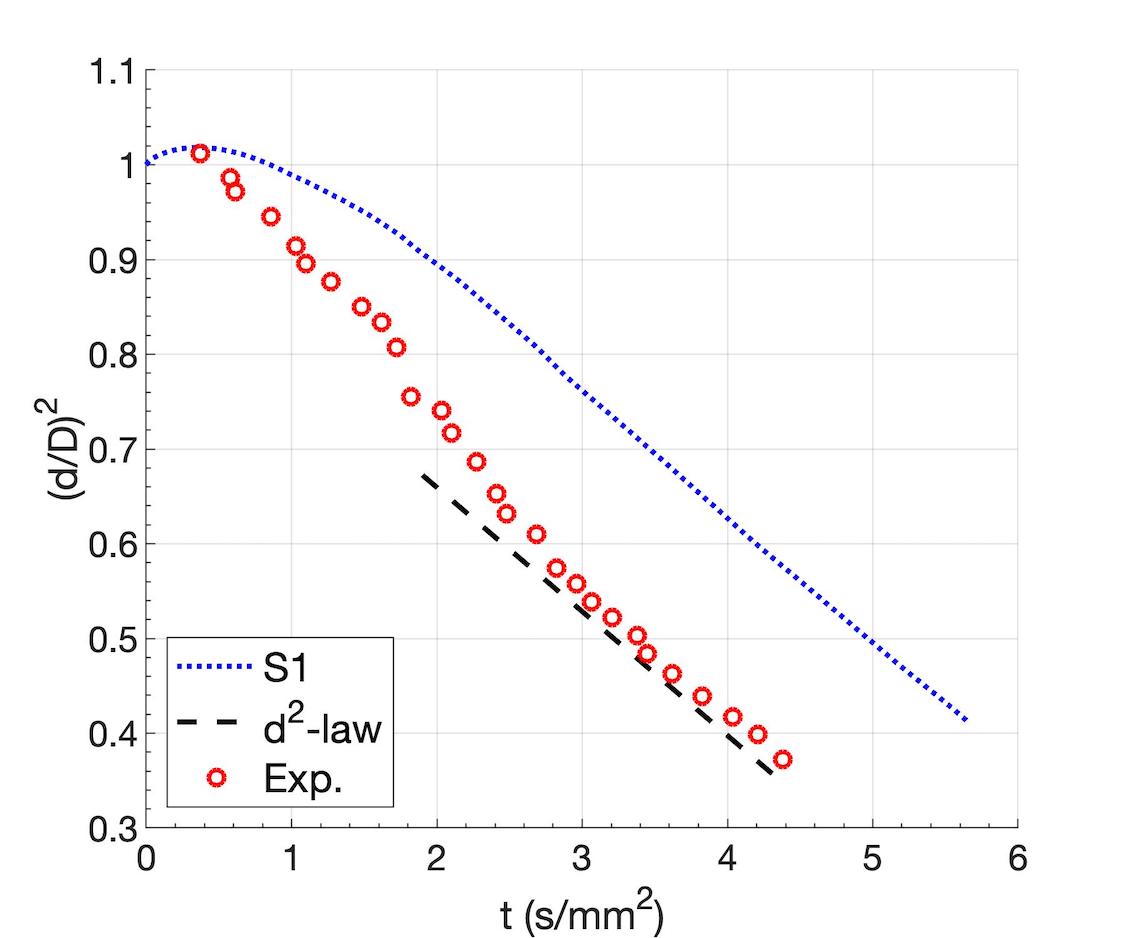}
  \caption{}
  \label{subfig:diam_493K}
\end{subfigure}%
\begin{subfigure}{.5\textwidth}
  \centering
  \includegraphics[width=1.0\linewidth]{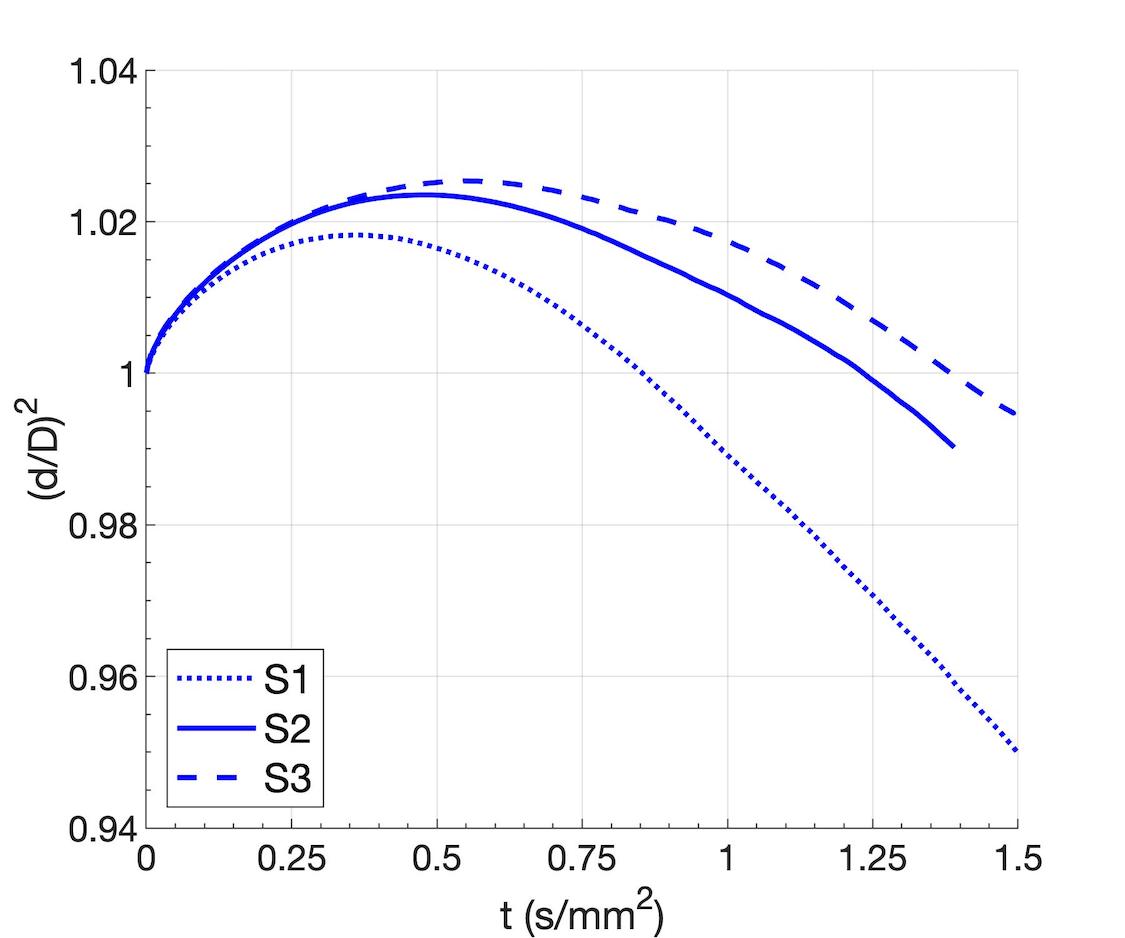}
  \caption{}
    \label{subfig:diam_493K_mesh}
\end{subfigure}%
\caption{Temporal evolution of \((d/D)^2\) for the three-dimensional vaporizing \textit{n}-heptane droplet. (a) \((d/D)^2\) evolution obtained with mesh S1 is compared to experimental data~\cite{nomura1996experimental}. The \(d^2\)-law slope agrees after the initial transient process; and (b) \((d/D)^2\) evolution during the early transient volume increase obtained with meshes S1, S2 and S3.}
\label{fig:3d_drop}
\end{figure}

Figure~\ref{fig:3d_drop} presents the evolution of the droplet diameter in a \((d/D)^2\) vs. time plot, where \(D\) is the initial droplet diameter defined earlier. Due to computational constraints, results for an extended time are only available for mesh S1. The instantaneous droplet diameter, \(d\), is obtained from the liquid volume data (i.e., \(V=\pi D^3/6\)) since the droplet remains spherical at all times. Spurious currents exist but do not affect the droplet's shape nor the surface regression under vaporization. A deviation between the numerical results and the experimental data is observed in Figure~\ref{subfig:diam_493K}. Initially, a transient process occurs where the droplet volume increases under thermal expansion and the enhanced dissolution of nitrogen, counteracting the surface regression due to vaporization. Then, the droplet heats and a classical \(d^2\)-law is recovered where the square of the diameter decreases linearly over time. \par

The apparent translation of the \((d/D)^2\) curve is also reported in Zhang~\cite{zhang2003evaporation} when validating against experimental data from Nomura et al.~\cite{nomura1996experimental}. It is concluded that the experimental setup induces enhanced vaporization of the droplet when it is moved from the generator to the test position. When including this displacement in the droplet model, excellent agreement with the experimental data is found~\cite{zhang2003evaporation}. On the other hand, the droplet model from Zhu and Aggarwal~\cite{zhu2000transient} using the SRK equation of state predicts a much better agreement. However, liquid volume expansion is not considered, which might overestimate the surface regression during the early times. \par

Note that the \(d^2\)-law slope compares to the experimental data, which further corroborates the accuracy of the phase change predictions using the thermodynamic model based on the volume-corrected SRK equation of state. With a less accurate thermodynamic model, such as one based on the RK equation of state, surface regression at transcritical conditions is overestimated and the \(d^2\)-law slope deviates considerably from the experimental data~\cite{zhu2000transient}. A similar discussion has been provided in Subsection~\ref{subsec:drop_res2D}. \par 

Lastly, Figure~\ref{subfig:diam_493K_mesh} shows the effect of mesh resolution during the early transient volume increase. Mesh refinement predicts more liquid volume expansion, and the onset of vaporization-dominated surface regression is delayed. This behavior is explained by our model better resolving the initial thin diffusion layers in the liquid phase as the mesh is refined. Mesh convergence of the evolution of \((d/D)^2\) is observed. \par

\subsection{Two-dimensional planar jet}
\label{subsec:jet_res2D}

A temporal study of a symmetric two-dimensional planar jet is performed to demonstrate the ability of the numerical model in capturing the interface deformation and the relevant high-pressure physics during the liquid injection. Sustained liquid surface deformation occurs in this test. Although planar jets might develop an antisymmetric behavior~\cite{zandian2018understanding}, the test problem is constrained to the symmetric configuration to limit the computational cost. \par 

The jet half-thickness is 10 \(\mu\)m and the interface is initially perturbed with a sinusoidal wave of 30 \(\mu\)m wavelength and 0.5 \(\mu\)m amplitude. The choice of wavelength is made following previous works in the incompressible and high-pressure weakly compressible frameworks~\cite{zandian2017planar,zandian2018understanding,poblador2019axisymmetric}. A thin velocity distribution of a few micrometers (i.e., about 6 \(\mu\)m) is imposed around the interface, where the streamwise velocity varies from 0 m/s in the liquid phase to 30 m/s in the gas phase with a hyperbolic tangent profile following the equation \(u(y)=15\big(\tanh{\big[6.5\times 10^{5}(y-10\times 10^{-6})\big]}+1\big)\). Periodic boundary conditions are imposed in the streamwise direction and outflow boundary conditions are imposed in the gas phase along the top boundary of the numerical domain. The top boundary is sufficiently far from the two-phase mixing region to not affect the results. The thermodynamic pressure is 150 bar and the liquid jet is initially composed of pure \textit{n}-decane at 450 K and the surrounding gas is pure oxygen at 550 K. Only one wavelength is enclosed in the computational domain and periodicity in the streamwise direction may be used to plot the results. \par 

The low-Mach-number domain is verified using the thermodynamic model. At this high pressure, the speed of sound in the gas phase obtained with the thermodynamic model is approximately 450 m/s. Therefore, a gas velocity of 30 m/s results in a Mach number of \(M \approx 0.0667\). The development of a compressible pressure equation (not the PPE utilized in this work) shows that compressible terms (i.e., wave-like equation) scale with \(M^2\approx 0.00444\sim\mathcal{O}(10^{-3})\), which can be reasonably neglected. Faster jet velocities up to 100 m/s could be analyzed in future works with \(M^2\sim\mathcal{O}(10^{-2})\) as the limit before the low-Mach-number model needs to be revised. \par

\subsubsection{Mesh resolution and mass conservation}
\label{subsubsec:2djet_mesh}

The results presented in Subsection~\ref{subsec:2Dres} show good mesh convergence and mass-conservation properties for configurations with small deformations. In other words, situations where the mesh resolves the shape of the liquid with high accuracy at all times. However, liquid atomization problems present a cascade process whereby smaller and more complex liquid structures develop before they break up into ligaments and droplets. \par

An analysis of mesh dependence both in the liquid surface deformation and mass-conservation properties is presented in this section. Three different uniform mesh sizes between M3 and M4 (see Table~\ref{tab:2Dwavemesh}) are analyzed. Their characteristics are presented in Table~\ref{tab:2Djetmesh}. \par 

\begin{table}[h!]
\centering
\begin{tabular}{ c c c c }
\hline
Mesh & J1 & J2 & J3 \\
 \hline
\(\Delta x\) (\(\mu\)m) & 1/10 & 1/15 & 1/20 \\
Cells/Wavelength & 300 & 450 & 600 \\
Cells/Amplitude & 5 & 7.5 & 10 \\
\hline
\end{tabular}
\caption{Mesh properties used in the analysis of a two-dimensional planar liquid jet at supercritical pressures. The number of cells per wavelength or amplitude refer to the initial configuration of the liquid-gas interface.}
\label{tab:2Djetmesh}
\end{table}

\begin{figure}[h!]
\centering
\begin{subfigure}{.33\textwidth}
  \centering
  \includegraphics[width=0.9\linewidth]{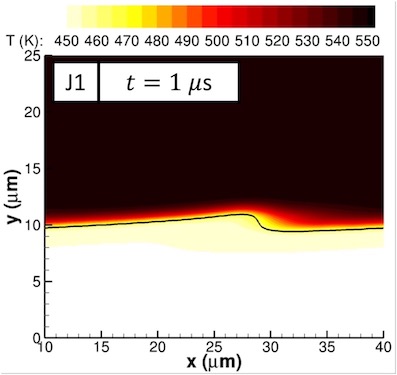}
  \label{subfig:C_1mus}
\end{subfigure}%
\begin{subfigure}{.33\textwidth}
  \centering
  \includegraphics[width=0.9\linewidth]{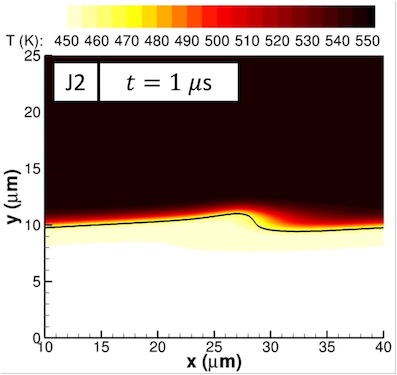}
    \label{subfig:M_1mus}
\end{subfigure}%
\begin{subfigure}{.33\textwidth}
  \centering
  \includegraphics[width=0.9\linewidth]{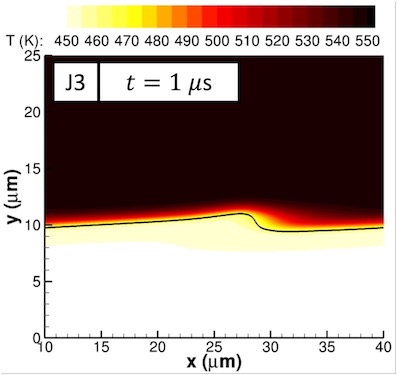}
    \label{subfig:F_1mus}
\end{subfigure}%
\\
\begin{subfigure}{.33\textwidth}
  \centering
  \includegraphics[width=0.9\linewidth]{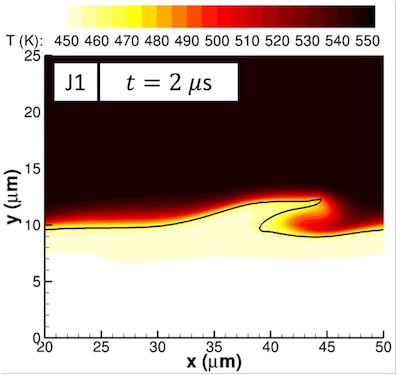}
  \label{subfig:C_1p55mus}
\end{subfigure}%
\begin{subfigure}{.33\textwidth}
  \centering
  \includegraphics[width=0.9\linewidth]{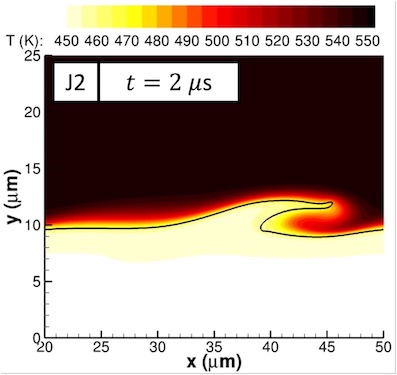}
    \label{subfig:M_1p55mus}
\end{subfigure}%
\begin{subfigure}{.33\textwidth}
  \centering
  \includegraphics[width=0.9\linewidth]{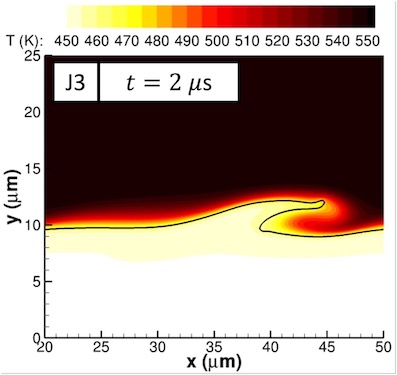}
    \label{subfig:F_1p55mus}
\end{subfigure}%
\\
\begin{subfigure}{.33\textwidth}
  \centering
  \includegraphics[width=0.9\linewidth]{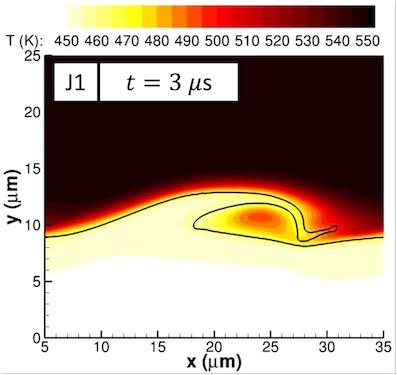}
  \label{subfig:C_2mus}
\end{subfigure}%
\begin{subfigure}{.33\textwidth}
  \centering
  \includegraphics[width=0.9\linewidth]{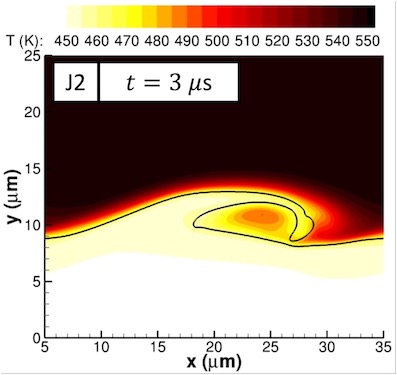}
    \label{subfig:M_2mus}
\end{subfigure}%
\begin{subfigure}{.33\textwidth}
  \centering
  \includegraphics[width=0.9\linewidth]{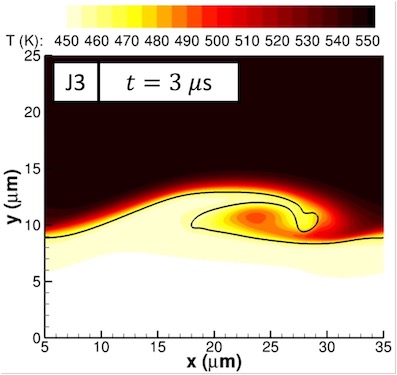}
    \label{subfig:F_2mus}
\end{subfigure}%
\\
\begin{subfigure}{.33\textwidth}
  \centering
  \includegraphics[width=0.9\linewidth]{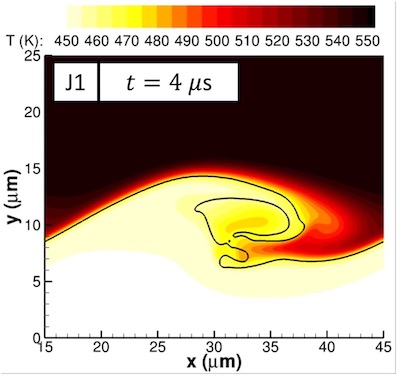}
  \label{subfig:C_2p55mus}
\end{subfigure}%
\begin{subfigure}{.33\textwidth}
  \centering
  \includegraphics[width=0.9\linewidth]{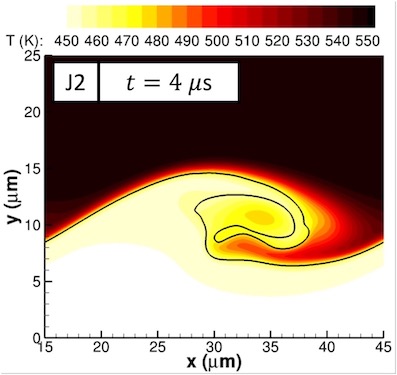}
    \label{subfig:M_2p55mus}
\end{subfigure}%
\begin{subfigure}{.33\textwidth}
  \centering
  \includegraphics[width=0.9\linewidth]{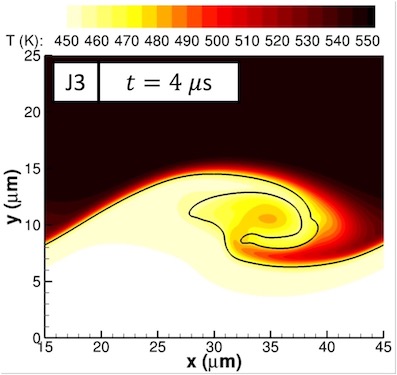}
    \label{subfig:F_2p55mus}
\end{subfigure}%
\caption{Temperature plots and interface deformation at 150 bar for the two-dimensional planar jet with three mesh sizes J1, J2 and J3. Plotted times correspond to \(1\) \(\mu\)s, \(2\) \(\mu\)s, \(3\) \(\mu\)s and \(4\) \(\mu\)s. The interface location is highlighted with a solid black curve representing the isocontour with \(C=0.5\).}
\label{fig:2djet_T_mesh}
\end{figure}

\begin{figure}[h!]
\centering
\begin{subfigure}{.33\textwidth}
  \centering
  \includegraphics[width=1.0\linewidth]{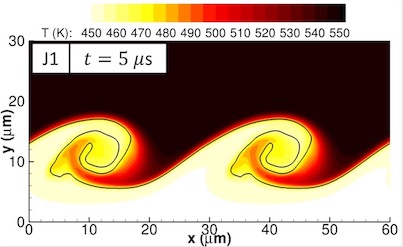}
  \label{subfig:C_5mus}
\end{subfigure}%
\begin{subfigure}{.33\textwidth}
  \centering
  \includegraphics[width=1.0\linewidth]{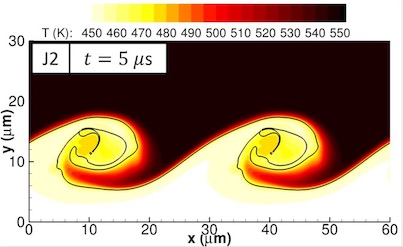}
    \label{subfig:M_5mus}
\end{subfigure}%
\begin{subfigure}{.33\textwidth}
  \centering
  \includegraphics[width=1.0\linewidth]{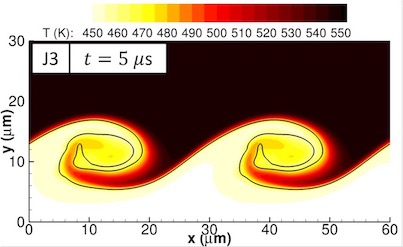}
    \label{subfig:F_5mus}
\end{subfigure}%
\\[-3ex]
\begin{subfigure}{.33\textwidth}
  \centering
  \includegraphics[width=1.0\linewidth]{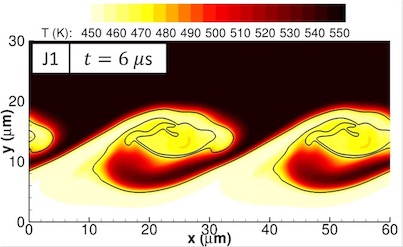}
  \label{subfig:C_6mus}
\end{subfigure}%
\begin{subfigure}{.33\textwidth}
  \centering
  \includegraphics[width=1.0\linewidth]{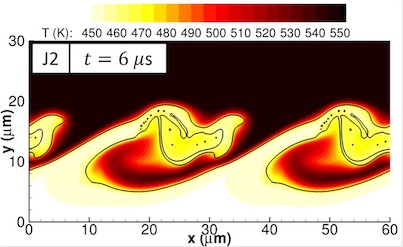}
    \label{subfig:M_6mus}
\end{subfigure}%
\begin{subfigure}{.33\textwidth}
  \centering
  \includegraphics[width=1.0\linewidth]{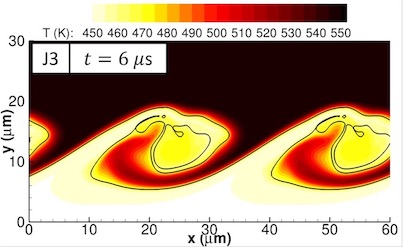}
    \label{subfig:F_6mus}
\end{subfigure}%
\\[-3ex]
\begin{subfigure}{.33\textwidth}
  \centering
  \includegraphics[width=1.0\linewidth]{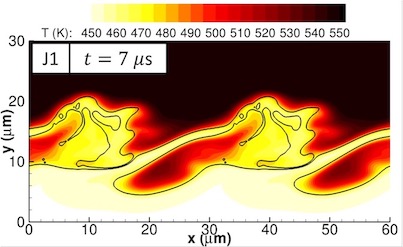}
  \label{subfig:C_7mus}
\end{subfigure}%
\begin{subfigure}{.33\textwidth}
  \centering
  \includegraphics[width=1.0\linewidth]{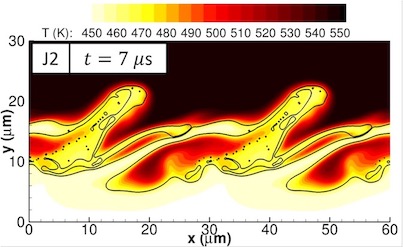}
    \label{subfig:M_7mus}
\end{subfigure}%
\begin{subfigure}{.33\textwidth}
  \centering
  \includegraphics[width=1.0\linewidth]{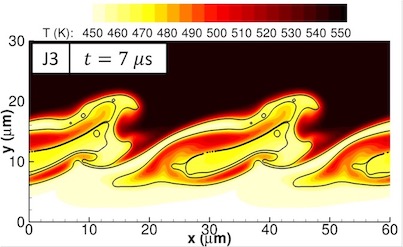}
    \label{subfig:F_7mus}
\end{subfigure}%
\\[-3ex]
\begin{subfigure}{.33\textwidth}
  \centering
  \includegraphics[width=1.0\linewidth]{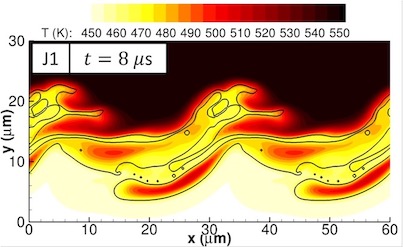}
  \label{subfig:C_8mus}
\end{subfigure}%
\begin{subfigure}{.33\textwidth}
  \centering
  \includegraphics[width=1.0\linewidth]{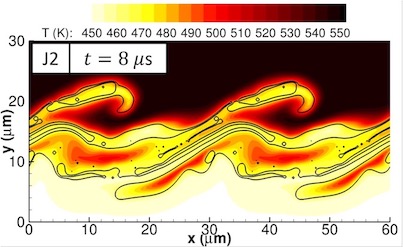}
    \label{subfig:M_8mus}
\end{subfigure}%
\begin{subfigure}{.33\textwidth}
  \centering
  \includegraphics[width=1.0\linewidth]{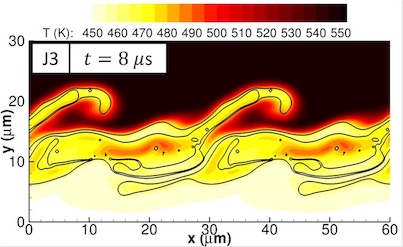}
    \label{subfig:F_8mus}
\end{subfigure}%
\caption{Temperature plots and interface deformation at 150 bar for the two-dimensional planar jet with three mesh sizes J1, J2 and J3. Plotted times correspond to \(5\) \(\mu\)s, \(6\) \(\mu\)s, \(7\) \(\mu\)s and \(8\) \(\mu\)s. The interface location is highlighted with a solid black curve representing the isocontour with \(C=0.5\).}
\label{fig:2djet_T_mesh3}
\end{figure}

Figures~\ref{fig:2djet_T_mesh} and~\ref{fig:2djet_T_mesh3} show the evolution of the temperature field and the interface deformation obtained with each mesh J1, J2 and J3. For the early times (i.e., \(t\leq 4.5\) \(\mu\)s), detailed close-ups of the interface perturbation are shown. A broader picture is presented once the liquid deformation becomes chaotic. The three meshes show almost an identical evolution up to \(t\approx 1.5\) \(\mu\)s. Then, the coarser mesh J1 starts to deviate from J2 and J3, as seen in the snapshots at 2 \(\mu\)s. Deviations between J2 and J3 start to appear after 2.5 \(\mu\)s. Overall, the qualitative evolution of the interface is very similar in J1, J2, and J3, even for longer times. Deviations are observable once small liquid structures, with respect to the mesh size, develop (e.g., high local curvature). These poorly-resolved regions suffer from less accurate interface geometrical and thermodynamic properties, as well as some simplifications to the numerical methodology that are implemented to allow for extensive temporal development. Therefore, the liquid surface may evolve differently once these mesh constraints appear. Another important consideration relates to the variation of fluid properties at these high pressures. Compared to grid-independence studies in two-phase incompressible flows, the resolution with which, for instance, the variable-density field is captured also has an impact in the dynamics of the liquid phase. Thus, many more factors contribute to the mesh performance. Some solutions to this problem exist, like using local mesh refinement near these under-resolved regions. However, they are not pursued here to avoid increased modeling complexity. Nevertheless, the mesh cannot be refined indefinitely because of computational and physical constraints. \par 

\begin{figure}[h!]
\centering
\begin{subfigure}{.5\textwidth}
  \centering
  \includegraphics[width=1.0\linewidth]{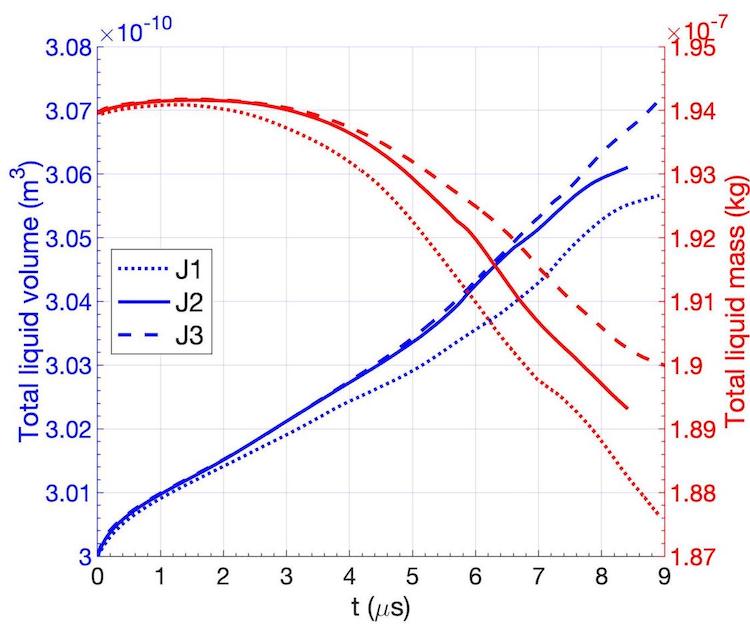}
  \caption{}
  \label{subfig:150mesh_totalvolume_totalmass_jet}
\end{subfigure}%
\begin{subfigure}{.5\textwidth}
  \centering
  \includegraphics[width=1.0\linewidth]{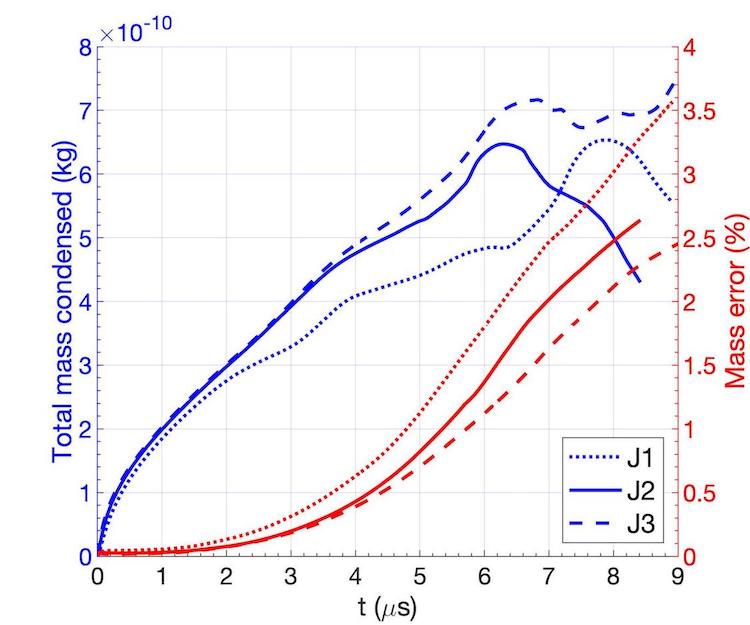}
  \caption{}
    \label{subfig:150mesh_totalmassexchange_masserror_jet}
\end{subfigure}%
\caption{Temporal evolution of the total liquid volume, the total liquid mass, the total net mass exchanged across the interface and the estimated mass error for the two-dimensional planar liquid jet at 150 bar. (a) total liquid volume and total liquid mass; and (b) total net mass exchanged and mass error.}
\label{fig:150mesh_totalvolumeandmass_jet}
\end{figure}

A mass conservation analysis is performed for the liquid jet. Mass errors are evaluated as defined in Subsection~\ref{subsubsec:2Dspatial}. A worse performance than the standing capillary wave problem is observed. The generation of smaller liquid structures typical of atomization problems is problematic for fixed meshes. Not only the evolution of the liquid surface is affected, but also mass-conservation properties suffer. J1 is the worst performing mesh, with considerable deviations from J2 and J3 early in the simulations. J2 and J3 present a similar performance until the liquid evolution starts to differ (see Figure~\ref{fig:150mesh_totalvolumeandmass_jet}). Total liquid volume remains nearly identical up to 6 \(\mu\)s. However, deviations in total liquid mass and total net mass condensed across the interface appear much sooner, approximately between 3 and 4 \(\mu\)s. Mass errors are below 0.5\% up to 4 \(\mu\)s approximately, and then they increase indefinitely as the deformation cascade develops. The mass errors, as defined in this work, are fairly negligible compared to the total liquid mass. During the analyzed times, the total mass exchanged across the interface only represents about 0.42\% of the liquid mass. \par 

Note that these errors are mainly related to a worse resolution of the liquid density field, the interface geometry and its equilibrium state. For example, as surface area grows and smaller structures form, more interface cells exist and their mass is approximated by considering the interface liquid density only. They do not limit the ability of the model to predict condensation and vaporization rates with reasonable accuracy, even for small liquid structures. As seen in Figure~\ref{fig:150mesh_totalvolumeandmass_jet}, the total mass exchanged across the interface results in net condensation up to 6 \(\mu\)s, with the liquid shape still being well-defined with none or minimal coalescence and breakup. Nevertheless, as thinner ligaments develop, the total liquid mass drops when it should increase. \par 

The analysis presented in this section suggests that achieving good mass-conservation properties and grid independence in the evolution of the liquid shape becomes a greater challenge than similar VOF codes used in incompressible two-phase flows. Computational costs and the physics limit how fine the mesh can be; thus, a compromise between numerical performance, physical coherence (i.e., avoiding a mesh size that enters the phase non-equilibrium transition layer across the interface) and simulation goals is required. If computational cost is not an issue, mesh J3 is preferred under this problem configuration. However, mesh J2 could be used in favor of cheaper three-dimensional computations with a similar problem configuration. \par 

\clearpage

\clearpage

\clearpage

\subsubsection{Supercritical pressure effects}
\label{subsubsec:2djet_pressure}

\begin{figure}[h!]
\centering
\begin{subfigure}{.5\textwidth}
  \centering
  \includegraphics[width=1.0\linewidth]{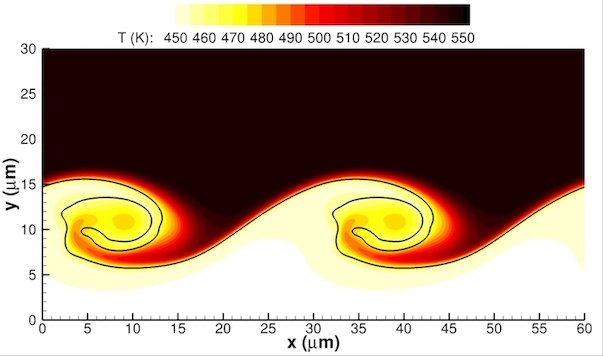}
  \caption{}
  \label{subfig:T_4p5mus_new}
\end{subfigure}%
\begin{subfigure}{.5\textwidth}
  \centering
  \includegraphics[width=1.0\linewidth]{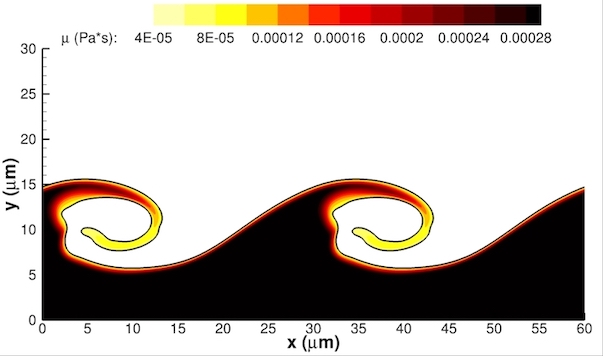}
  \caption{}
    \label{subfig:VIS_4p5mus_new}
\end{subfigure}%
\\
\begin{subfigure}{.5\textwidth}
  \centering
  \includegraphics[width=1.0\linewidth]{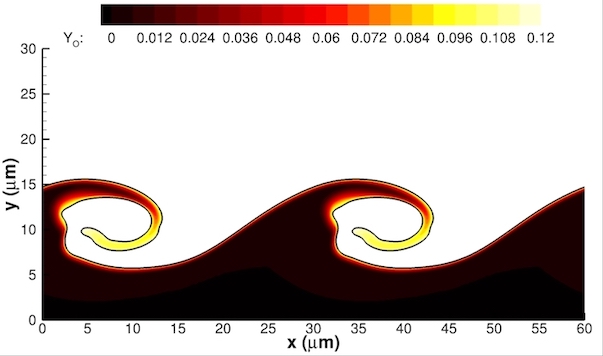}
  \caption{}
  \label{subfig:YOl_4p5mus_new}
\end{subfigure}%
\begin{subfigure}{.5\textwidth}
  \centering
  \includegraphics[width=1.0\linewidth]{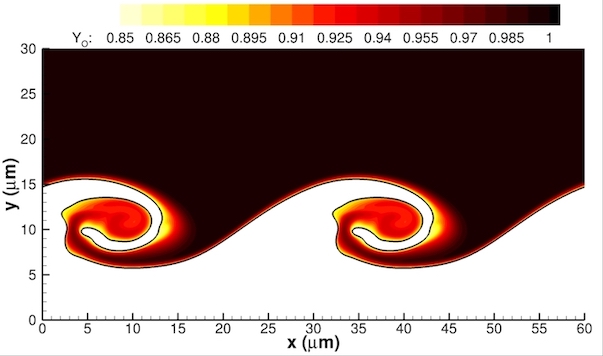}
  \caption{}
    \label{subfig:YOg_4p5mus_new}
\end{subfigure}%
\\
\begin{subfigure}{.5\textwidth}
  \centering
  \includegraphics[width=1.0\linewidth]{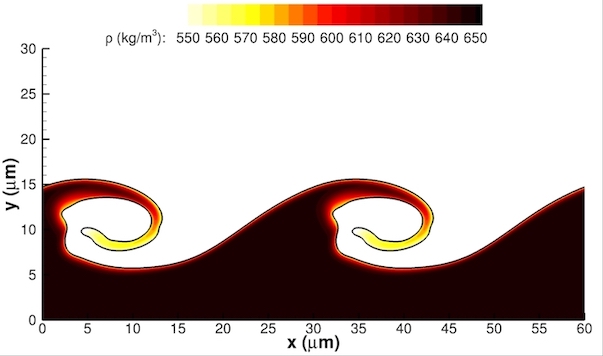}
  \caption{}
  \label{subfig:DENl_4p5mus_new}
\end{subfigure}%
\begin{subfigure}{.5\textwidth}
  \centering
  \includegraphics[width=1.0\linewidth]{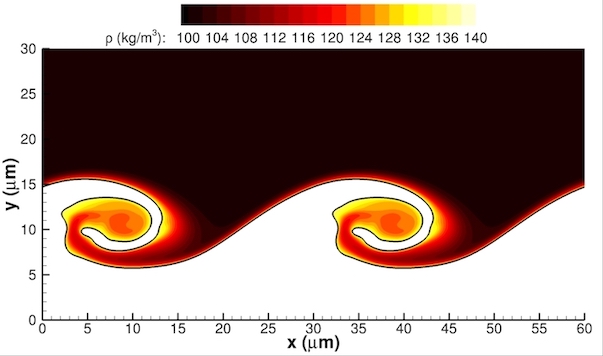}
  \caption{}
    \label{subfig:DENg_4p5mus_new}
\end{subfigure}%
\caption{Plots of various variables for the two-dimensional planar jet with mesh J3 at 150 bar and \(t=4.5\) \(\mu\)s. The interface location is highlighted with a solid black curve representing the isocontour with \(C=0.5\). The contour values are chosen to focus on a particular phase in some sub-figures. (a) temperature; (b) liquid viscosity; (c) oxygen mass fraction in the liquid phase; (d) oxygen mass fraction in the gas phase; (e) liquid density; and (f) gas density. }
\label{fig:2djet_J3}
\end{figure}

A summary of the main features of liquid injection at supercritical pressures is presented in this subsection. The results are obtained with mesh J3. Figure~\ref{fig:2djet_J3} presents the distributions of temperature, liquid viscosity, oxygen mass fraction, \(Y_O\), and density on each phase at \(t=4.5\) \(\mu\)s. The characteristics of the mixing process are seen where the swirling motion captures regions of hotter gas and higher oxygen concentration into the liquid structure. Moreover, the dissolution of the lighter oxygen into the liquid causes a decrease of the liquid viscosity to gas-like values near the interface and a substantial drop in liquid density with respect to liquid properties at the jet's core. These effects are more pronounced in the elongated liquid structures. This mixing process is responsible for the fast deformation of the liquid under the shear forces caused by the moving dense gas and for variations in the interface thermodynamic behavior along the interface. Later in Subsection~\ref{subsec:jet_res3D}, Figure~\ref{fig:150_30A_int_4mus} shows the variations of interface properties along the liquid surface in a three-dimensional configuration. \par 

Altogether, the fast growth of the surface instability at high pressures is apparent, which can be linked to faster atomization compared to injection at subcritical pressures. The time scales presented in this subsection are similar to those reported in temporal studies of axisymmetric liquid jets at high pressures~\cite{poblador2019axisymmetric}, although time scales may also be affected by the initial problem configuration. That is, the growth of surface instabilities will depend on the imposed perturbation and the initial velocity distribution across the interface. \par 

\newpage

\clearpage

\subsection{Three-dimensional planar jet}
\label{subsec:jet_res3D}

A temporal and symmetric three-dimensional planar jet is presented in this section to demonstrate further the ability of the numerical model to handle three-dimensional configurations with large surface deformations. Similar to the two-dimensional jet problem, the jet half-thickness is 10 \(\mu\)m and the interface is initially perturbed in the streamwise direction with a sinusoidal wave of 30 \(\mu\)m wavelength and 0.5 \(\mu\)m amplitude. Another sinusoidal perturbation in the spanwise direction is superimposed with a 20 \(\mu\)m wavelength and the amplitude ranges between 0.3-0.5 \(\mu\)m to enhance three-dimensional effects. \par 

Initially, the liquid is composed of \textit{n}-decane at 450 K and the gas is composed of oxygen at 550 K. The thermodynamic pressure is 150 bar and the streamwise velocity varies with the same hyperbolic tangent profile discussed in Subsection~\ref{subsec:jet_res2D} from 0 m/s in the liquid to 30 m/s in the gas. Periodic boundary conditions are imposed in the streamwise and spanwise directions and outflow boundary conditions are imposed in the gas domain top boundary away from the interface. Only one wavelength in each direction is enclosed in the computational domain and periodicity in the streamwise and spanwise directions may be used when plotting the results. \par 

\begin{figure}[h!]
\centering
\begin{subfigure}{.33\textwidth}
  \centering
  \includegraphics[width=1.0\linewidth]{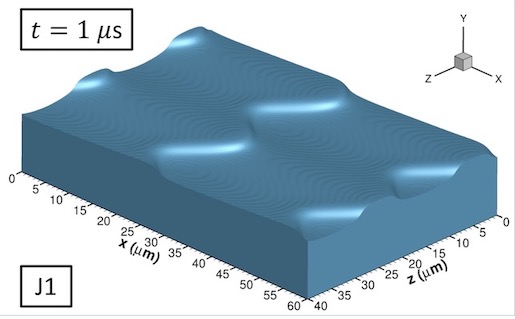}
  \label{subfig:150_A2_int_1mus_C}
\end{subfigure}%
\begin{subfigure}{.33\textwidth}
  \centering
  \includegraphics[width=1.0\linewidth]{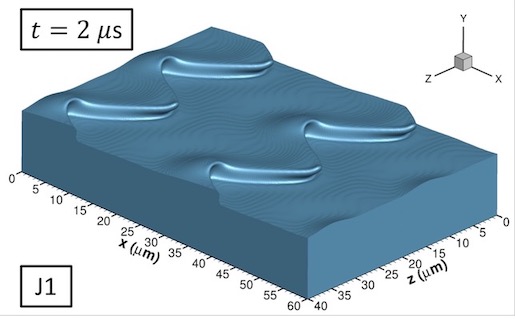}
    \label{subfig:150_A2_int_2mus_C}
\end{subfigure}%
\begin{subfigure}{.33\textwidth}
  \centering
  \includegraphics[width=1.0\linewidth]{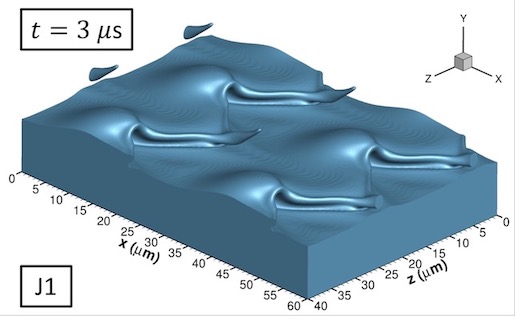}
    \label{subfig:150_A2_int_3mus_C}
\end{subfigure}%
\\[-3ex]
\begin{subfigure}{.33\textwidth}
  \centering
  \includegraphics[width=1.0\linewidth]{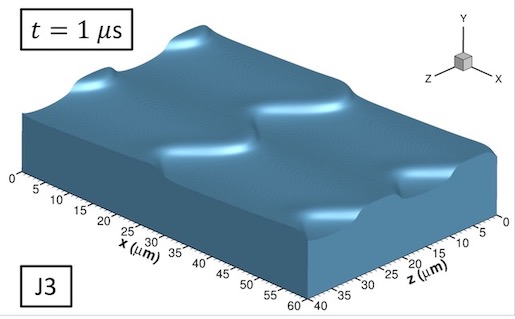}
  \label{subfig:150_A2_int_1mus}
\end{subfigure}%
\begin{subfigure}{.33\textwidth}
  \centering
  \includegraphics[width=1.0\linewidth]{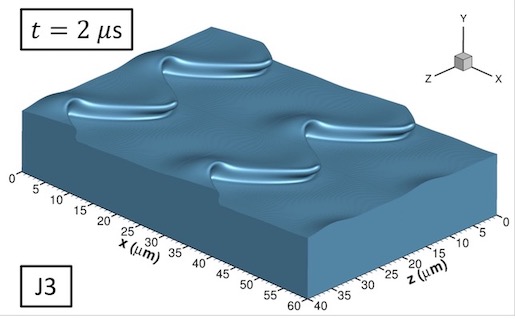}
    \label{subfig:150_A2_int_2mus}
\end{subfigure}%
\begin{subfigure}{.33\textwidth}
  \centering
  \includegraphics[width=1.0\linewidth]{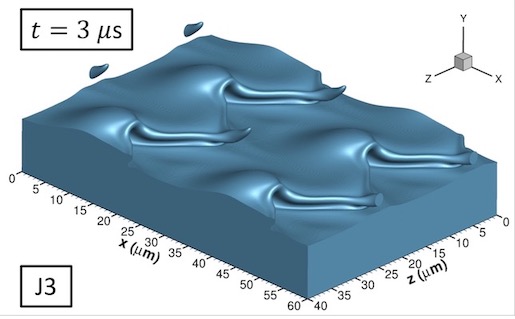}
    \label{subfig:150_A2_int_3mus}
\end{subfigure}%
\caption{Interface deformation at 150 bar for the three-dimensional planar jet with two meshes J1 and J3. The initial spanwise perturbation amplitude is 0.5 \(\mu\)m.}
\label{fig:3djet}
\end{figure}

Figure~\ref{fig:3djet} presents the interface deformation of the three-dimensional planar jet up to 3 \(\mu\)s in time with a spanwise perturbation amplitude of 0.5 \(\mu\)m. Two different meshes are used (i.e., J1 and J3) to estimate the effect of mesh size in the development of three-dimensional liquid structures. The liquid surface evolution is practically identical with J1 and J3. Only at 3 \(\mu\)s can some deviations be seen in the elongation of the ligament stretching from the tip of the lobes that form on the liquid surface. The coarser mesh predicts a thinner and longer ligament than the finer mesh due to a poorer mesh resolution near the ligament tip. The rest of the liquid surface looks very similar, which might justify using mesh J2 instead of mesh J3 to ease the computational cost of three-dimensional simulations. \par

\begin{figure}[h!]
\centering
\begin{subfigure}{.45\textwidth}
  \centering
  \includegraphics[width=1.0\linewidth]{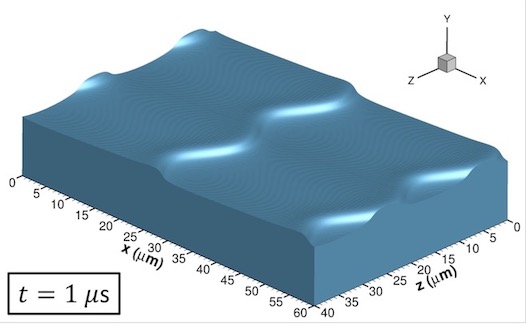}
  \label{subfig:150_30A_1mus_A}
\end{subfigure}%
\begin{subfigure}{.45\textwidth}
  \centering
  \includegraphics[width=1.0\linewidth]{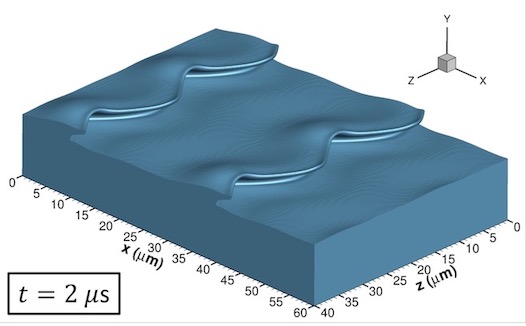}
    \label{subfig:150_30A_2mus_A}
\end{subfigure}%
\\[-3ex]
\begin{subfigure}{.45\textwidth}
  \centering
  \includegraphics[width=1.0\linewidth]{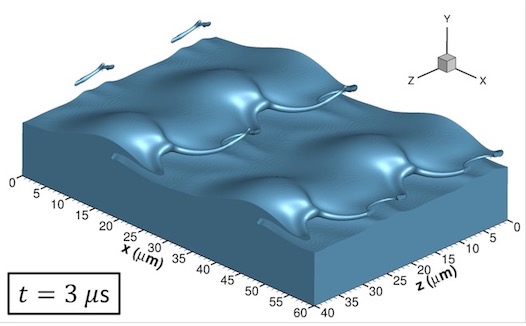}
  \label{subfig:150_30A_3mus_A}
\end{subfigure}%
\begin{subfigure}{.45\textwidth}
  \centering
  \includegraphics[width=1.0\linewidth]{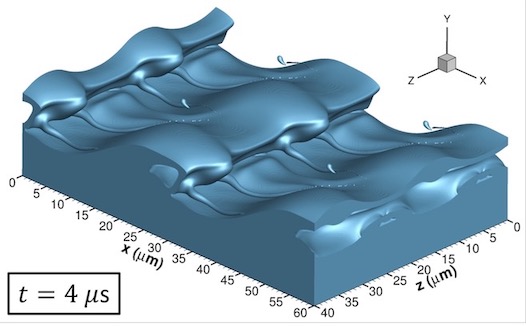}
    \label{subfig:150_30A_4mus_A}
\end{subfigure}%
\\[-3ex]
\begin{subfigure}{.45\textwidth}
  \centering
  \includegraphics[width=1.0\linewidth]{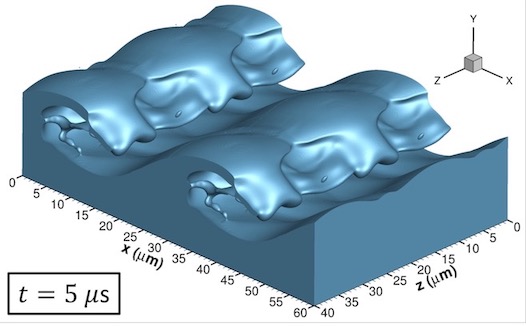}
  \label{subfig:150_30A_5mus_A}
\end{subfigure}%
\begin{subfigure}{.45\textwidth}
  \centering
  \includegraphics[width=1.0\linewidth]{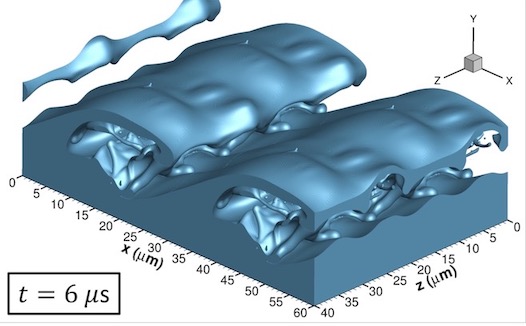}
    \label{subfig:150_30A_6mus_A}
\end{subfigure}%
\\[-3ex]
\begin{subfigure}{.45\textwidth}
  \centering
  \includegraphics[width=1.0\linewidth]{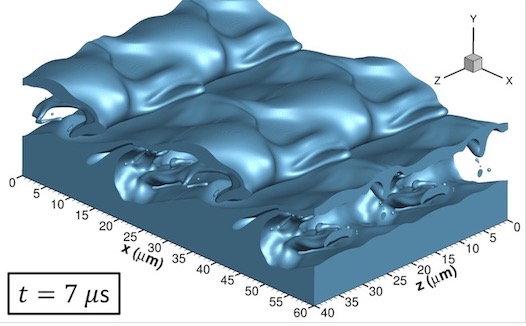}
  \label{subfig:150_30A_7mus_A}
\end{subfigure}%
\begin{subfigure}{.45\textwidth}
  \centering
  \includegraphics[width=1.0\linewidth]{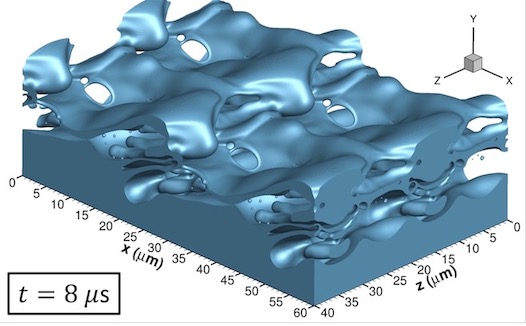}
    \label{subfig:150_30A_8mus_A}
\end{subfigure}%
\caption{Interface deformation at 150 bar for the three-dimensional planar jet with mesh J2. The initial spanwise perturbation amplitude is 0.3 \(\mu\)m.}
\label{fig:3djet_150_30A}
\end{figure}

\begin{figure}[h!]
\centering
\begin{subfigure}{.5\textwidth}
  \centering
  \includegraphics[width=1.0\linewidth]{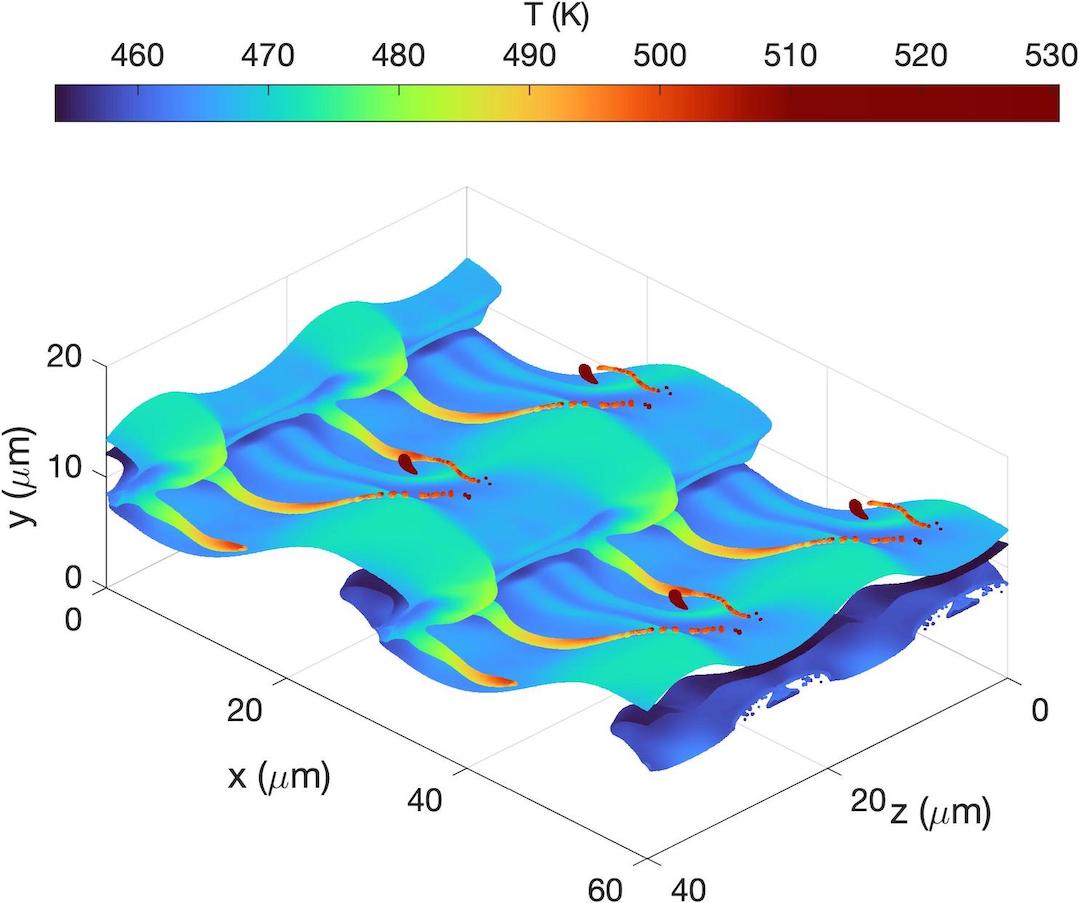}
  \caption{}
  \label{subfig:150_30A_int_T_4mus}
\end{subfigure}%
\begin{subfigure}{.5\textwidth}
  \centering
  \includegraphics[width=1.0\linewidth]{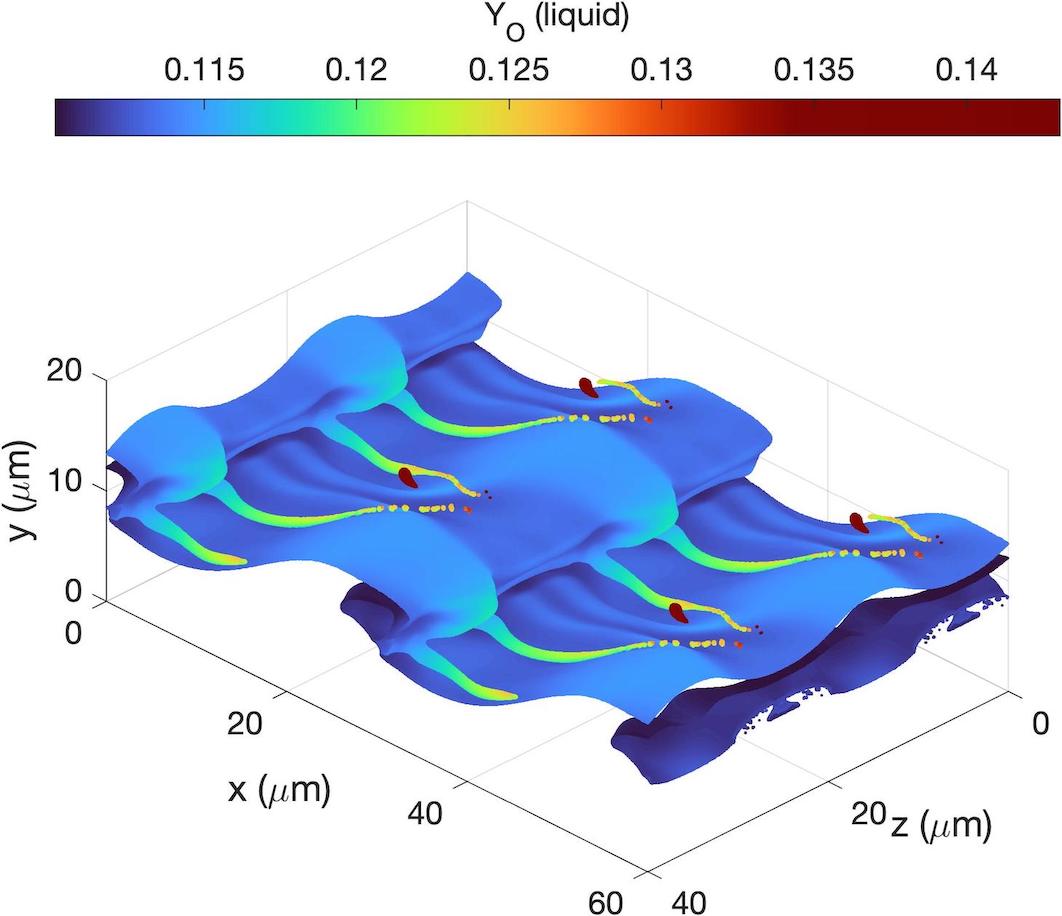}
  \caption{}
    \label{subfig:150_30A_int_YOl_4mus}
\end{subfigure}%
\\
\begin{subfigure}{.5\textwidth}
  \centering
  \includegraphics[width=1.0\linewidth]{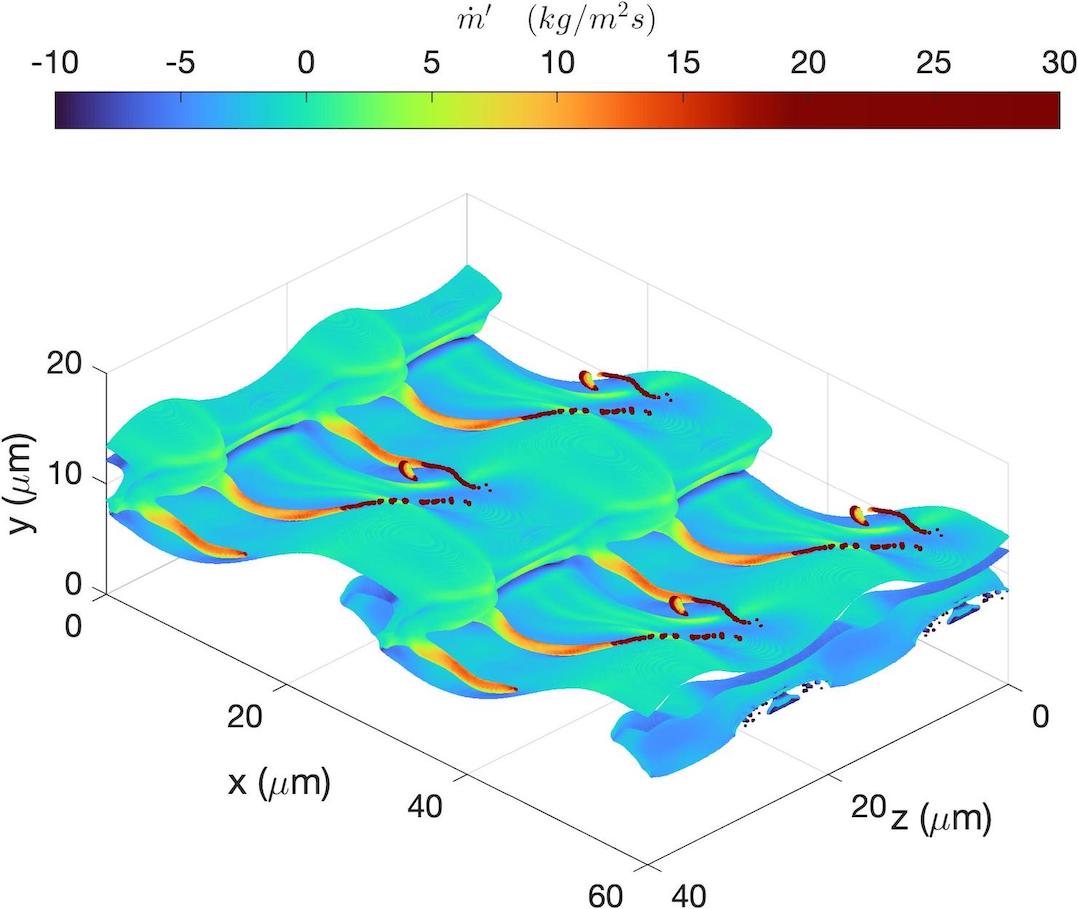}
  \caption{}
  \label{subfig:150_30A_int_mflux_4mus}
\end{subfigure}%
\begin{subfigure}{.5\textwidth}
  \centering
  \includegraphics[width=1.0\linewidth]{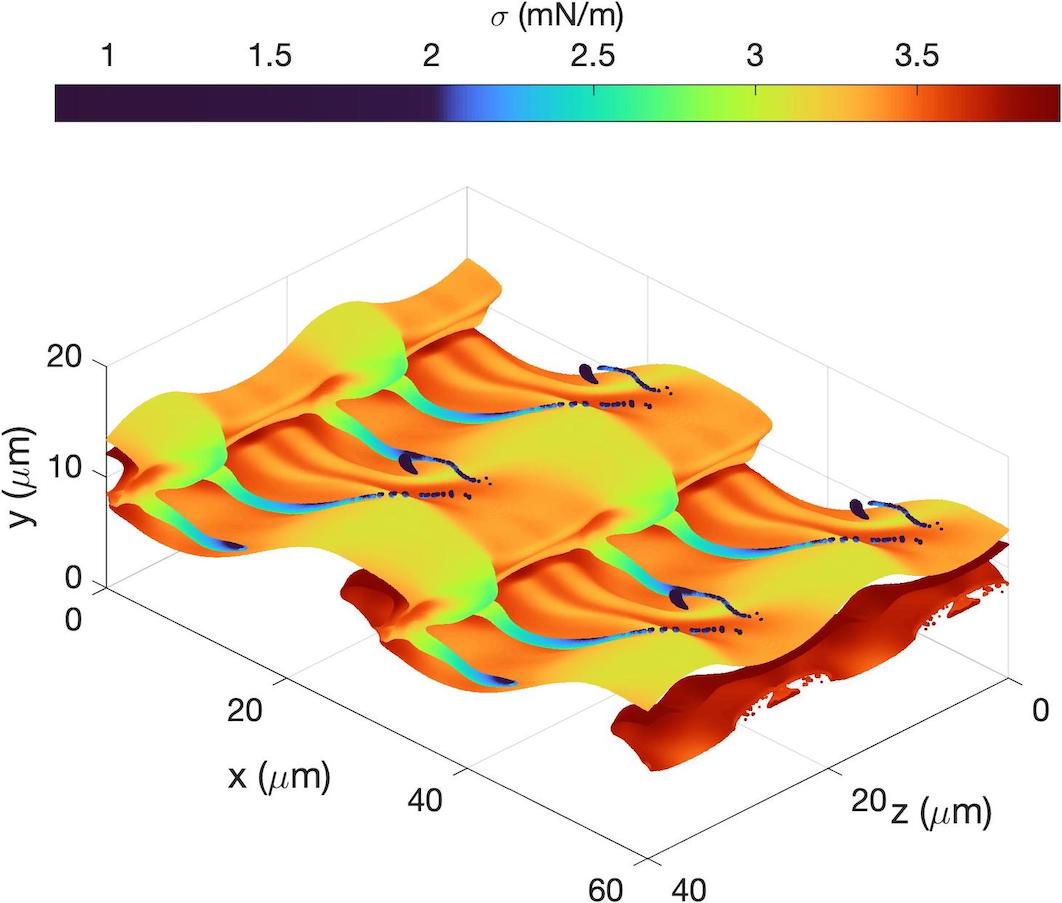}
  \caption{}
    \label{subfig:150_30A_int_sigma_4mus}
\end{subfigure}%
\caption{Interface properties for the three-dimensional planar jet with mesh J2 at 150 bar and \(t=4\) \(\mu\)s. The initial amplitude of the spanwise perturbation is 0.3 \(\mu\)m. The interface location is colored with the value of each variable. (a) temperature; (b) oxygen mass fraction in the liquid phase; (c) mass flux per unit area due to phase change; and (d) surface-tension coefficient.}
\label{fig:150_30A_int_4mus}
\end{figure}

Further three-dimensional results are presented in Figures~\ref{fig:3djet_150_30A} and~\ref{fig:150_30A_int_4mus}. In this case, mesh J2 is used to speed up the computation and the initial amplitude of the spanwise perturbation is set to 0.3 \(\mu\)m. As seen in Figure~\ref{fig:3djet_150_30A}, the liquid jet can deform substantially at relatively low velocities. Liquid sheets overlap onto each other as holes occasionally appear and some ligaments and droplets form. Interestingly, it becomes apparent that analysis of the interface solution along the surface is crucial (see Figure~\ref{fig:150_30A_int_4mus}). The interface deformation and the interaction with the surrounding fluid changes the LTE and jump conditions solution. Interface regions immersed or compressed toward the hotter gas present higher temperatures, which enhance the dissolution of oxygen into the liquid according to LTE (see Figure~\ref{subfig:pheq_diagram}). This increase in temperature affects the local mass exchange rate and there can be regions showing net vaporization (high interface temperature) while other regions show net condensation (low interface temperature). Simpler works using the same binary mixture but where the interface does not deform only showed net condensation at 100 bar and above~\cite{poblador2018transient,davis2019development,poblador2021selfsimilar}.\par 

Moreover, the surface-tension coefficient drops in the hotter interface. This phenomenon shows the importance of the variable surface-tension coefficient observed at high pressures, which affects how the interface deforms. On the other hand, the changes in the interface state are negligible at subcritical pressures cases, which simplifies the study of liquid atomization in that pressure range. \par 

Similarities are seen with the results of Jarrahbashi and Sirignano~\cite{jarrahbashi2014vorticity}, Jarrahbashi et al.~\cite{jarrahbashi2016early} and Zandian et al.~\cite{zandian2017planar,zandian2018understanding,zandian2019length} for incompressible fluids (e.g., formation of lobes, holes, bridges, and ligaments). The qualitative differences can be explained by the reduced surface tension at elevated pressures. However, in-depth analysis of the characteristics of liquid jets at high pressures is left for future works. Here, the focus is on showing the capabilities of the numerical method presented in this work to analyze the high-pressure liquid injection problem while predicting the relevant two-phase, high-pressure physics (e.g., variable interface state, enhanced mixing in the liquid phase). \par 

\clearpage

\section{Summary and conclusions}
\label{summary}

A new physical and numerical methodology has been presented to solve low-Mach-number compressible two-phase flows with phase change. The methodology addresses compressible liquids with phase change and its ultimate goal is the study of liquid fuel injection in the thermodynamic state where the pressure is supercritical for the fuel, but the temperature is subcritical for the resulting liquid mixture at the interface. These thermodynamic conditions are relevant in high-pressure combustion chambers of diesel, gas turbines and rocket engines using typical hydrocarbon-based liquid fuels, specifically near the fuel injectors before substantial heating occurs and the two-phase mixture transitions to a supercritical state. Therefore, the early atomization and fuel mixing may still be dominated by two-phase dynamics. \par

The advection of the liquid phase is performed by extending the VOF method from Baraldi et al.~\cite{baraldi2014mass} to compressible liquids undergoing phase change. Moreover, the low-Mach-number governing equations for two-phase flows, as well as their balancing across the interface, are coupled to a non-ideal thermodynamic model based on a volume-corrected SRK equation of state. Further details about this thermodynamic model are available in Davis et al.~\cite{davis2019development}. The complexity of the non-ideal physics at high pressures adds extra computational cost. To properly capture the interface properties and its displacement, jump conditions and LTE have to be solved at each interface cell and extrapolations of phase-wise fluid compressibilities and the corresponding velocity field have to be performed. This interface-resolved approach provides a comprehensive description of the variation of interface properties inherent of transcritical domains and how they affect the liquid deformation. A constant-coefficient PPE for low-Mach-number flows is developed based on Dodd et al.~\cite{dodd2014fast,dodd2021vof}. This PPE can be solved with a computationally-efficient FFT method to alleviate the increase in computational cost. \par

Various tests show the viability and accuracy of the numerical model presented in this work. A sufficiently fine mesh is needed to provide a smooth interface solution and to minimize numerical errors and mass errors. In the limit of incompressible liquid without phase change, the method recovers the mass-conserving properties from Baraldi et al.~\cite{baraldi2014mass}. The highly-coupled approach induces the generation of spurious currents around the interface and caution is needed to make sure they are not detrimental for the actual interface evolution. The VOF method alone is known to cause spurious currents, both due to the HF method used to evaluate curvature and due to the sharp volume-averaging of fluid properties at the interface when solving the one-fluid momentum equation. Additionally, the localized source terms related to mass exchange and different fluid compressibilities at the interface contribute further to the generation of spurious currents. \par 

Future work can consider improvements to the new methodology. A few ideas for the topic are provided below.
\begin{itemize}
\item Optimize the cost of the thermodynamic model and the extrapolations across the interface.
\item Reduce the generation of spurious currents around the interface while keeping a sharp interface.
\item Improve the accuracy of the numerical schemes used to discretize the governing equations.
\item Consider the interface transition from a sharp to a diffuse interface near and beyond the mixture critical point.
\end{itemize}

Ongoing analysis with the current code has the following goals:
\begin{itemize}
\item Analyze the early deformation of three-dimensional liquid jets at high pressures.
\item Characterize the interface deformation as a function of pressure and other parameters.
\item Use the interface-resolved methodology to study the thermodynamic state of the interface. That is, determine regions of high/low surface-tension coefficient, identify regions of condensation/vaporization, etc.
\item Determine the role of vortex dynamics in the early deformation of the liquid.
\end{itemize}

\section*{Acknowledgments}
The authors are grateful for the helpful discussions in developing this work with Prof. Antonino Ferrante and his student Pablo Trefftz-Posada, from the University of Washington. The incompressible VOF subroutines from his group shared with us are also appreciated. The authors are also grateful for the support of the NSF grant with Award Number 1803833 and Dr. Ron Joslin as Scientific Officer. \par

This work utilized the infrastructure for high-performance and high-throughput computing, research data storage and analysis, and scientific software tool integration built, operated, and updated by the Research Cyberinfrastructure Center (RCIC) at the University of California, Irvine (UCI). The RCIC provides cluster-based systems, application software, and scalable storage to directly support the UCI research community. https://rcic.uci.edu \par

\section*{Conflict of interest}
The authors declared that there is no conflict of interest.

\section*{Data availability}
The data that support the findings of this study are available from the corresponding author upon reasonable request.

\appendix

\section{Details on the thermodynamic model}
\label{apn:thermo}

\begin{figure}[h!]
\centering
\includegraphics[width=0.5\linewidth]{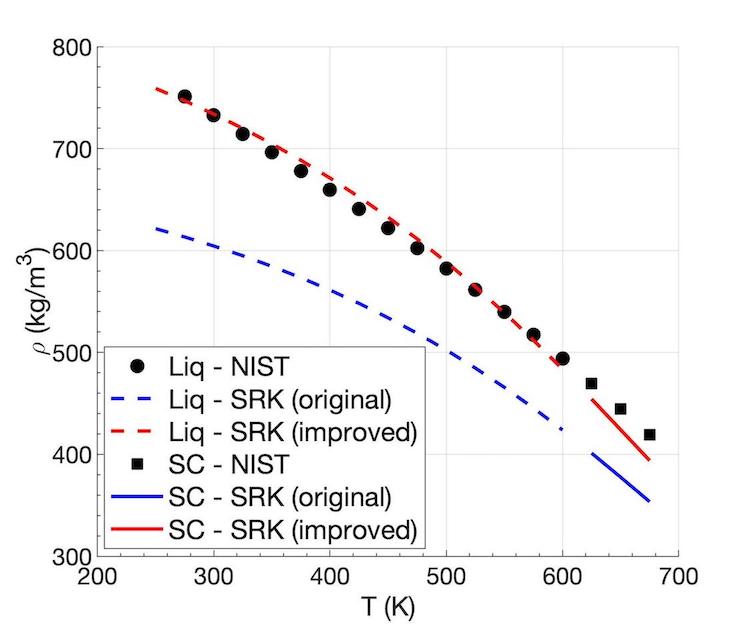}
\caption{Density of \textit{n}-decane at 100 bar obtained with the original SRK equation of state~\cite{soave1972equilibrium} and with the improved SRK equation of state~\cite{lin2006volumetric}. The volume-corrected model predicts the reference data available at NIST with higher accuracy as it transitions from the liquid state (Liq) to a supercritical state (SC) with the increase of temperature.}
\label{fig:DEN_EoS}
\end{figure}

The thermodynamic model implemented in this work is based on a volume-corrected SRK cubic equation of state~\cite{lin2006volumetric}, which is able to represent non-ideal fluid states for both the gas and liquid phases. The original SRK equation of state~\cite{soave1972equilibrium} presents density errors of up to 20\% when compared to experimental measurements whenever a dense fluid is being predicted~\cite{yang2000modeling,prausnitz2004thermodynamics} (i.e., liquid or high pressure gas). Therefore, it affects the prediction of fluid properties and transport coefficients using other models and correlations that rely on the fluid density. An example of the improvement in the density prediction is shown in Figure~\ref{fig:DEN_EoS}, where the density of \textit{n}-decane at 100 bar and various temperatures is plotted. Since the correction is implemented as a volume translation, other thermodynamic variables, such as fugacity or enthalpy, are equivalent between both the improved and the original SRK equation of state. \par 

The volume-corrected SRK equation of state is expressed in terms of the compressibility factor, \(Z\), as
\begin{equation}
\label{eqn:SRKEoS}
Z^3+(3B_{*}-1)Z^2+\big[B_{*}(3B_{*}-2)+A-B-B^2\big]Z+B_{*}(B_{*}^{2}-B_{*}+A-B-B^2)-AB=0
\end{equation}

\noindent
with
\begin{equation}
Z = \frac{p}{\rho RT} \quad ; \quad A = \frac{a(T)p}{R_{u}^{2}T^2} \quad ; \quad B = \frac{bp}{R_uT} \quad ; \quad B_{*} = \frac{c(T)p}{R_uT}
\end{equation}

The parameters of this equation of state are the following. \(a(T)\) is a temperature-dependent cohesive energy parameter, \(b\) represents a volumetric parameter related to the space occupied by the molecules, and \(c(T)\) is a temperature-dependent volume correction. \(R\) and \(R_u\) are the specific gas constant and the universal gas constant, respectively. Generalized quadratic mixing rules are used~\cite{soave1972equilibrium,lin2006volumetric}, which work well for nonpolar fluids. The binary interaction parameter, \(k_{ij}\), can be obtained from models correlated with vapor-liquid equilibrium experimental data. For instance, Soave et al.~\cite{soave2010srk} provide correlations specific for the SRK equation of state and different mixtures (e.g., nitrogen-alkane pairs). When data are not available, neglecting the interaction coefficients must be justified. For instance, \(k_{ij}\approx 0\) for nitrogen-alkane mixtures; thus, under the assumption that nitrogen and oxygen are similar components, the binary interaction coefficients could be neglected and set to zero for oxygen-alkane mixtures if no data are available. \par

The solution of this cubic equation provides the density of the fluid mixture, \(\rho\), and is computationally efficient since an analytical solution exists. Together with high-pressure correlations, the volume-corrected SRK equation of state is used to evaluate fluid properties and transport coefficients. The generalized multi-parameter correlation from Chung et al.~\cite{chung1988generalized} is used to evaluate viscosity and thermal conductivity. The mass diffusion coefficient is obtained from the unified model for non-ideal fluids presented in Leahy-Dios and Firoozabadi~\cite{leahy2007unified} simplified for binary mixtures and the surface-tension coefficient is estimated as a function of the interface properties and composition from the Macleod-Sugden correlation as suggested in Poling et al.~\cite{poling2001properties}. The Macleod-Sugden correlation is preferred for mixtures near the critical point since it provides the correct limit whereby the surface-tension coefficient drops to zero at the critical point. \par

Further details on the development of the necessary thermodynamic expressions to evaluate fluid properties for the non-ideal mixture (e.g., mixture enthalpy) based on the concept of a departure function from the ideal state~\cite{poling2001properties} and the practical implementation of this thermodynamic model can be found in Davis et al.~\cite{davis2019development} and the respective references mentioned in this appendix. For low-Mach-number compressible flows, the pressure throughout the thermodynamic model is assumed uniform and equal to the thermodynamic or ambient pressure. \par

\section{Implementation details}
\label{apn:implementation}

The code used for this work has been written in Fortran 90 and uses the message-passing interface (MPI) and OpenMP for multi-processor computing. The computational domain is partitioned in a pencil-like distribution with the contiguous memory along the jet's transverse direction (see the problem configurations presented in Subsections~\ref{subsec:jet_res2D} and~\ref{subsec:jet_res3D}). The computational resources are better utilized with this decomposition when capturing the interface and performing the various operations that scale with the surface area (e.g., local equilibrium state, extrapolations). The following external open-source libraries have been linked to the code: 2DECOMP\&FFT~\cite{li20102decomp} to perform the domain decomposition and FFTW3~\cite{frigo2005design} to solve the pressure Poisson equation, Eq.~(\ref{eqn:ppe2}), using Discrete Fourier Transforms. The simulations were carried out on the University of California Irvine's local HPC3 cluster.\par

\section{Discretization of the species and enthalpy transport equations}
\label{apn:discr}

Details regarding the discretization of the species and enthalpy transport equations in non-conservative form, Eqs.~(\ref{eqn:spcont_disc}) and~(\ref{eqn:energy_disc}), are provided in this appendix. As an example, the discretization of \(\nabla Y_O\) in the convective term  \(\vec{u}_{f}\cdot \nabla Y_O\) is shown here following the configuration presented in Figure~\ref{fig:scalar_disc}. The method shown here for derivatives with respect to the \(x\)-variable will serve as templates for the derivatives with respect to \(y\) and \(z\) and can also be implemented for the respective term in the enthalpy transport equation. For simplicity, the explicit notation has been dropped. This example covers the different scenarios that can be found when evaluating \(\partial Y_O/\partial x\) with an adaptive first/second-order upwinding scheme and can be easily extended to other directions and interface configurations. \par 

At cell \(i\), the gradient using gas values is found as follows. With a uniform mesh, for \((u_{f}^{i}+u_{f}^{i+1})/2<0\) (where the phase-wise velocity is \(f=g\)), \(\partial Y_O/\partial x\) becomes
\begin{equation}
\label{eqn:gradY1_cell_i_right}
\Bigg(\frac{\partial Y_O}{\partial x}\Bigg)^i=
\begin{cases}
\frac{Y_{O}^{i+2}-Y_{O}^{i}+(Y_{O}^{i}-Y_{O}^{i+1})(1+\Delta x_3/\Delta x_2)^2}{\Delta x_2 + \Delta x_3 - \frac{(\Delta x_2 + \Delta x_3)^2}{\Delta x_2}} & \text{\footnotesize if bounded second-order} \\
\frac{Y_{O}^{i+1}-Y_{O}^{i}}{\Delta x_2}& \text{\footnotesize if not bounded second-order} 
\end{cases}
\end{equation}

\noindent
and if \((u_{f}^{i}+u_{f}^{i+1})/2 \geq 0\), it becomes
\begin{equation}
\label{eqn:gradY1_cell_i_left}
\Bigg(\frac{\partial Y_O}{\partial x}\Bigg)^i=
\begin{cases}
\frac{Y_{O}^{i}-Y_{O,g}^{\Gamma}}{\Delta x_1} & \text{if $\Delta x_1 \geq 0.05\Delta x$} \\
\frac{Y_{O}^{i+1}-Y_{O,g}^{\Gamma}}{\Delta x_1+\Delta x_2} & \text{if $\Delta x_1 < 0.05\Delta x$}
\end{cases}
\end{equation}

\noindent
where \(Y_{O,g}^{\Gamma}\) is the gas mass fraction of the oxidizer species at the interface. The interface values used in the numerical stencils are averaged similar to Eq.~(\ref{eqn:faceavg}). If the distance between node \(i\) and the interface is very small (i.e., \(\Delta x_1 < 0.05 \Delta x\)), node \(i\) is skipped and the gradient is evaluated using the neighboring node \(i\)+1. This step becomes necessary to avoid a spike in the value of \(\partial Y_O/\partial x\) when the interface is very close to the grid node, which might cause unrealistic solutions near the interface. \par 

Another possible stencil is given in cell \(i\)+1 if \((u_{f}^{i+1}+u_{f}^{i+2})/2 \geq 0\). Here, \(\partial Y_O/\partial x\) is evaluated as
\begin{equation}
\label{eqn:gradY1_cell_ip1_left}
\Bigg(\frac{\partial Y_O}{\partial x}\Bigg)^{i+1}=
\begin{cases}
\frac{Y_{O,g}^{\Gamma}-Y_{O}^{i+1}+(Y_{O}^{i+1}-Y_{O}^{i})(-1-\Delta x_1/\Delta x_2)^2}{-(\Delta x_1 + \Delta x_2) + \frac{(-\Delta x_1 - \Delta x_2)^2}{\Delta x_2}} & \text{\footnotesize if bounded second-order and $\Delta x_1 \geq 0.05\Delta x$} \\
\frac{Y_{O}^{i+1}-Y_{O}^{i}}{\Delta x_2}& \text{\footnotesize if not bounded second-order solution or $\Delta x_1 < 0.05\Delta x$} 
\end{cases}
\end{equation}

\noindent
where a combination of the conditions given in Eqs.~(\ref{eqn:gradY1_cell_i_right}) and~(\ref{eqn:gradY1_cell_i_left}) is used to maintain numerical stability and boundedness. Away from the interface, the upwinded gradient \(\partial Y_O/\partial x\) is obtained with similar expressions as in Eqs.~(\ref{eqn:gradY1_cell_i_right}) and~(\ref{eqn:gradY1_cell_ip1_left}) only considering the boundedness condition. \par

Diffusive terms are discretized using second-order central differences, although near the interface the spatial accuracy might be reduced when including the interface. Similarly, we look at the discretization of \(\nabla \cdot (\rho D_m \nabla Y_O)\) in the \(x\) direction under the configuration presented in Figure~\ref{fig:scalar_disc}. However, its extension to other directions or interface configurations is straightforward. The generic discretization of the diffusive term is 
\begin{equation}
\label{eqn:diffY1_general}
\frac{\partial}{\partial x}\Bigg(\rho D_m \frac{\partial Y_O}{\partial x}\Bigg)= \frac{(\rho D_m \partial Y_O/\partial x)^E-(\rho D_m \partial Y_O/\partial x)^W}{\Delta x} 
\end{equation}

\noindent
with \(E\) and \(W\) referring to the East and West cell faces, respectively. \par 

Fluid properties at the cell face are obtained by linear interpolation between two neighboring grid nodes. If the interface is located between the grid node and its respective cell face, the interface replaces the cell face (e.g., the West face of node \(i\) in Figure~\ref{fig:scalar_disc}). Also, if the interface is located beyond a cell face but before a neighboring node, its equilibrium solution and position are used for the linear interpolation of fluid properties at the cell face and to evaluate \(\partial Y_O/\partial x\) (e.g., the East face of node \(i\)-1 in Figure~\ref{fig:scalar_disc}). \par 

For example, the discretization of the diffusive term at cell \(i\) results in
\begin{equation}
\label{eqn:diffY1_cell_i_a}
\frac{\partial}{\partial x}\Bigg(\rho D_m \frac{\partial Y_O}{\partial x}\Bigg)^i= \frac{(\rho D_m \partial Y_O/\partial x)^E-(\rho D_m \partial Y_O/\partial x)^W}{\Delta x_1 + \Delta x_2/2} 
\end{equation}

\noindent
with
\begin{subequations}
\label{eqn:diffY1_cell_i_b}
\begin{equation}
\Bigg(\rho D_m \frac{\partial Y_O}{\partial x}\Bigg)^E = (\rho D_m)^E\Bigg(\frac{Y_{O}^{i+1}-Y_{O}^{i}}{\Delta x_2}\Bigg)
\end{equation}
\begin{equation}
\Bigg(\rho D_m \frac{\partial Y_O}{\partial x}\Bigg)^W = 
\begin{cases}
(\rho D_m)_{g}^{\Gamma}\Bigg(\frac{Y_{O}^{i}-Y_{O,g}^{\Gamma}}{\Delta x_1}\Bigg) & \text{if $\Delta x_1 \geq 0.05\Delta x$} \\
(\rho D_m)_{g}^{\Gamma}\Bigg(\frac{Y_{O}^{i+1}-Y_{O,g}^{\Gamma}}{\Delta x_1+\Delta x_2}\Bigg) & \text{if $\Delta x_1 < 0.05\Delta x$}
\end{cases}
\end{equation}
\end{subequations}

\noindent
where \((\rho D_m)^E\) is the product of the linear interpolations of \(\rho\) and \(D_m\) at the East face and \((\rho D_m)_{g}^{\Gamma}\) is the product of the interface gas values of \(\rho\) and \(D_m\) at equilibrium. Similar to the discretization of the convective term, the approximation of \(\partial Y_O/\partial x\) skips node \(i\) if the interface is very close to it in order to avoid nonphysical spikes. \par 

Some other special considerations are needed to ensure a stable and physically-correct solution. Two extreme cases may exist, either when the interface is nearly touching the grid node (i.e., \(\Delta x_1 < 0.01 \Delta x\) in Figure~\ref{fig:scalar_disc}) or when the node changes phase (i.e., \(C^n < 0.5\) and \(C^{n+1}\geq 0.5\) or \(C^n \geq 0.5\) and \(C^{n+1}<0.5\)). In both scenarios, the interface value of the correct phase is assigned to a grid node based on its proximity. If the interface displacement in \(\Delta t\) is very small, this approximation is reasonable. Moreover, under-resolved regions (i.e., droplets or thin areas of the order of the mesh size) might generate incorrect solutions. To avoid this problem, if a nonphysical solution is detected because of this reason, the values for oxygen mass fraction or enthalpy assigned to the problematic cell are obtained from an average of the surrounding physically-correct values of the same fluid phase. \par

\section{Details on the techniques used to extrapolate the fluid compressibilities and phase-wise velocities}
\label{apn:extrap}

The equations presented in Aslam~\cite{aslam2004partial} used to extrapolate the fluid compressibility across the interface as detailed in Subsection~\ref{subsec:extrapolation} are summarized in this appendix. Generally, a linear extrapolation of the divergence of each phase is preferred. However, a constant extrapolation might be necessary for numerical stability depending on the problem configuration~\cite{palmore2019volume}. Using the extrapolation of the liquid-phase divergence as an example, Eqs.~(\ref{eqn:extrapconst}) and (\ref{eqn:extraplinear}) have to be solved. The gas-phase divergence extrapolation follows the same approach.
\begin{equation}
\label{eqn:extrapconst}
\frac{\partial g_n}{\partial \tau} + H(C)\big(\vec{n}\cdot\nabla g_n\big) = 0
\end{equation}
\begin{equation}
\label{eqn:extraplinear}
\frac{\partial g}{\partial \tau} + H(C)\big(\vec{n}\cdot\nabla g-g_n\big) = 0
\end{equation}

In Eqs.~(\ref{eqn:extrapconst}) and (\ref{eqn:extraplinear}), \(g=\nabla\cdot\vec{u}_l\) and \(g_n=\vec{n}\cdot\nabla g\). Here, \(\vec{n}\) is defined pointing toward the gas phase. These equations are solved to steady-state in a fictitious time, \(\tau\), which does not necessarily have units of time. First, the normal gradient of \(g\) (or \(g_n\)) is extrapolated in a constant fashion using Eq.~(\ref{eqn:extrapconst}) and, then, \(g\) is extrapolated linearly following \(g_n\) using Eq.~(\ref{eqn:extraplinear}). The liquid-phase extrapolation is only performed in the region defined by \(H(C)\), where \(H(C)=0\) if \(C=1\) and \(H(C)=1\) otherwise. That is, the extrapolation is done only at interface cells (i.e., \(0<C<1\)) and gas cells (i.e., \(C=0\)). When extrapolating gas-phase values, \(H(C)=0\) if \(C=0\) and \(H(C)=1\) otherwise. For practical purposes, it is sufficient to reach steady-state only within the extrapolation region defined in Figure~\ref{fig:extrap}. If a constant extrapolation is required for numerical stability, it is only necessary to solve Eq.~(\ref{eqn:extraplinear}) by setting \(g_n=0\). That is, a linear extrapolation might overestimate or underestimate the fluid compressibility within the extrapolation region, especially during initialization with sharp initial conditions. This problem might result in the development of an unstable or unrealistic solution. More information on the numerical discretization and solution of these two equations can be found in Aslam~\cite{aslam2004partial}. \par 

Within the VOF framework, the normal unit vector is defined only at interface cells.
Therefore, we need to populate the extrapolation region with an estimate of \(\vec{n}\) in order to solve Eqs.~(\ref{eqn:extrapconst}) and (\ref{eqn:extraplinear}). For that purpose, an inverse-distance weighted average is used to obtain each component \(n_m\) (i.e., \(m={x,y,z}\)) of \(\vec{n}\) at non-interface cells following \(n_m = \big(\sum_i \frac{n_{m,i}}{d_i}\big)/\big(\sum_i\frac{1}{d_i}\big)\). To evaluate this average, only the information of the closest set of \(i\) interface cells in the neighborhood of the node of interest is used. \(d_i\) is defined as the distance between the node and the centroid of the interface plane in cell \(i\). Once each component \(n_m\) is obtained, the vector is re-normalized to have \(|\vec{n}|=1\). \par 

Another method to extrapolate the fluid compressibility is presented by McCaslin et al.~\cite{mccaslin2014fast}. The same extrapolation equations presented in Aslam~\cite{aslam2004partial} are solved directly at steady-state using a fast marching method (FMM). The method is computationally more efficient since it does not require iterations on a pseudo-time. However, spatial convergence is less accurate for higher-order extrapolations (e.g., linear or quadratic). For a constant extrapolation, the FMM approach and the iterative method by Aslam~\cite{aslam2004partial} have the same spatial accuracy; thereby, the FMM approach has an advantage and has been used to speed the three-dimensional computations shown in Subsections~\ref{subsec:drop_3D} and~\ref{subsec:jet_res3D}.  \par 

Once the fluid compressibilities have been estimated across the interface for each phase, the extrapolation method presented in Dodd et al.~\cite{dodd2021vof} adapted to compressible flows is used to obtain the phase-wise velocities. Eq.~(\ref{eqn:extrapvel}) solves the extrapolation of \(\vec{u}_l\) and is directly solved at steady-state conditions (i.e., as \(\tau \rightarrow \infty\), \(\frac{\partial \vec{u}_l}{\partial \tau}\rightarrow 0\)) using an iterative solver until some desired tolerance is achieved. \(\tau\) is a fictitious time-like variable that does not have units of time.
\begin{equation}
\label{eqn:extrapvel}
\frac{\partial \vec{u}_l}{\partial \tau} + \nabla(\nabla\cdot\vec{u}_l) = \nabla g \rightarrow \nabla(\nabla\cdot\vec{u}_l) = \nabla g
\end{equation}

The two boundary conditions imposed in Eq.~(\ref{eqn:extrapvel}) are: (a) the velocity components at the boundary of the extrapolation region inside the real phase are fixed and are equal to the one-fluid velocity. That is, \(\vec{u}_l=\vec{u}\) in the example shown in Figure~\ref{fig:extrap}; and (b) the velocity components at the boundary of the extrapolation region in the ghost phase must satisfy the discrete divergence of the boundary cell (i.e., the extrapolated \(\nabla\cdot\vec{u}_l\)). The velocity components inside the extrapolation region are solved by discretizing Eq.~(\ref{eqn:extrapvel}) using second-order finite differences. This methodology ensures that the extrapolated staggered phase-wise velocity field satisfies the extrapolated phase-wise divergence field. Although Eq.~(\ref{eqn:extrapvel}) only shows the extrapolation of liquid phase-wise velocities, the phase-wise velocities related to the gas phase are obtained using the same approach. \par 

A note on the treatment of under-resolved regions must be provided here. As pointed out in Subsections \ref{subsec:VoFgeom} and \ref{subsec:contmom}, a minimum mesh resolution is needed to capture the interface geometry and surface-tension force accurately. The extrapolation of phase-wise fluid compressibilities and velocities becomes problematic in under-resolved areas, where poor convergence (or none at all) of the extrapolation equations may arise. A possible solution, but out of the scope of this work, would be to refine the mesh locally (i.e., use Adaptive Mesh Refinement or AMR). Here, a different approach is proposed. A group of nodes or ``block" is defined around under-resolved interface locations. The nodes inside these under-resolved domains are excluded from both the phase-wise fluid compressibility and velocity extrapolations. In these areas, it is assumed that the extrapolation region is incompressible and the phase-wise velocities correspond to the one-fluid velocity. To maintain consistency with the pressure solver, the divergence of the one-fluid velocity is zero in under-resolved interface cells despite the volume expansion or contraction that occurs under mass exchange across the interface. Regardless, phase change is still considered in the VOF method (i.e., Eq.~(\ref{subeqn:PCstep}) is still used). \par 

Under-resolved areas are identified in high-curvature regions where the local radius of curvature is below a certain resolution threshold. This threshold is defined as \(1/\kappa<3\Delta x\) in two dimensions, whereas in three dimensions three curvatures and thresholds are considered: the three-dimensional curvature of the surface with \(1/\kappa<6\Delta x\) and the curvatures in the two principal directions used to determine the three-dimensional curvature with \(1/\kappa_1<3\Delta x\) and \(1/\kappa_2<3\Delta x\). That is, the three-dimensional curvature may be twice the curvature of a two-dimensional surface with the same radii of curvature (e.g., a circle and a sphere with the same diameter) and it may also be zero if the two radii of curvature cancel each other locally. Similarly, under-resolved areas are also defined around thin liquid structures or gas pockets where two conflicting extrapolations coming from two different interfaces exist. \par 

The main goal here is to define a numerical approach that will keep the simulation advancing in time. The errors introduced with this treatment are expected to be minimal in regions where geometry errors already exist and might be dominant. The treatment of these under-resolved areas might also affect mass conservation since fluid compressibilities are neglected in the advection of the interface. For instance, the error introduced in the pressure solver when not accounting for the velocity jump due to mass exchange is related to the strength of \(\dot{m}'\). For the type of flows analyzed here, the velocity jump \((\vec{u}_g-\vec{u}_l) \cdot \vec{n}\) may be of the order of \(\mathcal{O}(10^{-1}-10^{-2}\) m/s) and the velocity field is of the order of \(\mathcal{O}(10^{1}-10^{2}\) m/s). \par

\section{Verification of two-phase incompressible flows}
\label{apn:nume_two}

\subsection{Standing capillary wave}

The analytical solution for the relaxation of small-amplitude, incompressible, two-dimensional capillary waves presented by Prosperetti~\cite{prosperetti1981motion} is used to validate the dynamical behavior of the liquid surface in the presence of surface tension. The analytical model addresses superposed fluids with infinite depth and lateral extension, which is then solved numerically by reducing the problem to two dimensions and employing a sufficiently wide domain in the surface transverse direction with open boundary conditions. In addition, periodic boundary conditions in the wave direction are applied. Here, only the continuity and momentum equations for two-phase flows are considered. \par 

\begin{figure}[h!]
\centering
\includegraphics[width=0.5\linewidth]{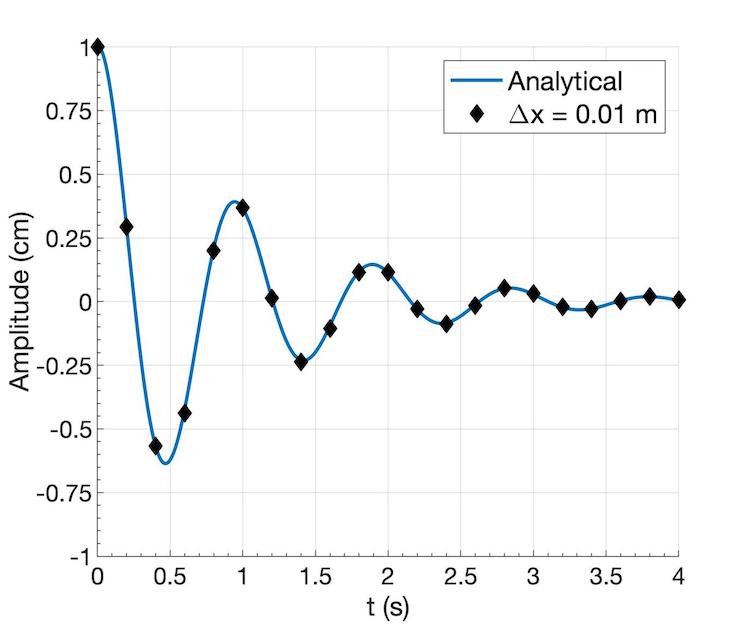}
\caption{Incompressible standing capillary wave. Showing wave amplitude vs. time. A uniform mesh is considered where \(\Delta x = \Delta y = 0.01\) m.}
\label{fig:PM_val3}
\end{figure}

Figure~\ref{fig:PM_val3} depicts the amplitude decay of an initial sinusoidal perturbation. We consider a sinusoidal wave with a wavelength of 1 m and an initial amplitude of 0.01 m. Each fluid has a 1.5 m depth. The liquid density is 1000 kg/m\(^3\), and the gas density is 100 kg/m\(^3\), with a liquid viscosity of 17.989 Pa\(\cdot\)s and a gas viscosity of 1.7989 Pa\(\cdot\)s. The surface-tension coefficient is 0.01 N/m, and gravity is taken into account by placing the lighter fluid on top of the denser fluid. \par 

\subsection{Liquid jet atomization}

The two-phase, incompressible limit is verified by comparing the current code's performance to previous codes used to analyze the atomization of incompressible planar liquid jets. The early deformation of a planar liquid jet defined in Table 1 of Zandian et al.~\cite{zandian2018understanding} as case D3a is analyzed for this purpose. Following the definition of the liquid Reynolds number, \(Re_L=\rho_L u_G h/\mu_L\), and gas Weber number, \(We_G = \rho_G u_{G}^{2} h/\sigma\), this case presents \(Re_L=5000\) and \(We_G=7250\). The jet thickness is \(h = 50\) \(\mu\)m and the freestream gas velocity is \(u_G=100\) m/s, while the liquid density is set at \(\rho_L = 804\) kg/m\(^3\) and the gas density is \(\rho_G = 402\) kg/m\(^3\). The initial perturbation wavelength is 100 \(\mu\)m. More details about the problem configuration are provided in Zandian et al.~\cite{zandian2018understanding}. Again, only the solution of the continuity and momentum equations is addressed here. \par

The computational setup between both codes is different. The numerical model introduced in this paper is based on a VOF approach, while the numerical method used in Zandian et al.~\cite{zandian2018understanding} is based on the LS method with an artificially diffuse interface. Therefore, both codes are substantially different in how they capture the interface and address its sharpness. Moreover, the uniform mesh size used in Zandian et al.~\cite{zandian2018understanding} is \(\Delta x = 1.25\) \(\mu\)m with the planar jet centered in the domain, while the results obtained with the present model use \(\Delta x = 0.667\) \(\mu\)m with a reduced computational domain that includes fewer wavelengths of the initial perturbation and that considers symmetry boundary conditions at the center plane of the jet. The time step value is consistent with the respective numerical method. \par 

\begin{figure}[h!]
\centering
\begin{subfigure}{.5\textwidth}
  \centering
  \includegraphics[width=1.0\linewidth]{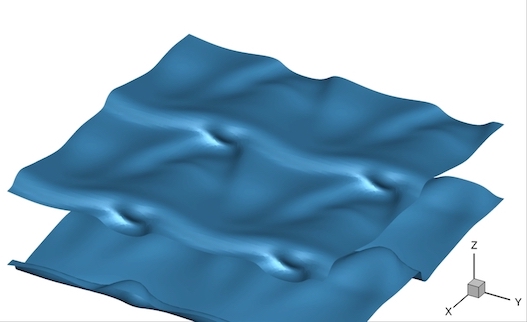}
  \caption{}
\end{subfigure}%
\begin{subfigure}{.5\textwidth}
  \centering
  \includegraphics[width=1.0\linewidth]{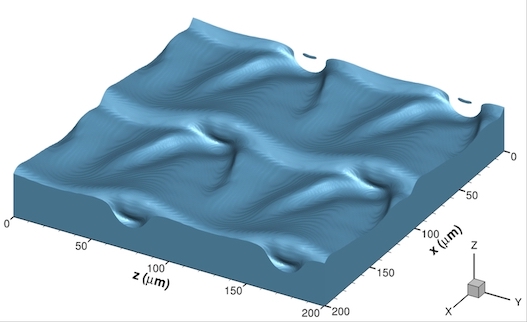}
  \caption{}
\end{subfigure}%
\\
\begin{subfigure}{.5\textwidth}
  \centering
  \includegraphics[width=0.8\linewidth]{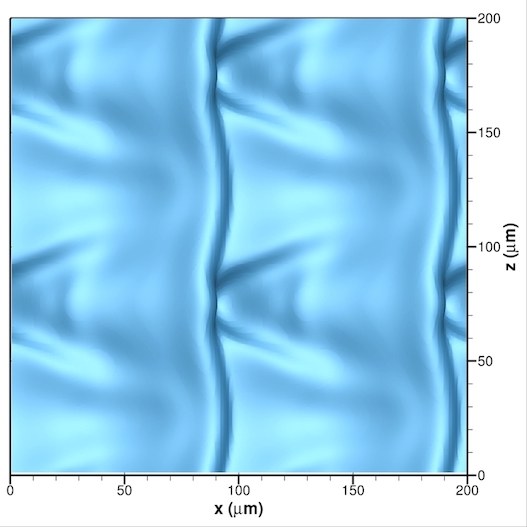}
  \caption{}
\end{subfigure}%
\begin{subfigure}{.5\textwidth}
  \centering
  \includegraphics[width=0.8\linewidth]{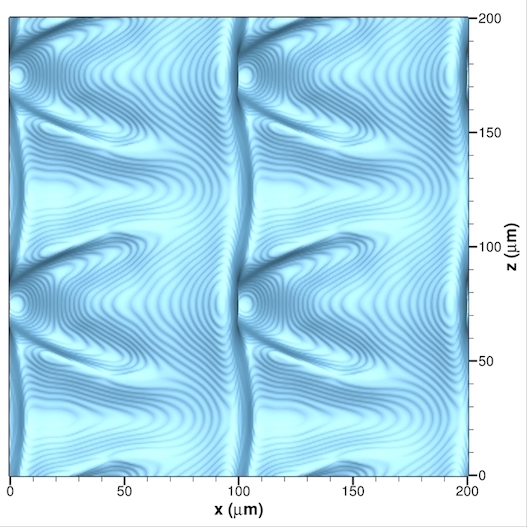}
  \caption{}
\end{subfigure}%
\caption{Incompressible planar liquid jet injection (a). Surface deformation comparison at \(t=30\) \(\mu\)s between a numerical approach based on the Level-Set (LS) method~\cite{zandian2018understanding} and the present numerical approach based on the Volume-of-Fluid (VOF) method. The LS method identifies the liquid surface as the iso-surface with \(\theta=0\) and the VOF method identifies the liquid surface as the iso-surface with \(C=0.5\). (a) 3D view (LS); (b) 3D view (VOF); (c) top view (LS); and (d) top view (VOF).}
\label{fig:PM_val4c}
\end{figure}

\begin{figure}[h!]
\centering
\begin{subfigure}{.5\textwidth}
  \centering
  \includegraphics[width=1.0\linewidth]{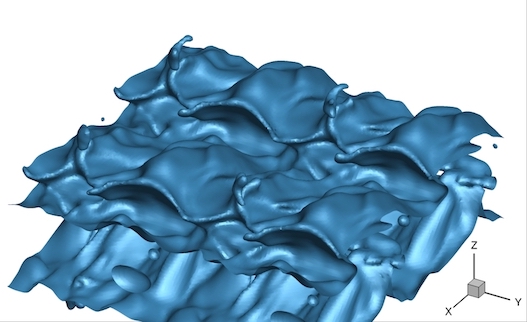}
  \caption{}
\end{subfigure}%
\begin{subfigure}{.5\textwidth}
  \centering
  \includegraphics[width=1.0\linewidth]{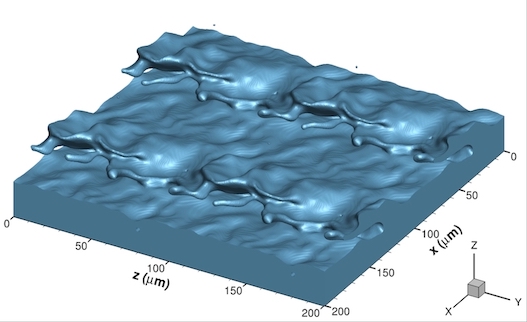}
  \caption{}
\end{subfigure}%
\\
\begin{subfigure}{.5\textwidth}
  \centering
  \includegraphics[width=0.8\linewidth]{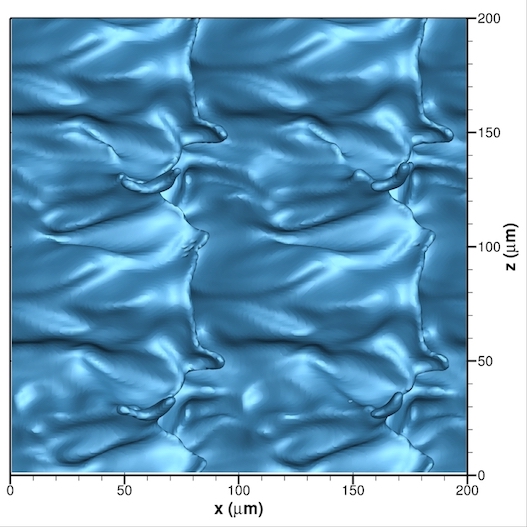}
  \caption{}
\end{subfigure}%
\begin{subfigure}{.5\textwidth}
  \centering
  \includegraphics[width=0.8\linewidth]{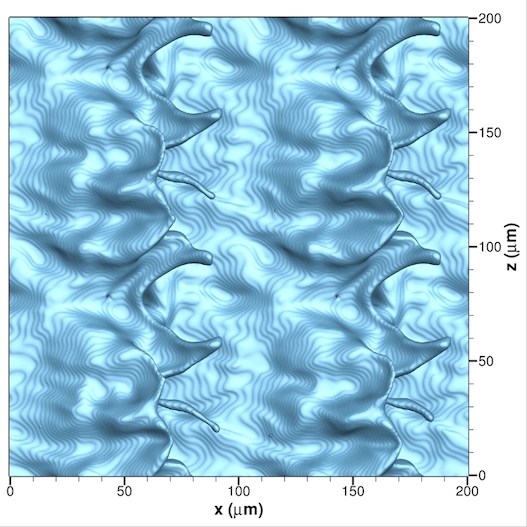}
  \caption{}
\end{subfigure}%
\caption{Incompressible planar liquid jet injection (b). Surface deformation comparison at \(t=50\) \(\mu\)s between a numerical approach based on the Level-Set (LS) method~\cite{zandian2018understanding} and the present numerical approach based on the Volume-of-Fluid (VOF) method. The LS method identifies the liquid surface as the iso-surface with \(\theta=0\) and the VOF method identifies the liquid surface as the iso-surface with \(C=0.5\). (a) 3D view (LS); (b) 3D view (VOF); (c) top view (LS); and (d) top view (VOF).}
\label{fig:PM_val4e}
\end{figure}

Figures~\ref{fig:PM_val4c} and~\ref{fig:PM_val4e} show a comparison of the two approaches at different times. Both methods predict nearly identical surface deformation at 30 \(\mu\)s, which validates the implemented interface capturing approach and momentum solver against a previously validated code that has been used in numerous works~\cite{jarrahbashi2014vorticity,jarrahbashi2016early,zandian2017planar,zandian2018understanding,zandian2019length,zandian2019vorticity}. There are minor differences that become apparent later in time (i.e., 50 \(\mu\)s).
That is, the current code keeps a sharper interface and employs a finer mesh than the results obtained by Zandian et al.
~\cite{zandian2018understanding}. As a result, it can resolve smaller structures and high-curvature regions more effectively.
Note that once the surface has deformed significantly, the symmetry boundary condition may affect the comparison between the two simulations. Additionally, the VOF results display what appears to be surface wrinkles with a very short wavelength. These wrinkles originate from the plotting software when representing the iso-surface with \(C=0.5\) of a non-smooth data set and do not exist in the actual computation where the proper interface reconstruction algorithms are used (i.e., PLIC). \par 

\clearpage

\clearpage

\bibliography{journal_bib}

\end{document}